\documentclass[11pt]{article}
\usepackage[margin=0.9in]{geometry}
\usepackage{times}
\usepackage{setspace}
\usepackage{amsfonts, amssymb, bm, amsmath, amsthm, mathtools}
\usepackage{xcolor}
\usepackage{subfig}
\usepackage{enumerate}
\usepackage{booktabs}

\usepackage{natbib}
\usepackage[inline]{enumitem}
\usepackage{makecell}

\usepackage[color=orange!20]{todonotes} \presetkeys{todonotes}{inline}{}

\usepackage{tikz}
\usetikzlibrary{positioning,chains,fit,shapes,calc}
\tikzset{fit margins/.style={/tikz/afit/.cd,#1,
    /tikz/.cd,
    inner xsep=\pgfkeysvalueof{/tikz/afit/left}+\pgfkeysvalueof{/tikz/afit/right},
    inner ysep=\pgfkeysvalueof{/tikz/afit/top}+\pgfkeysvalueof{/tikz/afit/bottom},
    xshift=-\pgfkeysvalueof{/tikz/afit/left}+\pgfkeysvalueof{/tikz/afit/right},
    yshift=-\pgfkeysvalueof{/tikz/afit/bottom}+\pgfkeysvalueof{/tikz/afit/top}},
    afit/.cd,left/.initial=2pt,right/.initial=2pt,bottom/.initial=2pt,top/.initial=2pt}
    
\usepackage{float}

\theoremstyle{plain}
\newtheorem{theorem}{Theorem}
\newtheorem{proposition}{Proposition}
\newtheorem{corollary}{Corollary}

\theoremstyle{definition}
\newtheorem{definition}{Definition}
\newtheorem{assumption}{Assumption}

\newtheorem{remark}{Remark}
\newtheorem{example}{Example}
\newtheorem{specialcase}{Special Case}

\renewenvironment{proof}[1][\proofname]{{\noindent \bfseries{\textit{#1.}}}}{\qed}

\usepackage[citecolor=blue,colorlinks=TRUE,linkcolor=blue]{hyperref}
\usepackage[capitalize,noabbrev]{cleveref}
\crefformat{assumption}{Assumption #1}
\crefformat{specialcase}{Special Case #1}

\usepackage{titlesec}
\titleformat*{\section}{\large\bfseries}
\titleformat*{\subsection}{\normalsize\bfseries}
\titleformat{\subsubsection}[runin]
  {\normalfont\normalsize\bfseries}{\thesubsubsection}{1em}{}
\titleformat*{\paragraph}{\normalsize\bfseries}
\titleformat*{\subparagraph}{\normalsize\bfseries}

\title{Causal Inference  when  Intervention Units \\ and Outcome Units Differ
}
\author{Georgia Papadogeorgou${}^1$  \hspace{0.5cm}  Zhaoyan Song${}^2$  \\ Guido Imbens${}^3$ \hspace{0.5cm} 
Fabrizia Mealli${}^4$  \\
\small ${}^1$University of Florida, gpapadogeorgou@ufl.edu, ${}^2$University of Florida, zhaoyan.song@ufl.edu, ${}^3$Stanford \\ 
\small University, imbens@stanford.edu, ${}^4$European University Institute, University of Florence, fabrizia.mealli@eui.eu
}

\date{\vspace{.8 cm} }
\newcommand{\E}{\mathbb{E}}
\newcommand{\Var}{\mathbb{V}}
\newcommand{\commentG}[1]{{\color{teal} GP: #1}}
\newcommand{\commentGWI}[1]{{\color{orange} \noindent GI: #1}}

\newcommand{\calg}{{\cal G}}
\newcommand{\caln}{{\cal N}}
\newcommand{\calm}{{\cal M}}
\newcommand{\calw}{{\cal W}}

\newcommand\numberthis{\addtocounter{equation}{1}\tag{\theequation}}
\newcommand{\iteg}{{\it e.g.}}

\allowdisplaybreaks

\begin{document}

%\definecolor{myblue}{RGB}{0,29,100}
%\definecolor{mygreen}{RGB}{80,160,80}

\definecolor{myblue}{rgb}{1.0, 0.13, 0.32}
\definecolor{mygreen}{rgb}{0, 0.5, 1}

\maketitle

\vspace{-50pt}

\begin{abstract}
    
We study causal inference in settings characterized by interference with a \emph{bipartite} structure. There are two
distinct sets of units: \emph{intervention} units to which an intervention can be applied and \emph{outcome} units on which the outcome of interest can be measured.  Outcome units may be affected by interventions on some, but not all, intervention units, as captured by a bipartite graph. Examples of this setting can be found  in analyses of the impact of pollution abatement in plants on health outcomes for individuals,  or the effect of transportation network expansions on regional economic activity. We introduce and discuss a variety of old and new causal estimands for these bipartite settings.   We do not impose restrictions on the functional form of the exposure mapping and the potential outcomes, thus allowing for heterogeneity, non-linearity, non-additivity, and potential interactions in treatment effects.  We propose unbiased weighting estimators for these  estimands from a design-based perspective, based on  the knowledge of the bipartite network under general experimental designs.  We derive their variance and prove consistency  for increasing number of outcome units. Using the Chinese high-speed rail construction study, analyzed in \cite{borusyakhull2023}, we discuss non-trivial positivity violations that depend on the estimands, the adopted experimental design, and the structure of the bipartite graph. 
%We also show how some estimands that contrast average potential outcomes under specific (stochastic) interventions can be estimated more precisely when ``closer" to the actual treatment allocation.
\end{abstract}

%\listoftodos

% \clearpage

\section{Introduction}

Methods to evaluate policies and interventions are often challenged by connections between units which may interact in clusters ({\emph{e.g.},} schools) or across networks ({\emph{e.g.},} social networks). This raises the problem of interference or spillovers, where a unit's outcome might depend not only on their treatment but also on the treatments of other units \citep[{\emph{e.g.},}][]{sobel2006randomized, hudgens_toward_2008, forastiere_identification_2021}.

This paper studies
estimation of causal effects in experimental settings with interference characterized by a distinction between the population of units to which the intervention is applied, termed \textit{intervention} units, and the population of units on which the outcomes are measured, termed \textit{outcome} units. Following \cite{zigler2021bipartite} we refer to these as {\it bipartite interference} settings.
Examples of the bipartite interference setting can be found across many disciplines; they arise  in social sciences because networks are often bipartite, with edges connecting exclusively nodes of different types \citep{LoOlivellaImai2023}. The bipartite setting can be thought of as generalizing the standard causal inference under interference setup where the two populations are identical, \emph{e.g.}, we assign individuals to health interventions (say, vaccines) and measure results (say, disease contraction) on the same set of units \citep{imbens2015causal}.
The bipartite setting also generalizes  hierarchical structures of observational units, where typically interventions are applied to outcome units (say, students or households) and those are hierarchically clustered into a second type of units (say, classes or villages), which form the basis for clustered experiments \citep[{\emph{e.g.},}][]{hudgens_toward_2008, papadogeorgou2019causal}. Specific examples of bipartite interference include the following:

\begin{itemize}[leftmargin=0pt]
    \item[]\textbf{Air pollution epidemiology:}
In air pollution epidemiology, the intervention units could be pollution emitters such as power plants which are possible to install a filter on their smokestack or not, and the outcome units could correspond to geographical units affected by air pollution. In \cite{zigler2023air}, the structure of bipartite interference is dictated by atmospheric processes governing how pollution emitted from a particular source is transformed and transported across space: interventions at a particular pollution source may impact air quality and health at different  locations and air quality and health at any given location may be  impacted by interventions at several pollution sources; \cite{chen2024differenceindifferencesbipartitenetworkinterference} extends the analysis to a longitudinal setting.

 \item[]\textbf{Waste disposal facility locations:}
In  analyses of the effects of waste disposal facilities, the value of houses at different locations (outcome units) may be affected by cleaning
up nearby contaminated hazardous-waste disposal facilities (intervention units)  \citep{stock1989nonparametric}.

 \item[]\textbf{Education economics:} In education economics, the intervention units could be neighborhoods which may be exposed to the opening of charter schools, and the outcome units could be  nearby traditional public schools 
\citep{Crema2022}, whose enrollment may be affected by the opening of charter schools in surrounding neighborhoods.

\item[]\textbf{Transportation literature:} In the transportation literature, the intervention units could be segments of an inter-regional railroad network, on some of which new railway lines are opened; the outcome of interest could be the commute time for residents or employment growth in an area \citep{borusyakhull2023};
Relatedly, the interventional units could be locations of a newly-constructed train line, and the outcome units businesses that could be affected by the construction or operation of the train \citep{grossi2020synthetic, grossi2025spatial}.

  \item[]\textbf{Public health:}
The presence of a supermarket or grocery store in an area (intervention unit) might affect cardiovascular health of the surrounding population of potential customers (outcome units) \citep{kelli2019living,Schnell2020mitigating};
 \item[]\textbf{Forest conservation:} In a forest conservation experiment, the intervention units are villages that can host a conservation intervention, for example forest-owning households in treated villages receive an annual payments  if they conserved their forest (treatment); outcomes (change in land covered by trees) are measured in geographical areas in, around and at long distance from villages \citep{Jayachandranetal2017};

\item[]\textbf{E-commerce:}
In online marketplaces, like  Airbnb, intervention units can be listings offering a price reduction or a specific booking feature, outcome units can be potential renters, and the outcome might be whether they rented on that marketplace \citep{bajari2023experimental}. 
\end{itemize}

The examples above are all observational. Following a long tradition in causal inference,  this manuscript studies  estimation of causal estimands in experimental settings with bipartite interference. These results then motivate the analyses of observational studies such as these examples, where we can identify the distinctive features of the bipartite interference settings we consider in this paper:
\begin{enumerate*}[label=(\roman*)]
\item there are two distinct populations of units: the first of which contains units that are potentially exposed the intervention, and the second of which contains units on which the outcomes are measured, 
\item these two populations of units may have a different nature, and they do not typically (although they may) represent a partition of a population of units into two sets;
 
\item an outcome unit's outcome is possibly affected by the treatment level of multiple intervention units;
\item researchers know the bipartite graph, that is, they know for each outcome unit the set of intervention units that may impact the outcome of that outcome unit.

%\item the population  of intervention units cannot be clustered of outcome units and clusters of intervention units so that outcome units are only affected by the .
\end{enumerate*}

In this paper, we are agnostic about potential exposure mappings \citep{aronow2017estimating}: we allow bipartite interference to be general and not restricted to partial (clustered) interference. Hence, our work imposes minimal limitations on the potential outcomes.
We focus on settings where the presence of links in the bipartite graph is known, although we do not need to know the strength of the link between an intervention and an outcome unit. 
This is also the starting point in \cite{lu2025design}; however they focus on a single estimand, the total treatment effect that we name \emph{all-or-none} effect below, and derive estimation and inference results under the specific design of Bernoulli randomization only.
Instead, we first discuss Fisher randomization testing for sharp null hypotheses,  which has not yet been rigorously discussed in bipartite settings.
Then, we establish \emph{new} causal estimands for bipartite settings under a variety of policies that may be of interest.
We start from the {\it all-or-none} estimand,  a causal estimand  that contrasts what would happen on average across the outcome units if all intervention units where switched from control to treatment. 
A related causal estimand contrasts the average outcome of outcome units under a universal treatment, against the status quo. This estimand, which we refer to as the {\it all-vs-status quo}, is closely related to the Average Effect on the Controls in the causal inference literature without interference. We also introduce the symmetric {\it status quo-vs-none} estimand which is closely related to the Average Effect on the Treated units and contrasts the average outcome of outcome units under the status quo against treating none of the intervention units.  We propose estimation from a design-based perspective where the uncertainty comes from the stochastic nature of the interventions. We introduce inverse-propensity score-weighting  (IPW) estimators for the {\it all-or-none}, {\it all-vs-status quo} and  {\it status quo-vs-none} estimands and prove unbiasedness, derive the design-based variance for \emph{general} designs, and prove consistency.

The all-or-none and the status quo effects can be challenging ones to estimate, because they may be quite far away from the current allocation. This shows up in large variances associated with their estimators. Therefore, we introduce new policy-relevant estimands. Some are defined with respect to stochastic interventions \citep{diaz2012population, papadogeorgou2022causal} over the treatment assignment of the intervention and outcome units.
Other estimands we introduce  describe the  effects of switching one (or $k$ or a fraction)  additional randomly drawn control intervention unit(s) to treatment. 
Because these policies move units by relatively small amounts of exposure, the variances for the associated estimators are often smaller.
Unbiased weighting estimators for these estimands are proposed.  We derive expression of their variance and prove consistency under general designs and specific asymptotic regimes for the intervention and the outcome units.

We employ our randomization testing framework and estimate our proposed estimands in the context of  the Chinese high-speed rail construction study, analyzed in \cite{borusyakhull2023}. 

%\commentFM{We will then comment on main results, positivity etc.}

\subsection{Related work on interference and bipartite interference}

Interference is a violation of the Stable Unit Treatment Value assumption, SUTVA \citep{rubin1980randomization}.
Most of the existing work on assessing causal effects under interference is formalized in the context of one set of observational units on which an intervention is applied and outcomes are measured and the treatment of one unit can affect the outcome of others \citep{sobel2006randomized, halloran_causal_1995, Rosenbaum2007, hudgens_toward_2008}.
Contributions in causal inference under interference range from randomized experiments under partial interference \citep{hudgens_toward_2008, baird2018optimal, basse2018analyzing, bassefellertoulis} to observational studies on networks where interference is not restricted to disjoint groups \citep{liu_inverse_2016, forastiere_identification_2021, forastiere_estimating_2022, ogburn_causal_2022, papadogeorgou2022causal}, and combinations of these features \citep{athey2018exact,EcklesKarrerUgander+2017, aronow2017estimating,BasseAiroldi2018, Savje2021, Leung2022, LiWager2022, chattopadhyay2023design, Viviano2022, wang2023design}.
 % GP: Commented the following out -- not necessary -- added a couple citations elsewhere.
% The spatial interference literature addresses a setting in which units are locations in space, and treatment effects can emanate across space in a manner that is often dictated by spatial proximity \citep{verbitsky-savitz_causal_2012,giffin_generalized_2020, zirkle_addressing_2021, papadogeorgou2022causal, wang2023design, pollmann2020causal}.
Fisher's exact randomization approach \citep{Fisher1935}  has been adapted and extended to study interference specifically (see, \iteg, \cite{Rosenbaum2007,aronow2012,luo_inference_2012,bowers2013reasoning,choi2017networkexp,athey2018exact,basse2019randomization,puelzbassefellertoulis2022}).

Causal inference with interference often employs {\it exposure mappings} \citep{aronow2017estimating, Savje2024} for defining or estimating causal effects, or both. Exposure mappings are functions that describe the effective treatment that the intervention on everyone implies on each unit \citep{manski2013identification}. Their origin can be found in \cite{halloran_causal_1995} and \cite{sobel2006randomized} where exposures describe the proportion of treated units in disjoint clusters. \cite{aronow2017estimating} and  \cite{forastiere_identification_2021} define exposures based on arbitrary summaries of the nominal treatments received by the single units.
%The idea of exposure mapping has facilitated the development of inferential tools because it essentially implies that a unit's potential outcomes can be indexed only using its exposure mapping values. 

Bipartite interference was formally introduced in \cite{zigler2021bipartite}, where inverse probability weighted estimators for some bipartite estimands are introduced under a bipartite version of partial interference. Partial interference is also assumed in \cite{chattopadhyay2023design} to define estimands and investigate design-based methods for estimation and inference. The bipartite interference literature has also largely employed exposure mapping assumptions for defining and estimating causal effects, and exposure mappings are used, often implicitly, in many empirical studies where a model at the outcome units' level is specified as a function of a specific combination of the treatments applied to intervention units. Employing an exposure mapping assumption, \cite{zigler2023air} extend prior work to bipartite network settings without clusters. 
Under a linear exposure assumption, \cite{Harshawetal2023} introduce the Exposure Reweighted Linear estimator for the causal effect of treating all versus none of the intervention units, and study its properties.
\cite{pouget-abadie_variance_2019} and \cite{Brennanetal2022} discuss experimental designs in bipartite interference settings that improve efficiency of estimators and reduce bias due to interference, respectively.
\cite{doudchenko_causal_2020} focus on exposure mappings that are a function of the treatment assignment and the bipartite graph.
\cite{song2024bipartite} introduce a framework for bipartite interference with time series data.

As noted in \cite{Savje2024}, most of the methods that rely on exposure mapping require to use the same exposure mappings to both define  the estimands of interest and to restrict the interference structure. The two roles of exposure mappings should be separated: exposures can be used  to define effects without necessarily assuming that they are capturing the complete causal structure in the experiment, which could otherwise be subject to misspecification.
As highlighted in \cite{aronow2017estimating} and in \cite{borusyakhull2023}, where the related concepts of \emph{formula treatments}  are introduced, exposure mappings are typically function of the assignment vector as well as of unit-specific attributes (\iteg, some units' characteristics or entries in a network adjacency matrix). Typically only the first ones are under experimental control (when, for example, treatments are randomized or shocks are exogenous), but the second ones may compromise the ignorability of the \emph{effective treatments}.
This has several consequences. First, simple na\"ive solutions, such as including exposure as a regressor in a regression model, lead to biased estimates of causal effects, due to essentially omitted variable bias \citep{borusyakhull2023}. Several solutions to this problems have been proposed which include inverse probability  weighting estimators \citep{aronow2017estimating, papadogeorgou2019causal}, generalized propensity score methods \citep{forastiere_identification_2021}, recentering-type regression adjustments \citep{borusyakhull2023, borusyak2025design}, control function approaches \citep{wooldridge2015control}. See also the literature on Bartik or shift-share instruments \citep{goldsmith2020bartik, borusyak2025practical}.
Second, all these estimators are valid under the assumption that the exposure mapping is correctly specified. Correct specification of exposure mapping requires correct assumptions on the interference structure. 
%An exposure mapping is correctly specified if a  modified SUTVA assumption holds: exposure mapping captures all causal information pertaining to the nominal treatments; two nominal treatments (assignment vectors) leading to the same level of the exposure mapping are associated to the same potential outcome.  
Third, sources of confounding are related to both decisions taken at the level of the nominal treatments for intervention units and to features of the network structure \citep{doudchenko_causal_2020, song2024bipartite}. This implies that ignorability type assumptions at the level of the exposure mapping  for outcome units may not be trivial  \citep[see discussion in][]{zigler2023air}.
The only work that, like ours here, does not rely on modelling the interference structure through strong restrictions on the potential outcomes in bipartite settings is \cite{lu2025design}; like us, they take a design-based approach, but focus only on the all-or-none estimand under Bernoulli experimental designs.

\section{Setup}

\subsection{Two Populations and Potential Outcomes}

The setting we consider is characterized by the presence of two populations. The first population is composed of $N$ \textit{intervention} units (also referred to as  \textit{diversion} units in the literature \citep{Harshawetal2023}), $\caln = \{1, 2, \dots, N\}$, where unit  $n \in \caln$ is eligible to receive a treatment.
For most of the discussion we focus on the case with a binary treatment, $W_n\in{\cal W}=\{0,1\}$,
with $W_n = 1$ denoting that intervention unit $n$ receives the active treatment level, and $W_n = 0$  denoting that it receives the control level.
% GP: Taking this footnote out because it doesnt seem that important and I cant figure out how to double-space it!
%\footnote{ Many of the insights and results carry over directly to the discrete treatment case.}   
$\mathbf{W}\in {\cal W}^N$ denotes the $N$-vector of treatment assignments.
We denote the set of indices for the treated units as $\caln^t = \{n \in \caln: W_n = 1\}$ and the set of indices for the control units as $\caln^c = \caln \setminus \caln^t$.

The second population is composed of $M$ \textit{outcome} units, $\mathcal{M} = \{1, 2, \dots, M\}$. Outcomes for these outcome units may be affected by the treatments applied to the intervention units.\footnote{As a trick to remember notation: we use variations of the letter \textit{n} for i\textit{\textbf{n}}terventional units, and variations of the letter \textit{m} for outco\textit{\textbf{m}}e units. We use calligraphic notation for sets, capital notation for the number of units in a set, and lower-case notation for a unit.}
At the most  general level an outcome unit $m$ could be influenced by the treatment level assigned to all $N$ intervention units. Then, the outcome unit $m$ has $2^N$ potential outcomes $Y_m(\bm w) \in \mathbb{R}$, for $\bm w \in {\cal W}^N$. The potential outcomes for all outcome units under the $N$-component treatment vector $\bm w$ is denoted by the $M$-component vector $\bm Y(\bm w) = (Y_1(\bm w), Y_2(\bm w), \dots, Y_M(\bm w)) \in \mathbb{R}^M$.

\subsection{Restricting potential outcomes according to a bipartite graph}

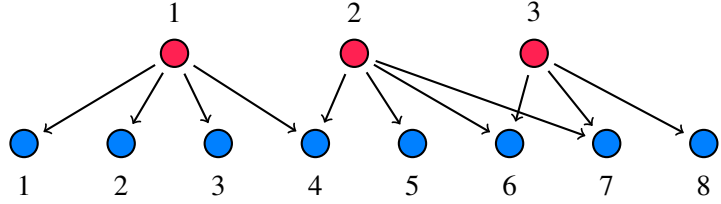
\begin{figure}[!t]
\centering

\begin{tikzpicture}[thick,
  every node/.style={draw,circle},
  fsnode/.style={fill=myblue},
  ssnode/.style={fill=mygreen},
  every fit/.style={rectangle,draw},
  -,shorten >= 3pt,shorten <= 3pt
]

% the vertices of U
\begin{scope}[start chain=going right,node distance=20mm]
\foreach \i in {1,2,3}
  \node[fsnode,on chain] (f\i) [label=above: \i] {};
\end{scope}

% the vertices of V
\begin{scope}[xshift=-2cm,yshift=-1.2cm,start chain=going right,node distance=9mm]
\foreach \i in {1,2,...,8}
  \node[ssnode,on chain] (s\i) [label=below: \i] {};
\end{scope}

% the set U
%\node [myblue,fit=(f1) (f1),label=above:Interventional Units] {};
% the set V
%\node [mygreen,fit=(s1) (s15),label=below:Outcome units] {};

% the edges
\foreach \i in {1,2,3,4} \draw[->] (f1) edge (s\i);
\foreach \i in {4,5,6,7} \draw[->] (f2) edge (s\i);
\foreach \i in {6,7,8} \draw[->] (f3) edge (s\i);
\end{tikzpicture}

\caption{General bipartite interference graph with 3 interventional units and 8 outcome units.}
\label{fig:full_graph}

\end{figure}

We focus on the setting where the dependence of the potential outcomes on the treatments of the intervention units is restricted through a bipartite graph. 
We assume there exists a bipartite graph $\calg$ with two sets of nodes, the intervention units $\caln$ and the outcome units $\calm$, and edges which can only exist between an intervention and an outcome unit. Then, this graph can be described by $\calg = \{ \caln, \calm, A\}$ where $A$ is an $N \times M$ binary matrix, $A_{nm} = 1 $ if nodes $n, m$ are connected with respect to $\calg$, and $A_{nm} = 0$ otherwise. We consider  the graph $\calg$  a fixed quantity that characterizes the populations. An example is given in \cref{fig:full_graph}. We do not assume that we know the strength of the connection between nodes, often described through a weighted adjacency matrix, but only the existence of a connection between an outcome and an intervention unit (\iteg, \cite{zigler2023air, pouget-abadie_variance_2019}).

%\guido{do we want to say that the bipartite graph is known, but that a 1 indicates some potential effect, and that this therefore allows for weighted graphs? Not sure I like the terminology ``partiall known'' that suggests there are edges we dont know.}

Next, we introduce two concepts that help organize the bipartite graphs we study. First, we define the {\it intervention set} of an outcome unit. 

\begin{definition}{\sc (Intervention Set for Outcome Unit $m$)}
Let $\caln_m = \{n \in \caln: A_{nm} = 1\}$ be the intervention set for unit $m$, that is, the set of indices of interventional units that are connected to outcome unit $m$ through $\cal G$. Let also $N_m$ denote the size of the intervention set of outcome unit $m$, as $N_m = |\caln_m|$.
\label{def:interventional_set}
\end{definition}
Given an intervention set $\caln_m$, let  $\bm w_{\caln_m}$ denote the subvector
of the treatment vector $\bm w \in \calw^N$  
containing only the elements of $\bm w$ corresponding to the indices in $\caln_m$. 
As an example, in \cref{fig:full_graph} the intervention set for outcome unit $3$ is $\{1\}$ with $\bm w_{\caln_3} = w_1$, and the intervention set for outcome unit $4$ is $\{1,2\}$ with $\bm w_{\caln_4} = (w_1, w_2)$.

As a mirror to the intervention set in \cref{def:interventional_set}, we define the {\it outcome set} of an intervention unit.

\begin{definition}{\sc (Outcome Set for Intervention Unit $n$)}
The outcome set for intervention unit $n$, denoted by
 $\calm_n = \{m \in \calm: A_{nm} = 1\}$,  is the set of indices of outcome units that are connected to intervention unit $n$ through $\calg$. 
\label{def:outcome_set}
\end{definition}

In \cref{fig:full_graph} the outcome set  for intervention  unit $1$ is $\{1,2,3,4\}$, and the outcome set for unit $3$ is $\{6, 7, 8\}$.

% -- GP: I commented the following out because we do not use this concept anywhere and it feels unnecessary:
%
% Following \cite{zigler2021bipartite}, two outcome units ($m, m'$) are said to be subject to interference if they have overlapping  intervention sets, that is, if an intervention unit $n$ exists such that $A_{nm}=A_{nm'}= 1$; analogously  two intervention units ($n, n'$) are said to be subject to interfere if they have overlapping outcome sets, that is, if an outcome unit $m$ exists such that $A_{nm}=A_{n'm}= 1$.\footnote{In observational settings, these features of intervention and outcome units are important when formalizing the  different types of confounding. See discussion in \cite{zigler2023air}.}

Next, we define a {\it consistent experience} for an outcome unit $m$ under two treatment vectors, if the intervention units with which it is connected receive the same treatment under the two assignments.

\begin{definition}{\sc (Consistent Experience for Outcome Unit $m$)}
Given assignment vectors $\bm w, \bm w'$, an outcome unit $m$ has a consistent experience if $w_n=w_n'$ for all $n$ such that $A_{nm}=1$; that is, if $\bm w_{\caln_m} = \bm w'_{\caln_m}$.%
\label{def:constant_experience}
\end{definition}

The definition of a consistent experience in \cref{def:constant_experience} plays a crucial role in reducing the dimensionality of the potential outcomes, by restricting the potential outcomes according to the bipartite graph, as in the following statement.

\begin{assumption}[{\sc B-SUTVA: Bipartite Stable Unit Treatment Value Assumption}]
% The potential outcomes for outcome unit $m$ agree with the bipartite graph $\calg$: 
If the assignment vectors $\bm w, \bm w' \in \{0, 1\}^N$ provide unit $m$ a consistent experience, it holds that $Y_m(\bm w) = Y_m(\bm w')$. Then, potential outcomes for unit $m$ can be denoted using only the treatment level of its intervention set as $Y_m(\bm w_{\caln_m}$).
\label{ass:b-sutva}
\end{assumption}

\noindent

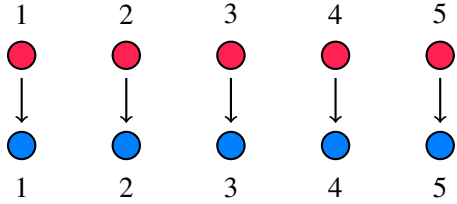
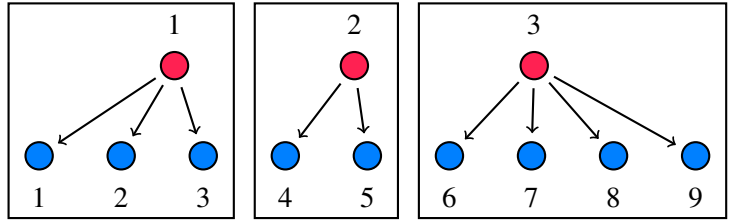
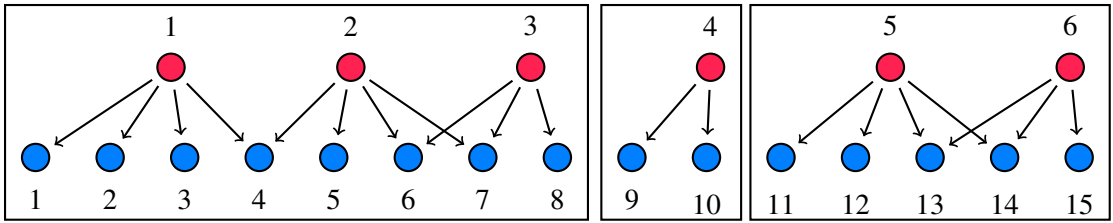
\begin{figure}[!t]
\centering

\subfloat[][\cref{case:iid}: The same units receive the treatment and experience the outcome, and a unit's outcome can be driven by their own treatment only.]{
\makebox[0.9\textwidth][c]{
\begin{tikzpicture}[thick,
  every node/.style={draw,circle},
  fsnode/.style={fill=myblue},
  ssnode/.style={fill=mygreen},
  every fit/.style={rectangle,draw},
  -,shorten >= 3pt,shorten <= 3pt
]

\begin{scope}[start chain=going right,node distance=10mm]
\foreach \i in {1,2,...,5}
  \node[fsnode,on chain] (f\i) [label=above: \i] {};
\end{scope}
\begin{scope}[yshift=-1.2cm,start chain=going right,node distance=10mm]
\foreach \i in {1,2,...,5}
  \node[ssnode,on chain] (s\i) [label=below: \i] {};
\end{scope}

\foreach \i in {1,2,...,5} \draw[->] (f\i) edge (s\i);

\end{tikzpicture}
\label{fig:case_iid}
}}
\vspace{10pt}

\subfloat[][\cref{case:1_cluster}: Each outcome unit can be affected by a single intervention unit, but the same intervention unit can affect multiple outcome units.]{
\makebox[0.9\textwidth][c]{
\begin{tikzpicture}[thick,
  every node/.style={draw,circle},
  fsnode/.style={fill=myblue},
  ssnode/.style={fill=mygreen},
  every fit/.style={rectangle,draw},
  -,shorten >= 3pt,shorten <= 3pt
]

% the vertices of U
\begin{scope}[start chain=going right,node distance=20mm]
\foreach \i in {1,2,3}
  \node[fsnode,on chain] (f\i) [label=above: \i] {};
\end{scope}

% the vertices of V
\begin{scope}[xshift=-1.8cm,yshift=-1.2cm,start chain=going right,node distance=7mm]
\foreach \i in {1,2,...,9}
  \node[ssnode,on chain] (s\i) [label=below: \i] {};
\end{scope}

% the set U
%\node [myblue,fit=(f1) (f1),label=above:Interventional Units] {};
% the set V
%\node [mygreen,fit=(s1) (s15),label=below:Outcome units] {};

% the edges
\foreach \i in {1,2,3} \draw[->] (f1) edge (s\i);
\foreach \i in {4,5} \draw[->] (f2) edge (s\i);
\foreach \i in {6,7,8,9} \draw[->] (f3) edge (s\i);

\node[draw,fit margins={left=3pt,right=3pt,bottom=9pt,top=9pt},fit=(f1) (s1) (s3)] (c1) {};
\node[draw,fit margins={left=3pt,right=3pt,bottom=9pt,top=9pt},fit=(f2) (s4) (s5)] (c2)  {};
\node[draw,fit margins={left=3pt,right=3pt,bottom=9pt,top=9pt},fit=(f3) (s6) (s9)] (c3) {};

\end{tikzpicture}
}
\label{fig:case_1_cluster}
}
\vspace{10pt}

\subfloat[][\cref{case:partial_graph}: Partial bipartite interference network with 6 interventional units and 15 outcome units organized in three clusters]{
\makebox[0.9\textwidth][c]{
\begin{tikzpicture}[thick,
  every node/.style={draw,circle},
  fsnode/.style={fill=myblue},
  ssnode/.style={fill=mygreen},
  every fit/.style={rectangle,draw},
  -,shorten >= 3pt,shorten <= 3pt
]

% the vertices of U
\begin{scope}[start chain=going right,node distance=20mm]
\foreach \i in {1,2,...,6}
  \node[fsnode,on chain] (f\i) [label=above: \i] {};
\end{scope}

% the vertices of V
\begin{scope}[xshift=-1.8cm,yshift=-1.2cm,start chain=going right,node distance=6mm]
\foreach \i in {1,2,...,15}
  \node[ssnode,on chain] (s\i) [label=below: \i] {};
\end{scope}

% the set U
%\node [myblue,fit=(f1) (f1),label=above:Interventional Units] {};
% the set V
%\node [mygreen,fit=(s1) (s15),label=below:Outcome units] {};

% the edges
\foreach \i in {1,2,3,4} \draw[->] (f1) edge (s\i);
\foreach \i in {4,5,6,7} \draw[->] (f2) edge (s\i);
\foreach \i in {6,7,8} \draw[->] (f3) edge (s\i);
\foreach \i in {9,10} \draw[->] (f4) edge (s\i);
\foreach \i in {11,12,13,14} \draw[->] (f5) edge (s\i);
\foreach \i in {13,14,15} \draw[->] (f6) edge (s\i);

% \node[draw,fit margins={left=3pt,right=3pt,bottom=9pt,top=9pt},fit=(f1) (f3) (s1) (s8)] (c1) [label={[font=\small, label distance = -5mm]below:{Cluster 1}}] {};
% \node[draw,fit margins={left=3pt,right=3pt,bottom=9pt,top=9pt},fit=(f4) (s9) (s10)] (c2) [label={[font=\small, label distance = -5mm]below:{Cluster 2}}] {};
% \node[draw,fit margins={left=3pt,right=3pt,bottom=9pt,top=9pt},fit=(f5) (f6) (s11) (s15)] (c3) [label={[font=\small, label distance = -5mm]below:{Cluster 3}}] {};
\node[draw,fit margins={left=3pt,right=3pt,bottom=9pt,top=9pt},fit=(f1) (f3) (s1) (s8)] (c1) {};
\node[draw,fit margins={left=3pt,right=3pt,bottom=9pt,top=9pt},fit=(f4) (s9) (s10)] (c2)  {};
\node[draw,fit margins={left=3pt,right=3pt,bottom=9pt,top=9pt},fit=(f5) (f6) (s11) (s15)] (c3) {};
\end{tikzpicture}
}
\label{fig:partial_graph}
}

\caption{Special cases of bipartite interference graphs.}
\label{fig:graphs}

\end{figure}

\noindent \cref{ass:b-sutva}, like the original SUTVA \citep{rubin1980randomization}, is a type of exclusion restriction in that some potential outcomes are assumed to be equal. It could be viewed as a very flexible form of exposure mapping in that only some features of the treatment vector (namely the treatment of an outcome unit's intervention set) is relevant for the potential outcome. B-SUTVA is however not very restrictive, as it allows any type of heterogeneity, nonlinearities and interactions for how the intervention units' treatments are affecting the outcomes.
If B-SUTVA holds for every outcome unit, then the population potential outcomes can be denoted as $\bm Y (\bm w) = (Y_1(\bm w_{\caln_1}), Y_2(\bm w_{\caln_2}), \dots, Y_N(\bm w_{\caln_N}))$.
Below we describe some special cases of bipartite graphs, which are in turn depicted in \cref{fig:graphs}.

\begin{specialcase}
Most of the causal literature is concerned with the case where there is a one-to-one relation between intervention units and outcome units, and the treatment of a unit can only affect its own outcome. This case corresponds to $N=M$, and $A$ is the identity matrix. This case is depicted in \cref{fig:case_iid}.
\label{case:iid}
\end{specialcase}

\begin{specialcase}
% [\sc Single Intervention Unit Clustering] Georgia: Why name it? We do not need this name anywhere else and we do not wanna name things that are not important.
There are at least as many outcome units as intervention units, $M \geq N$, and each outcome unit is affected by only one intervention unit, $\sum_n A_{nm}=1$ for all $m$. Therefore, the outcome units can be partitioned in groups with a single corresponding intervention unit. This case is depicted in \cref{fig:case_1_cluster}.
\label{case:1_cluster}
\end{specialcase}

\begin{specialcase}%[\sc Generalized Clustering / Partial bipartite interference] 
We can partition the set of intervention units and the set of outcome units so that each subset of intervention units only affects a corresponding subset of outcome units (depicted in \cref{fig:partial_graph}). So, ${\cal M}=\cup_{r=1}^R {\cal M}_r$, non-overlapping, ${\cal N}=\cup_{r=1}^R {\cal N}_r$, nonoverlapping, and
\[ A_{nm}=1 \Longrightarrow \exists \text{ unique } r\ \  \textrm{s.t. } m\in {\cal M}_r, n \in{\cal N}_r.\]
This is a case of partial bipartite interference.
\label{case:partial_graph}
\end{specialcase}

\noindent In order to illustrate the concepts introduced in this section, we return to some of the examples discussed in the introduction. 

\begin{itemize}[leftmargin=0pt]

\item[]\textbf{Air pollution epidemiology (cont.):}
In \cite{zigler2023air} the intervention units are the $N=472$ coal-burning power plants operating in $2005$ and the outcome units are the   $M = 25,553$ ZIP codes in the United States. 
%where $21,577,552$ Medicare beneficiaries reside.
The binary treatment is the indicator of having (or not)  scrubbers installed during at least half of $2005$.
The outcome is the number of Ischemic Heart Disease (IHD) hospitalizations over Medicare beneficiaries residing in each ZIP code.
The bipartite graph is characterized by a $472 \times 25,553$ {\it weighted} adjacency matrix $A$ \citep{henneman_characterizing_2019,henneman_accountability_2019}, denoting how much air mass originating from each power plant travels to each ZIP code.
% The authors define the key-associated plant for each ZIP code $m$, $n^*_{(m)}$, based on the plant exhibiting the highest influence on ZIP code $m$ during 2005. The authors then assume a bivariate exposure mapping that reflects the notion of direct and indirect effects in a bipartite settings. Specifically, they define a  key-associated treatment, $Z_m = w_{n^*_{(m)}}$, and an upwind treatment, $G_m$, defined as a linear function of the treatment statuses of all power plants but power plant $n^*_{(m)}$, weighted by the elements of the interference mapping: $G_m = \sum_{n \ne n^*_{(m)}} A_{nm}w_n$.  
 
 \item[]\textbf{Economics of housing (cont.):}
In \cite{stock1989waste}, the intervention units are the $N=11$ waste sites in Boston identified before 1982, while the outcome units are $M=324$ single-family homes in the Boston suburbs sold between 1978 and 1981. The treatment  for intervention unit $n$ by the time of the sale of unit $m$, $w_{nm}$, takes on value 1 if the site had been discovered as hazardous by the time of the sale, and $0$ otherwise.\footnote{Note that the vector $\bm w$ in this example changes over time in a staggered way; hence takes on different values for the different houses, depending on the time of the sale.}  The outcome for a home $m$ is the sale's price.  The bipartite graph is characterized by an $11 \times 324$ matrix $A$, including the distances from each house $m$ to each site $n$. 
% The author assumes a bivariate exposure mapping (that he defines as two proxy variables of risk). Those are the two  weighted sum of the treated sites with weights 1) equal to the  inverse of the square of the distance to the site and 2) equal to the size of the site times the  inverse of the square of the distance to the site. An hedonic pricing equation is then specified for the sales price as a (semiparametric) function of the bivariate exposure mapping and some other house  attributes and town fixed effects.

\item[] \textbf{Education economics (cont.):} In \cite{Crema2022}, the intervention units are the $N=115$ school districts in North Carolina, $97$ of which were treated if they experienced an elementary charter school opening between 1997 and 2015.\footnote{The treatment here changes over time in a staggered way as well; hence it takes on different values at different times. The same holds for the outcome variables measured on TPS's. We omit the time index here to simplify description of the example, so this may represent the treatment status and the outcomes at a specific point in time.} The outcome units are the $M=1,250$ elementary traditional public schools (TPS) in North Carolina. The bipartite graph is characterized by an $115 \times 1250$ matrix $A$, with entries $A_{nm}$ equal $1$ if the distance from neighboorhood $n$ to TPS $m$ is less than $5$ miles and $0$ otherwise. The outcome is racial segregation across classrooms in TPS. 
%\guido{do we not know what the total number of intervention units is? A little weird to use $N$ here and the actual number in the other examples.}
% The author implicitly assumes an exposure mapping with \emph{effective} binary treatment for a TPS defined as $G_m=I(\sum_n w_nA_{nm}>0)$. That is, a TPS is exposed to a new charter school opening if at least one charter school opens in a nearby (less than $5$ miles away) neighboorhood - school districts. 

% \item[]\textbf{Transportation literature (cont.):} %In the transportation literature, the intervention units could be segments of the road network that could be extended to include an additional lane, and the outcome of interest could be the commute time for residents of the area \citep{borusyak2020non}.
%Relatedly, the interventional units could be locations of a newly-constructed train line, and the outcome units businesses in the area that could be affected by the construction or operation of the train \citep{grossi2020synthetic}.

 %\item[]\textbf{Public health (cont.):}
%The presence of a supermarket or grocery store in an area (intervention unit) might affect cardiovascular health of the surrounding population of potential customers (outcome units) \citep{kelli2019living,Schnell2020mitigating}.
 %\item[]\textbf{Forest conservation (cont.):} %In a forest conservation experiment, the intervention units are villages that can host a conservation intervention (treatment) or not, and outcomes are measured at the forest level (forest cover outcomes) in and around villages \citep{Jayachandranetal2017}.
\end{itemize}

\section{Randomization tests for the null hypothesis of no effect in bipartite experiments}
\label{sec:randomization_tests}

%Fisher's exact randomization approach (Fisher, 1935) employs sharp null hypotheses that stipulates
%the outcomes of all units under all assignments. The most common such hypothesis is the Fisher's sharp null that assumes treatment has no effect on any unit. As this implicitly assumes that both the primary (direct) effect and spillover effects do not exist, the approach
%tests for the existence of both types of effects simultaneously. The test has recently
%been adapted and extended to study interference specifically (see, \iteg, \cite{Rosenbaum2007,aronow2012,luo_inference_2012,bowers2013reasoning,choi2017networkexp,athey2018exact,basse2019randomization,puelzbassefellertoulis2022}).

We discuss how to assess null hypotheses of no effect. We focus  on the case where assignment is random, with
\( \bm W \sim \pi(\cdot),\)
with $\pi:{\cal W}^N \mapsto [0,1]$ known. The Fisher sharp null hypothesis of no effect can be formalized as follows:
\[ H_0:\ \bm Y(\bm w)=\bm Y(\bm w')\ \forall \ \bm w,\bm w',\]
\noindent and we are interested in testing it against the alternative hypothesis
\[ H_a: \exists \bm w,\bm w'\ \ \textrm{such that } \bm Y(\bm w)\neq \bm Y(\bm w').\]

\noindent Fisher's null of no effect says that every outcome unit would exhibit the same response under any treatment assignment of the intervention units. So, if this hypothesis is true there is no effect from any of the treatments applied to any of the intervention units. Consequently, Fisher's randomization test of no effect has the correct level for any non trivial test statistic. This was also observed for randomized experiments with unit-to-unit interference \citep{Rosenbaum2007}.
%the null hypothesis implicitly assumes that both the primary (direct) effect and spillover effects do not exist, hence the testing procedure tests for the existence of both types of effects simultaneously.
Different test statistics would have varying power based on the underlying truth on the bipartite structure. 

Some remarks pertaining the bipartite settings are necessary. First, there are no obvious test statistics that one can use and therefore no clear way on how randomization tests can be performed in the completely agnostic case, where there is no notion of \emph{individual} treatment for outcome units. This is different from the usual unit-to-unit interference case where, even without specifying the structure of interference, under the sharp null of no effect  \citep{Rosenbaum2007}, simple test statistics can be used based on the individual treatment, \iteg, difference in sample averages of treated and control units. Second, one can design a test statistic if an interference structure can be plausibly assumed, which allows us to link each outcome unit to one or a few intervention units that are hypothesized to be most influential. This essentially amounts to knowing or assuming a bipartite graph and some form of B-SUTVA.
%For example, \cite{zigler2023air} assumed that the biggest power plant that emits over a zip code is the most influential for that zip code.
%Under B-SUTVA, we can design test statistics  at the outcome unit level. 

A test statistic $T$ is a known real value function of the observed vector of assignment $\bm W$, the vector of observed outcomes $\bm Y$, and possibly the pre-treatment variables which include the bipartite matrix $A$.
Test statistics can be considered at the level of the outcome units, or at the level of the intervention units. Focusing on outcome unit test statistics first, the notion of  {\it total experience} and {\it average experience} can be useful:
\begin{definition}{\sc (total and average experience)}
Given an observed assignment vector $\bm W$, the total experience for an outcome unit $m$ is
\[
{W}_m^{tot}=\sum_{n=1}^N A_{nm} W_n,
\]
and the average experience for an outcome unit $m$ is
% \end{definition}

% \begin{definition}{\sc (average experience)}
% Given an observed assignment vector $\bm W$, the average experience for an outcome unit $m$ is
\[
\overline{W}_m=\frac{1}{N_m} \sum_{n=1}^N A_{nm} W_n.
\]
\end{definition}

Natural choices for the statistic 
$T$ are the OLS estimates of the coefficient $\beta$ in one of the following two linear regressions:
\begin{equation}
    Y_m=\alpha+\beta {W}_m^{tot} +\epsilon_m
\quad \quad \text{and} \quad \quad
    Y_m=\alpha+\beta \overline{W}_m +\epsilon_m
\label{eq:random-outcome-level}
\end{equation}
These tests statistic are attractive if the most plausible alternative hypotheses correspond to treatment effects that are additive and no interaction among intervention units or saturation is plausible. 
%More powerful test statistics can be found depending on assumptions about alternative specific deviations from the null. 

We can also work at the intervention unit level. For each intervention unit, $n$, we can define a  summary $f$ (\iteg, average) of the outcomes of units in its outcome set (\cref{def:outcome_set}), and then use some of the usual test statistics under randomization of the intervention units. Let $\bm Y(\bm w)$ be the population potential outcome under treatment vector $\bm w$, and let $\bm Y_{\mathcal{M}_n}(\bm w)$ denote the subvector containing only the elements corresponding to the indices in $\mathcal{M}_n$.
Under Fisher's null all intervention unit responses $Y_n(\bm w)= f(\bm Y_{\mathcal{M}_n}(\bm w))$ will be the same for all assignments $\bm w$, and therefore Fisher's randomization test would have the correct level. %(despite dependence of the responses). 
A simple test statistics could then be $\overline y(1)-\overline y(0)$ where:
 \begin{equation}
\overline y(0) = \frac{1}{N_0} \sum_{n = 1}^N (1-W_n) f(\bm Y_{\calm_n}) \quad \text{and} \quad
\overline y(1) = \frac{1}{N_1} \sum_{n = 1}^N W_n f(\bm Y_{\calm_n}),
\label{eq:y_teststatistics_n}
\end{equation}

\noindent and $N_0$, $N_1$ are the number of control and treated intervention units respectively.

One might think that test statistics at the level of outcome units are more powerful than the ones at the level of intervention units, with the argument that such test statistics are using ``more observations." However, this is not necessarily true. The power of any given test statistic depends on the interference structure of each specific study and the bipartite graph. In fact, test statistic at the outcome unit level may ``dilute" the effects of treatment of intervention units by the fact that multiple (treated and not treated) intervention units are connected with a single outcome unit. Therefore, in certain cases, test statistics at the intervention unit might be more powerful.

 %\fabri{Discuss here about inverting these tests for making confidence statements, maybe along the lines of Rosenbaum 2007? Right now I doubt we can target any meaningful estimands by doing this.}

 %\fabri{F: Consider taking a look and citing this paper and bicliques/conditional randomization tests "A graph-theoretic approach to randomization tests of causal effects under interference" by 
%with Puelz, D., Basse, G., Feller, A., Toulis P.(2022). Journal of the Royal Statistical Society, Series B, 84(1), pp.174-204. NOW CITED}

%\commentG{I think I had put in a comment: Do we have a reason to choose between outcome-based and intervention-based test statistics? I think that outcome-based *might* be more powerful because regression with a categorical predictor has more power than regression with binary predictor? In any case, can we do both in the application, if we haven't done so already?}

\section{Estimands}

%\georgia{
%Currently we define all or nothing effect (section 4.1), all or status quo (section 4.2), effects under stochastic interventions (section 4.3), effect of treating one extra unit (section 4.4).
%The all or nothing effect is a special case of stochastic intervention effects. So estimators and asymptotics can be acquired in a unified manner for these. However, I think that these cannot be acquired in the same manner for the +1 estimand, and I think it's because the {\it intervention} depends on the observed treatment. In any case, as written, things are too long and complicated. Need to somehow unify.}

\noindent We now introduce causal estimands, that is, comparisons of  outcomes under various policies that determine the treatments assigned to the intervention units.

\subsection{The All-or-None Effect}

The simplest policy  is the one where all intervention units are treated. A causal effect of such an intervention could be defined by contrasting the average outcome of outcome units under the all-treated scenario with  the average outcome of outcome units under the scenario where none of the intervention units is treated. This would correspond, in the unit-to-unit interference setting, to the \emph{all-or-none} or \emph{all-or-nothing}  effect \citep{Savje2021, BasseAiroldi2018}, which under the no-interference assumption coincides with the conventional Average Treatment Effect (ATE).

Define the average potential outcome under treatment vector $\bm w$ as
\begin{equation}
\overline Y(\bm w) = \frac{1}{M}\sum_{m = 1}^M Y_m(\bm w).
\label{eq:y_estimand_general_w}
\end{equation}
Then, 
% Consider the average potential outcome across the population when nobody is treated, and when everyone is treated, defined as
% \begin{equation}
% \overline Y(0) = \frac{1}{M} \sum_{m = 1}^M Y_m(\bm 0_N) \quad \text{and} \quad
% \overline Y(1) = \frac{1}{M} \sum_{m = 1}^M Y_m(\bm 1_N),
% \label{eq:y_estimand_all_or_nothing}
% \end{equation}
% respectively, where $\bm 0_N$ is a vector of length $N$ with all 0s, and similarly for $\bm 1_N$. 
we define the \emph{all-or-none} causal effect as the difference in the average potential outcome when everyone is treated versus when nobody is treated as
\begin{equation}
\tau^\textrm{aon} = \overline Y(\bm 1_N) - \overline Y(\bm 0_N) =
\frac{1}{M} \sum_{m = 1}^M Y_m(\bm 1_N) -
\frac{1}{M} \sum_{m = 1}^M Y_m(\bm 0_N),
\label{eq:tau_estimand_all_or_nothing}
\end{equation}
where $\bm 0_N$ is a vector of length $N$ with all 0s, and similarly for $\bm 1_N$.

\subsection{The  Status Quo Effects}

It is often also of interest to study the effect of the realized treatment versus withholding treatment from everyone, defined as
\begin{equation}
\tau^\textrm{sq}_1 = \overline Y(\bm W) - \overline Y(\bm 0_N).
\label{eq:tau_estimand_status_quo}
\end{equation}

\noindent We refer to this as the \emph{status quo-vs-none} effect;
this effect has the Average Treatment Effect on the Treated units (ATT) as counterpart in the single-level setting without interference. It describes what would happen on average to the outcome units if all treated intervention units where not treated. Analogously, we can define the \emph{all-vs-status quo} effect:

\begin{equation}
\tau^\textrm{sq}_0 = \overline Y(\bm 1_N) - \overline Y(\bm W).
\label{eq:tau_estimand_status_quocontrol}
\end{equation}

\noindent
This case is symmetric to the one discussed above: it has the Average Treatment Effect on the Control units (ATC) as counterpart in the single-level setting without interference. It describes what would happen on average to the outcome units if all control intervention units where treated.
From a design-based perspective, both these status quo estimands, $\tau^\textrm{sq}_1, \tau^\textrm{sq}_0$  can be viewed as random estimands \citep{sekhon2020inference}  because they  depend on the random $\bm W$.

% Obviously, the status-quo effect depends on the realized treatment, though we refrain from making this dependence explicit in the definition, and maintain $\tau^{sq}$ as the notation. 

%\guido{GWI: The two estimands above are hard to estimate because they may be quite far away from current allocations. It would be good to think of estimands that are easier to estimate given, say, complete randomization. The discussion in section 6 about alternative estimands is in the right direction. What would be the average effect of increasing the number of treated units by one. This could be tied to the cost of treating units. It is plausible in many cases that the effect of moving from $n$ randomly selected treated units to $n+1$ randomly selected treated units is decreasing in $n$ at least beyond some point. What can we say about this average effect at the current level of treatment, and possibly about its derivative? (the latter may be much harder. But I think the formal results will be a bit different from those for the current estimands, and more different from those in the literature, so rather than leaving this to the end, we should make this more central and discuss the estimands here. We may also be able to say something about the heterogeneity of additional treated units for outcome units with lots of links and for units with few links.}

%This is what is now done 

\subsection{Stochastic intervention effects}
\label{subsec:stochastic_effects}

The effects defined above can be hard to estimate. As we will see in later sections, depending on the experimental design, the necessary positivity assumptions for estimating these quantities will often be violated. Furthermore, estimators of these effects that are assumption-free for the form of the outcome will generally have high variance and they might use only a small proportion of the outcome units if the realized treatment is very different from the treatment assignment specified in the estimand, $\bm 1_N$ and $\bm 0_N$. Here, we discuss alternative effects whose positivity assumptions for estimation can be guaranteed to hold by design, and which might be able to be estimated more efficiently. These effects are defined as contrasts of average potential outcomes under stochastic interventions on the treatment assignment.

We start by defining a stochastic intervention. For outcome unit $m \in \calm$, we hypothesize a distribution $h_{m,\alpha}: \calw^{N_m} \mapsto [0, 1]$ with parameter $\alpha$ over the treatment assignment of its intervention set, $\caln_m$.
We define the average outcome for unit $m$ under this hypothesized assignment on its intervention set, as
\begin{equation*}
\overline Y_{m, h_{m,\alpha}}
= \sum_{\bm w_{\caln_m} \in \calw^{N_m}} Y_m(\bm w_{\caln_m}) h_{m,\alpha}(\bm w_{\caln_m}). 
\label{eq:stochastic_outcome_m}
\end{equation*}
Then, under stochastic interventions for all the outcome units, we define the average outcome in the population for the collective stochastic intervention $h_\alpha = (h_{1, \alpha}, h_{2, \alpha}, \dots, h_{M, \alpha})$ as
\begin{equation}
\overline Y_{h_\alpha} =
\frac1M \sum_{m = 1}^M \overline Y_{m, h_{m,\alpha}}.
\label{eq:stochastic_outcome}
\end{equation}
We define the causal effect of switching the hypothetical distribution from parameter $\alpha$ to $\alpha'$ as
\begin{equation}
\tau(\alpha, \alpha') = \overline Y_{h_{\alpha'}} - \overline Y_{h_\alpha}. 
\label{eq:stochastic_effect}
\end{equation}
The parameter $\alpha$ can be high-dimensional and include information on the outcome unit. 

The definition of the stochastic intervention drives the interpretation of the corresponding estimand.
By varying the choice of $h_{\alpha}$ this class of estimands encompasses a wide variety of causal quantities of interest which are commonly targeted in the literature, such as direct and spillover effects of treatment, or effects under specified exposure mappings. We provide specific examples below.

\begin{specialcase}[Bernoulli intervention]
For a Bernoulli intervention over the outcome unit's intervention set with probability of treatment $\alpha$, we have $h_{m, \alpha} (\bm w_{\caln_m}) = \prod_{n \in \caln_m} \alpha^{w_n} (1 - \alpha)^{1 - w_n}.$ 
Larger values of $\alpha$ imply larger values for the expected number of treated interventional units. This stochastic intervention has been described previously for causal inference with partial interference in the unipartite setting \citep{tchetgen_causal_2012,liu_inverse_2016}.
\label{case:stochastic_Bernoulli}
\end{specialcase}

\begin{specialcase}    [Completely randomized intervention]
Similarly, the stochastic intervention $h_{m, \alpha}$ might describe a completely randomized experiment over the intervention set of outcome unit $m$. In this case, $\alpha$ will describe the number of units that are treated under the intervention. Since the size of the intervention set might vary across outcome units, $\alpha$ might include information pertaining to the number of treated intervention units for each outcome unit. Similar estimands in the unipartite setting with partial interference were described in \cite{hudgens_toward_2008}.
\label{case:stochastic_CR}
\end{specialcase}

\begin{specialcase}[Direct and spillover effects]
In the bipartite interference setting of \cite{zigler2021bipartite}, each outcome unit $m$ has a ``key-associated'' intervention unit that is believed to be the most influential, denoted as $n_m^*$. Consider a stochastic intervention $h_{m, \alpha}$ that fixes the treatment value of $n_m^*$ to $a\in\{0, 1\}$, while the treatment of the rest are Bernoulli draws with a fixed probability $p$, specified as
\( h_{m,\alpha}(\bm w_{\caln_m}) = I(w_{n_m^*} = a) 
\prod_{\substack{n \in \caln_m \\ n \neq n_m^*}} p^{w_n} (1 - p)^{1 - w_n} . \)
Then, causal estimands under these interventions would allow evaluation of direct and indirect effects of the treatment for a change in the treatment of the key-associated intervention unit through $a$, or the others through $p$, respectively. Again, here the parameter $\alpha$ is high-dimensional and includes the information on the key-associated intervention unit for all outcome units.
\label{case:stochastic_keyassociated}
\end{specialcase}

\begin{specialcase}
The all-or-none effect in \cref{eq:tau_estimand_all_or_nothing} can be expressed based on stochastic interventions where $h_{m,\alpha}(\bm 1_{N_m}) = 1$ and 0 otherwise, or $h_{m,\alpha}(\bm 0_{N_m}) = 1$ and 0 otherwise. Therefore, the interventions $h_{\alpha}$ need not necessarily be stochastic, and they can be specified to be point-mass distributions over a specific treatment vector.
\end{specialcase}

The stochastic intervention framework is very general, and it includes estimands under targeted exposures based on exposure mappings \citep{aronow2017estimating, forastiere_identification_2021}, which express the effect that a certain function of the intervention units' treatment assignment has on the outcome of the outcome units. This is discussed in the following special case. 

First, we provide a definition for the design's implied treatment assignment on a set of interventional units.

\begin{definition} \label{def:implied_trt_strategy}
Let $\pi: \mathcal{W}^N \mapsto [0,1]$ describe a distribution over the treatment assignment of all intervention units, and let $\mathcal{S} \subset \caln$ be a subset of size $S$. Then, the implied distribution over the treatment assignment of units in $\mathcal{S}$ is denoted by
$\pi_{\mathcal{S}}: \mathcal{W}^{S} \mapsto [0,1]$. For $\bm a_{\mathcal{S}} = (a_1, a_2, \dots, a_S) \in \mathcal{W}^{S}$, $\pi_{\mathcal{S}}(\bm a_{\mathcal{S}})$ describes the probability of assigning treatment levels $\bm a_{\mathcal{S}}$ over the units in $\mathcal{S}$, and it is defined as
\[
\pi_{\mathcal{S}}(\bm a_{\mathcal{S}}) =
\sum_{\substack{\bm w \in \mathcal{W}^N \\
\bm w_{\mathcal{S}} = \bm a_{\mathcal{S}}}} \pi(\bm w).
\]
\end{definition}

For example, if the treatment assignment $\pi$ is a Bernoulli allocation over the $N$ intervention units, then $\pi_{\mathcal{S}}$ is also a Bernoulli allocation over the units in $\mathcal{S}$ with the same probability of treatment.
A specific choice of $\mathcal{S}$ that will be useful is $\mathcal{S} = \caln_m$, the intervention set for outcome unit $m$. Then, $\pi_{\caln_m}$ denotes the implied intervention strategy over outcome unit $m$'s intervention set.
% and $\pi_{\caln_m}(\bm W_{\caln_m})$ denotes the probability of their realized treatment vector.

Now, we are ready to illustrate how estimands under target 
exposure fall within the stochastic intervention framework.

\begin{specialcase}[Estimands under target exposures]
Specifically, consider an $N \times M$ adjacency matrix $A^*$ over the set of intervention and outcome units (it could be the same as the adjacency matrix $A$ or not).
%, as long as not-connected units under $A$ are also not connected under $A^*$, $A_{nm} = 0 \Rightarrow A_{nm}^* = 0$. 
Suppose that we are interested in the effect of the outcome units' exposure to a function of the intervention units' treatments, which for outcome unit $m$ is defined as $e_m = f_m(A^*, \bm w)$ for some function $f_m$. For example, $f_m(A^*, \bm w) = \sum_n A_{nm}^* w_n / \sum_n A_{nm}^*$ describes the proportion of treated units connected to $m$.

To study this question within the stochastic intervention framework, we can design an appropriate collective stochastic intervention $h_\alpha$. For $m \in \calm$, we define $h_m^*$ as a uniform distribution over the treatment assignment of all intervention units that lead to the targeted exposure. Specifically, set $h_m^*: \calw^N \mapsto [0,1]$ such that $h_m^*(\bm w) = 1 / c_m$, for all $\bm w \in \calw^N$ for which $f_m(A^*, \bm w) = e_m$ where $c_m = | \{\bm w \in \calw^N : f_m(A^*, \bm w) = e_m \}|,$ and 0 otherwise. Then, the stochastic intervention over outcome unit $m$'s intervention set, $h_{m, \alpha}$ is the one implied from $h_m^*$,  according to \cref{def:implied_trt_strategy}.
Note that estimands based on targeted exposures could capture a relevant part of the causal effect of the interventional units' treatment, or not. 

Even though estimands are defined based on an exposure mapping definition, the estimation discussed in \cref{sec:estimation} does not rely on an assumption that the exposure mapping is a correct specification for how potential outcome vary \citep{Savje2024}.
\end{specialcase}

The estimands in all of these special cases can provide interesting and useful insights on how treatment effects manifest and propagate across units. However, they do not necessarily provide an {\it implementable} policy over the whole population of intervention units. The reason is that stochastic interventions across different outcome units can impose different, non-compliant information on the treatment assignment of the complete set of intervention units. As an example, consider outcome units $m, m'$ that share at least one intervention unit, $\caln_m \cap \caln_{m'} \neq \emptyset.$ Assume that a stochastic intervention for $m$ specifies that its intervention set is fully treated, whereas the stochastic intervention on $m'$ specifies that its intervention set is fully under control. Then, there is no implementable policy over the whole set of intervention units that would satisfy both. Even though this is an extreme example, similar issues arise under stochastic interventions that are not point-mass distributions, such as in \cref{case:stochastic_CR} of completely randomized experiments when intervention sets are of different sizes, or in \cref{case:stochastic_keyassociated} in the presence of units with different key-associated intervention units.

The following definition introduces the concept of implementable policies over the set of intervention units.

\begin{definition}[Implementable policies]
A stochastic intervention $h_\alpha = (h_{1, \alpha}, h_{2,\alpha}, \dots, h_{M, \alpha})$ is an implementable policy if there exists distribution $g_\alpha: \calw^N \mapsto [0,1]$ over the treatment assignment of all intervention units whose implied distribution over $\caln_m$ is $h_{m,\alpha}$ for all $m \in \calm$, i.e.,
\[
h_{m, \alpha}(\bm a_{\caln_m}) = \sum_{\substack{\bm w \in \calw^N \\ \bm w_{\caln_m} = \bm a_{\caln_m}}} g_\alpha(\bm w) ,
\quad
\forall m \in \calm, \text{ and } \bm a_{\caln_m} \in \calw^{N_m}.
\]
\end{definition}

For an implementable policy, the average effect over the population of outcome units in \cref{eq:stochastic_outcome} can also be written as
\[
\overline Y_{h_\alpha} 
= \sum_{\bm w \in \calw^N} \Big( \frac1M \sum_{j = 1}^m Y_j(\bm w) \Big) h_\alpha(\bm w) 
=
\sum_{\bm w \in \calw^N} \overline Y(\bm w) h_\alpha(\bm w),
\]
where $\overline Y(\bm w)$ is defined in \cref{eq:y_estimand_general_w}.

\begin{specialcase}[The Bernoulli allocation is an implementable policy]
The Bernoulli allocation in \cref{case:stochastic_Bernoulli} is an implementable policy with $g_\alpha(\bm w) = \prod_{n \in \caln} \alpha^{w_n}(1 - \alpha)^{1 - w_n}$.
\end{specialcase}

\begin{specialcase}[An implementable completely randomized policy]
Even though the stochastic intervention in \cref{case:stochastic_CR} is not an implementable policy in general, we can consider implementable policies under completely randomized designs. Consider a design that specifies that $\alpha$ intervention units are treated by specifying a distribution of the treatment assignment that is equal to $g_\alpha(\bm w) = 1 / {N \choose \alpha}$ when $\sum_n w_n = \alpha$, and 0 otherwise.
Then, $\overline Y_{h_\alpha}$ describes the average potential outcome over outcome units if the treatment of the intervention units was assigned randomly according to $g_\alpha$. Therefore, by setting $\alpha' = \alpha + 1$, the effect $\tau(\alpha, \alpha')$ describes the effect on the outcome units for increasing the number of treated interventional units by 1.
\label{spec_case:implementable_CRD}
\end{specialcase}

The estimand in \cref{spec_case:implementable_CRD} can be informative for the effect of increasing the number of treated units. However, in reality, the treatment is already applied (or rolled out) on the treated intervention units. Since this estimand averages over a treatment assignment distribution, it is not directly informative of what we would expect to observe if one or more of the control units were given the treatment, in addition to the ones already treated. We address this in the following section.

\subsection{The effect of treating an additional number of units}

Researchers might be interested in understanding the effect that treating an {\it additional} intervention unit will have on the outcome units, where it is only the additional intervention units that is chosen randomly from the set of available controls, and the treated units in the sample remain treated. Here, we introduce estimands that represent exactly the effect of treated one or more additional interventional units.

Define the average potential outcome on all outcome units if, out of the control units  $\caln^c$, $K$ were to be randomly chosen to be treated as 
\[
\overline Y^\textrm{+K}=\frac{1}{\binom{N^c}{K}}\sum_{\bm n }
\frac1M \sum_{m\in{\cal M}} Y_m(\bm W+e_{\bm n}),
\]
where $\bm n = (n_1, n_2, \dots, n_K)$ is a vector of distinct control units $n_i \in \caln^c$ with $n_i \neq n_{i'}$, and $e_{\bm n}$ is the $N$-vector with all elements equal to 0 other than the ones in $\bm n$ which are equal to one.
For example, for $K = 1$ we have: 
\[
\overline Y^\textrm{+1}=\frac{1}{N^c}\sum_{n \in \caln^c}
\frac1M \sum_{m\in{\cal M}} Y_m(\bm W+e_{\bm n}),
\]
where $e_n$ is the vector of all zeros, except for the $n^{th}$ element which is equal to 1.
Then, the effect of intervening on additional  $K$  units  is defined as
\[ 
\tau^\textrm{+K}=\overline Y^\textrm{+K} - 
\frac1M \sum_{m \in \calm} Y_m(\bm W).
\]
These effects, as the status quo estimands, are random estimands from a design-based perspective \citep{sekhon2020inference}, because they depend on the random $\bm W$.

\section{Estimation}
\label{sec:estimation}

\subsection{Additional Notation}

For two outcome units $m$ and $m'$ we  use $\pi_{\caln_m, \caln_{m'}}(\bm a, \bm b) = \pi(\bm W_{\caln_m} = \bm a, \bm W_{\caln_{m'}} = \bm b)$ to denote the probability that the intervention set of outcome unit $m$ receives treatment $\bm a$ and the intervention set of outcome unit $m'$ receives treatment $\bm b$, simultaneously, under the experimental design. If the two outcome units have overlapping intervention sets and $\bm a, \bm b$ specify a different assignment for at least one of their common units, this probability will be equal to 0. We also use $\bm a_k$ denote the vector of length $k$ repeating the value $a$, $\bm a_k = (a, a, \dots, a)$.

Furthermore, we use $\sum\limits_{m, m'} = \sum\limits_{m = 1}^M \sum\limits_{m' = 1}^M$ to denote the sum over all pairs of outcome units, and $\sum\limits_{m \neq m'} = \sum\limits_{m = 1} ^ M \sum\limits_{\substack{m' = 1 \\ m' \neq m}}^M$ to denote the double summation over all {\it ordered} pairs of distinct outcome units. 
Similarly, we use $\sum\limits_{\bm w_{\caln_m} \neq \bm w_{\caln_m}'}$ to denote the double summation over ${\bm w_{\caln_m} \in \calw^{N_m}}$ and 
$\bm w_{\caln_m}' \in \calw^{N_m} \setminus \{ \bm w_{\caln_m} \}$ 
and $ \sum\limits_{\bm w_{\caln_m}, \bm w_{\caln_{m'}}}$ to denote the double summation over $\bm w_{\caln_m} \in \calw^{N_m}$ and $\bm w_{\caln_{m'}} \in \calw^{N_{m'}}$. We note that {\footnotesize $\displaystyle \sum_{m \neq m'}$} includes each pair of distinct outcome units {\it twice} (for example $(m,m') = (1,2)$ and $(m,m')=(2,1)$), and similarly for {\footnotesize $\displaystyle \smashoperator{ \sum_{\bm w_{\caln_m} \neq \bm w_{\caln_m}'} }$}.

%\georgia{I have some issues with notation that I hate. We have been using $Y_m(\bm 0)$ to denote potential outcomes under $\bm w = (0, 0, \dots, 0) = \bm 0_N$ for all units, and under $\bm w_{\caln_m} = (0, 0, \dots, 0)$ for the intervention set of unit $m$. These two potential outcomes are equal to one another, so fine, it has not been a real issue. [There are some issues with this notation in the +k estimand but it's minor.] However, in the following theorem, we have $\pi_{\caln_m}(\bm a)$ and $\pi_{\caln_{m'}}(\bm a)$ appearing, for $\bm a$ that is equal to $\bm 0$ or $\bm 1$. However, the size of $\bm a$ is different when we look at $\caln_m$ and $\caln_{m'}$ because the two intervention sets can be of different size. So I want to use a subscript of some sort to denote the length of the vector. So $\bm 0_k$ denotes the vector that repeats 0 $k$ times, etc. But then the notation becomes tedious as well, since quantities like $\bm a_{N_m}, \bm a_{N_{m'}}, \bm a_{|\caln_m \cup \caln_m|}$ will all show up. Ideas on a way to make notation better?
%}
%\guido{yes, notation is very tricky here. Let me think about that. We could have a table with notation to make it clear, but at the moment it is quite cumbersome}

\subsection{Estimation of all-or-none and status quo effects}

Estimating $\overline Y(\bm 0_N), \overline Y(\bm 1_N)$ would suffice for estimating both the all-or-none effect $\tau$ in \cref{eq:tau_estimand_all_or_nothing} and the status quo effects $\tau_0^{sq}$ and  $\tau_1^{sq}$ in \cref{eq:tau_estimand_status_quo} and \cref{eq:tau_estimand_status_quocontrol} (since $\overline Y(\bm W)$ is the observed average outcome).

To do so, we consider a weighting estimator, where the weight for each outcome unit involves the probability of the realized treatment for the unit's intervention set.
Since $\pi$ denotes the treatment assignment over all interventional units, $\pi_{\caln_m}: \calw^{N_m} \mapsto [0, 1]$ is the implied strategy on the intervention set of outcome unit $m \in \calm$, according to \cref{def:implied_trt_strategy}.
Consider estimators for the average potential outcomes defined as
\begin{equation}
\widehat Y(\bm 0_N) =
\frac1M \sum_{m = 1}^M
\frac{I(\bm W_{\caln_m} = \bm 0_{N_m})}
{\pi_{\caln_m}(\bm W_{\caln_m})} 
Y_m,
\quad \text{and} \quad
\widehat Y(\bm 1_N) =
\frac1M \sum_{m = 1}^M
\frac{I(\bm W_{\caln_m} = \bm 1_{N_m})}
{\pi_{\caln_m}(\bm W_{\caln_m})} 
Y_m,
\label{eq:forwards_estimator}
\end{equation}
where $I(\bm W_{\caln_m} = \bm 0_{N_m})$ is the indicator that all elements in $\bm W_{\caln_m}$ are equal to zero, and
$I(\bm W_{\caln_m} = \bm 1_{N_m})$ is the indicator that all elements in $\bm W_{\caln_m}$ are equal to one. Consider also estimators of the all-or-none or status quo effects that are defined as
$\widehat \tau^\text{aon} = \widehat Y(\bm 1_N) - \widehat Y(\bm 0_N)$, $\widehat \tau_1^\text{sq} = \overline Y - \widehat Y(\bm 0_N)$, and $\widehat \tau_0^\text{sq} = \widehat Y(\bm 1_N) - \overline Y$, where $\overline Y = M^{-1} \sum_m Y_m$ is the average observed outcome.

\begin{theorem}
Assume that $\pi$ is a treatment allocation strategy with $\pi_{\caln_m}(\bm 0_{N_m}) > 0$ and $\pi_{\caln_m}(\bm 1_{N_m}) > 0$ for all $m \in \calm$. Then, for $\bm a = \bm 0_N, \bm 1_N$ and $\bm a_{N_m} = \bm 0_{N_m}, \bm 1_{N_m}$ , we have
\begin{enumerate}[leftmargin=*,label=(\alph*)]
\item \label{item:bias_yhat}

$\E \Big[ \widehat Y(\bm a) \Big] = \overline Y(\bm a)$. As a result, 
$\widehat \tau^\text{aon}$, $\widehat \tau_1^\text{sq}$ and $\widehat \tau_0^\text{sq}$ are unbiased for $\tau^\text{aon}$, $\tau_1^\text{sq},$ and $\tau_0^\text{sq}$, respectively.
\item \label{item:var_y0_y1}
$\displaystyle \Var \Big[ \widehat Y(\bm a) \Big] = \frac1{M^2}
\smashoperator{\sum_{m, m'}}
c_{m,m'}
%\text{Cov} \left(I(\bm W_{\caln_m} = \bm a), I(\bm W_{\caln_{m'}} = \bm a) \right)
\frac{Y_m(\bm a_{N_m})Y_{m'}(\bm a_{N_{m'}})} {\pi_{\caln_m}(\bm a_{N_m}) \pi_{\caln_{m'}}(\bm a_{N_{m'}})}
$,
where $c_{m, m'}$ is equal to $\pi_{\caln_m \cup \caln_{m'}}(\bm a_{|\caln_m \cup \caln_{m'}|}) - $ $ \pi_{\caln_m}(\bm a_{N_m}) \pi_{\caln_{m'}}(\bm a_{N_{m'}})$.
For $m = m'$, $c_{m, m'}$ simplifies to $c_{m, m} = \pi_{\caln_m}(\bm a_{N_m}) \left( 1 - \pi_{\caln_m}(\bm a_{N_m}) \right)$.

% Variance of tau hat
\item \label{item:var_aon}
$\Var \left( \widehat \tau^\text{aon} \right) = \Var \Big[ \widehat Y(\bm 0_N) \Big] + \Var \Big[ \widehat Y(\bm 1_N) \Big] - 2 \mathrm{Cov}\Big[ \widehat Y(\bm 0_N), \widehat Y(\bm 1_N) \Big]$, where $M^2 \mathrm{Cov}\Big[ \widehat Y(\bm 0_N), \widehat Y(\bm 1_N) \Big]$ is equal to
\begin{align*}
\smashoperator{\sum_{\substack{m, m': \\ \pi_{\caln_m, \caln_{m'}}(\bm 0_{N_m}, \bm 1_{N_{m'}}) \neq 0}}} 
    \left[ \pi_{\caln_m, \caln_{m'}}(\bm 0_{N_m}, \bm 1_{N_{m'}}) - \pi_{\caln_m}(\bm 0_{N_m}) \pi_{\caln_{m'}}(\bm 1_{N_{m'}}) \right] \frac{Y_m(\bm 0_{N_m}) Y_{m'}(\bm 1_{N_{m'}})}{\pi_{\caln_m}(\bm 0_{N_m}) \pi_{\caln_{m'}}(\bm 1_{N_{m'}})}
    \ &- \\[-5pt] \smashoperator{\sum_{\substack{m, m': \\ \pi_{\caln_m, \caln_{m'}}(\bm 0_{N_m}, \bm 1_{N_{m'}}) = 0}}}
    Y_m(\bm 0_{N_m}) & Y_{m'}(\bm 1_{N_{m'}}).
\label{eq:cov_aon}
\end{align*}
\end{enumerate}
\label{theorem:unbiasedness}
\end{theorem}

\noindent 
The result in \cref{theorem:unbiasedness} generalizes the results in \cite{aronow2017estimating} from a unipartite setting under an exposure mapping assumption to our bipartite setting without exposure mappings. In principle, for the status-quo effect estimators, we only require one arm of the positivity assumption: $\pi_{\caln_m}(\bm 0_{N_m}) > 0$ to estimate $\tau_1^\text{sq}$, and $\pi_{\caln_m}(\bm 1_{N_m}) > 0$ to estimate $\tau_0^\text{sq}$, for all $m \in \calm$.

\begin{remark}
   A necessary condition for $\pi_{\caln_m, \caln_{m'}}(\bm 0_{N_m}, \bm 1_{N_{m'}}) \neq 0$ in the variance of the all-or-none estimator is that the two units $m$ and $m'$ have non-overlapping intervention sets.
   If the experimental design specifies that the treatment assignment is independent for the non-overlapping intervention sets of two outcome units (a condition satisfied by, for example, a Bernoulli design), then
   the corresponding term in \cref{theorem:unbiasedness}\ref{item:var_aon} would be equal to zero, and the pair would not contribute to the covariance term.
\end{remark}

These theoretical design-based variances involve products of potential outcomes for the same or different outcome units, under the same or different treatment assignments on their intervention sets. When these variances involve products of potential outcomes that are possible to observe simultaneously, simple inverse weighting estimators for the variance can be used. However, as we discuss in \cref{sec:application}, in the presence of interference and complex experimental designs, positivity assumptions are often violated. For this reason, we propose variance estimators that are conservative when pairs of units cannot simultaneously experience a given treatment on their intervention sets \citep{aronow2013conservative}. Specifically, for the variance of the average potential outcome, we define the estimator
\begin{align*}
\widehat \Var \left[ \widehat Y(\bm a) \right] &= \frac1{M^2} \Bigg[ \sum_{m = 1}^M 
\frac{I(\bm W_{\caln_m} = \bm a_{N_m})}
{\pi_{\caln_m}(\bm a_{N_m})^2}
\left( 1 - \pi_{\caln_m}(\bm a_{N_m}) \right) Y_m^2 + \\
& \hspace{40pt} + \ \ 
\smashoperator{\sum_{\substack{m \neq m': \\ \pi_{\caln_m \cup \caln_{m'}}(\bm a_{|\caln_m \cup \caln_{m'}|}) \neq 0}}} \ \ 
% \sum_{m \neq m'} 
\frac{
I(\bm W_{\caln_m \cup \caln_{m'}} = \bm a_{|\caln_m \cup \caln_{m'}|})}
{
\pi_{\caln_m}(\bm a_{N_m}) \pi_{\caln_{m'}}(\bm a_{N_{m'}})
}
\left(1 - 
\frac{\pi_{\caln_m}(\bm a_{N_m}) \pi_{\caln_{m'}}(\bm a_{N_{m'}})}{\pi_{\caln_m \cup \caln_{m'}}(\bm a_{|\caln_m \cup \caln_{m'}|})} \right)
Y_m Y_{m'} \\
& \hspace{40pt} + \ \ \frac12 \ \ 
\smashoperator{\sum_{\substack{m \neq m': \\ \pi_{\caln_m \cup \caln_{m'}}(\bm a_{|\caln_m \cup \caln_{m'}|}) = 0}}} \ \ 
\left( 
\frac{I(\bm W_{\caln_m} = \bm a_{N_m}) Y_m^2}{\pi_{\caln_m}(\bm a_{N_m})} +
\frac{I(\bm W_{\caln_{m'}} = \bm a_{N_{m'}}) Y_{m'}^2}{\pi_{\caln_{m'}}(\bm a_{N_{m'}})} 
\right)
\Bigg].
\end{align*}
Similarly, we introduce the following estimator for the variance of the all-or-none estimator 
\begin{align*}
\widehat \Var\left( \widehat \tau^\text{aon} \right) &=
\widehat \Var \left[ \widehat Y(\bm 0_N) \right] + \widehat \Var \left[ \widehat Y(\bm 1_N) \right] -
2 \widehat{\mathrm{Cov}}^{\mathrm{LB}}\Big[ \widehat Y(\bm 0_N), \widehat Y(\bm 1_N) \Big]
\end{align*}
where
$\widehat{\mathrm{Cov}}^{\mathrm{LB}}\Big[ \widehat Y(\bm 0_N), \widehat Y(\bm 1_N) \Big]$ is defined as
\begin{align*}
& \hspace{20pt} \frac1{M^2}
\smashoperator{\sum_{\substack{m, m': \\ \pi_{\caln_m, \caln_{m'}}(\bm 0_{N_m}, \bm 1_{N_{m'}}) \neq 0}}}
\frac
{
I(\bm W_{\caln_m} = \bm 0_{N_m}) I(\bm W_{\caln_{m'}} = \bm 1_{N_{m'}})
}
{
\pi_{\caln_m}(\bm 0_{N_m}) \pi_{\caln_{m'}}(\bm 1_{N_{m'}})
}
    \left[ 1 - \frac{\pi_{\caln_m}(\bm 0_{N_m}) \pi_{\caln_{m'}}(\bm 1_{N_{m'}})}{\pi_{\caln_m, \caln_{m'}}(\bm 0_{N_m}, \bm 1_{N_{m'}})} \right] 
    Y_m Y_{m'}  \\
& \hspace{20pt} -
\frac1{2M^2}
\smashoperator{\sum_{\substack{m, m': \\ \pi_{\caln_m, \caln_{m'}}(\bm 0_{N_m}, \bm 1_{N_{m'}}) = 0}}}
\left( 
\frac{I(\bm W_{\caln_m} = \bm 0_{N_m}) Y_m^2}{\pi_{\caln_m}(\bm 0_{N_m})} +
\frac{I(\bm W_{\caln_{m'}} = \bm 1_{N_{m'}}) Y_{m'}^2}{\pi_{\caln_{m'}}(\bm 1_{N_{m'}})} 
\right).
\end{align*}

\begin{proposition}[Variance estimation]
\label{prop:var_estimation}
Assume that the conditions of \cref{theorem:unbiasedness} hold. Then, we have that $\E \left\{ \widehat \Var \left[ \widehat Y(\bm a) \right] \right\} \geq \Var \left [\widehat Y(\bm a) \right]$ and
$\E \left[ \widehat \Var \left( \widehat \tau^\text{aon} \right) \right] \geq \Var \left( \widehat \tau^\text{aon} \right)$.
Moreover, if $\pi_{\caln_m \cup\caln_{m'}}(\bm a_{|\caln_m \cup\caln_{m'}|}) > 0$ for all pairs $(m, m')$, then $\E \left\{ \widehat \Var \left[ \widehat Y(\bm a) \right] \right\} = \Var \left [\widehat Y(\bm a) \right]$.
\end{proposition}

The result in \cref{prop:var_estimation} holds irrespective of the experimental design and the form of the outcome. However, if all potential outcomes have the same sign, excluding the third term in $\widehat \Var \left[ \widehat Y(\bm a) \right]$ results in a sharper upper bound of the variance, hence we recommend using that.

%Note that \cref{prop:var_estimation} requires that all pairs of units are possible to be observed as simultaneously having their intervention sets being all treated or all control. If this condition is not satisfied, the estimator of the variance for $\widehat Y(a)$ is not unbiased. However, if all potential outcomes have the same sign, it is guaranteed that $\E \left\{ \widehat \Var \left[ \widehat Y(a) \right] \right\} \geq \Var \left [\widehat Y(a) \right]$, which would imply that $\E \left[ \widehat \Var \left( \widehat \tau^\text{aon} \right) \right] \geq \Var \left( \widehat \tau^\text{aon} \right)$ still holds. Alternatively, a correction term can be added to $\widehat \Var \left[ \widehat Y(a) \right]$ to ensure that it is positively biased, even without an assumption on the sign of the potential outcomes. 
%These results draw from  \cite{aronow2013conservative}. See also \cite{chattopadhyay2024neymanian}.

\begin{remark}[Variance of estimators for the status quo effects]
As already noted, from a design-based perspective the status quo estimands are random estimands \citep{sekhon2020inference} since they depend on the random treatment assignment $\bm W$. \cref{theorem:unbiasedness} establishes the unbiasedness of our status quo estimators. For their variance, we can consider the variance of $\widehat{\tau}^\text{sq}_a$ around $\tau^\text{sq}_a$, namely $\Var (\widehat{\tau}^\text{sq}_a -\tau^\text{sq}_a)$, which reduces to $\Var (\widehat Y(\bm 1_N- \bm a) - \overline Y(\bm 1_N - \bm a)) = \Var (\widehat Y(\bm 1_N - a))$ since $\overline Y(1 - \bm a)$ is constant with respect to the treatment assignment. Therefore, the variance estimators $\widehat \Var(\widehat Y(\bm a))$ provided in \cref{prop:var_estimation} can be used for quantifying the variability of the status quo estimators.
\end{remark}

To study the asymptotic behavior of the estimators, we assume that the potential outcomes for all units are bounded, as formalized in the following assumption.

\begin{assumption}
All potential outcomes for all outcome units are bounded, \iteg, there exists $B \in \mathbb{R}^+$ such that $|Y_m (\bm w_{\caln_m})| \leq B$ for all $m$ and $\bm w \in \calw^{N_m}$. 
\label{ass:po_bounded}
\end{assumption}

\begin{theorem}[Consistency of the IPW estimator of the all-or-none estimand] 
Consider a sequence of bipartite graphs $\calg_s = \{\caln_s, \calm_s, A_s\}$ for which the treatment assignment over the intervention units is $\pi_s$, $s = 1, 2, \dots$.
Assume that $\pi_s$ satisfies the conditions of \cref{theorem:unbiasedness}, and that potential outcomes for units in $\calm_s$ satisfy \cref{ass:po_bounded}.
Let
$M_s = |\calm_s|$ be the number of outcome units,
$\gamma_{s, a} = \min_{m \in \calm_S} \pi_{s, \caln_m}(\bm a_{N_m})$ be the minimum probability across outcome units that their intervention set has constant treatment $a$, 
$\mathcal{K}_s = \{ (m, m'): m \neq m' \text{ and } \pi_{s,\caln_m \cup \caln_{m'}}(\bm a_{|\caln_m \cup \caln_{m'}|}) \neq \pi_{s,\caln_m}(\bm a_{N_m}) \pi_{s,\caln_{m'}}(\bm a_{N_{m'}}) \}$ be the set of pairs of outcome units whose intervention sets' treatment assignment is correlated with cardinality $\kappa_s = |\mathcal{K}_s|$, 
and $\gamma_{s, a}^* = \max_{(m, m') \in \mathcal{K}_s} \big| \pi_{s,\caln_m \cup \caln_{m'}}(\bm a_{|\caln_m \cup \caln_{m'}|}) / [\pi_{s,\caln_m}(\bm a_{N_m}) \pi_{s,\caln_{m'}}(\bm a_{N_{m'}})] - 1 \big| $ be a measure of dependence in the treatment assignments of outcome unit pairs. 
If $M_s \rightarrow \infty$,
$\gamma_{s, a}^{-1} = o(M_s)$,
and
$\kappa_s \gamma_{s, a} ^ * = o(M_s^2)$, 
then $\widehat Y(\bm a)$ is consistent for $\overline Y(\bm a)$.
\label{theorem:consistency}
\end{theorem}

% \begin{remark}[Interpreting the conditions for consistency in terms of study design]
We can interpret the conditions for consistency in terms of the study design.
The first condition, which we refer to as a {\it positivity condition}, specifies that the positivity assumption holds for each outcome unit, which makes the average potential outcome estimable. If $\gamma_{s, a}$ converges to 0 faster than $M_s^{-1}$, the experimental design does not satisfy that all units could have been observed with the desired treatment level.
The second condition is referred to as a {\it correlation condition}, and it restricts the correlation of treatment assignments across outcome unit pairs' intervention sets, through the number of pairs with correlated assignments and how strong this correlation is allowed to be. As we will see below, this assumption is closely related to the number of shared units in outcome units' intervention sets. 
% \end{remark}

These conditions become more obvious when investigating a simple experimental design, such as the Bernoulli assignment over the intervention units with fixed probability of treatment. The following corollary establishes sufficient conditions for consistency of the causal estimator under this simplified design.

\begin{corollary}[Consistency of the IPW estimator of all-or-none under a Bernoulli experimental design]
     In the setting of \cref{theorem:consistency}, assume that the treatment assignment on the interventional units for graph $s$, $\pi_s$, is a Bernoulli design with probability of treatment $p \in (0, 1)$. Then, $\kappa_s$ is equal to the number of pairs of outcome units with overlapping intervention sets, $\kappa_{s} = \sum_{m\neq m'} I(\caln_{s, m} \cap \caln_{s, m'} \neq \emptyset)$.
     Let $d_{s, o} = \max_{m \in \calm_s} |\caln_{s, m}|$, denote the maximum size of the intervention set for all outcome units, and $d_{s, \kappa} = \max_{m \neq m'} |\caln_{s, m} \cap \caln_{s, m'}|$ the maximum number of common interventional units across all outcome unit pairs. If $M_s \rightarrow \infty$, $d_{s, o} = o (\log M_s)$, and $d_{s, \kappa} = o(\log(M_s^2 / \kappa_s))$, then $\widehat Y(\bm  a)$ is consistent for $\overline Y( \bm a)$.
\label{cor:consistency_bernoulli}
\end{corollary}

% \begin{remark}[Interpreting the conditions for consistency under a Bernoulli design]
We investigate the interpretation of the conditions for consistency in the Bernoulli design.
The positivity condition allows the maximum size of intervention sets, $d_{s, o}$, to grow slower than $\log(M_s)$. Therefore, it ensures that the minimum probability that all units in an intervention set receive treatment $a$ is not too small relative to the sample size of outcome units, and that all units are possible to be represented in the sample under the desired treatment.

For Bernoulli designs, correlation of treatment assignments across intervention sets can only arise when the intervention sets are overlapping. Therefore, the correlation condition limits the number of outcome units that have overlapping intervention sets and the size of this overlap, through $\kappa_s$ and $d_{s, \kappa}$, respectively. This is not necessarily true for other experimental designs where the treatment over non-overlapping intervention sets can be correlated (such as a completely randomized experiment).
If the maximum size of overlap in the intervention sets, $d_{s, \kappa}$, is bounded, the correlation condition that $d_{s, \kappa} = o(\log(M_s^2 / \kappa_s))$ can be replaced by $\kappa_s = o(M_s^2)$, limiting the rate with which pairs of outcome units have overlapping intervention sets. In general, the size of overlap in two outcome units' intervention sets can grow with $M_s$, however, in this case, this condition would impose a stronger restriction on the growth of $\kappa_s$ as a function of the number of outcome units.
% \end{remark}

% \begin{remark}
% Let's investigate the condition $d_{s, \kappa} = o(\log(M_s^2 / \kappa_s))$.
Furthermore, since $\kappa_s$ describes the number of {\it pairs} of outcome units with overlapping intervention sets, $\kappa_s$ grows {\it at most} as fast as $M_s^2$. If we further restrict $\kappa_s = O(M_s)$ to grow linearly with the number of outcome units, then the correlation condition becomes $d_{\kappa_s} = o (\log(M_s))$. Since $d_{\kappa_s}$ describes the maximum number of {\it common} intervention units for two outcome units, this has to be smaller than the maximum degree of outcome units, and $d_{s, \kappa} \leq d_{s, o} = o (\log(M_s))$ is satisfied. Therefore, as a special case, consistency of the IPW estimator under a Bernoulli design is achieved when $d_{s, o} = o (\log(M_s))$ and $\kappa_s = O(M_s)$.
% \end{remark}

\begin{remark}[Relative growth of intervention and outcome unit sample sizes]
    For Bernoulli designs, the maximum size of intervention sets across outcome units, $d_{s,o}$, introduced in  \cref{cor:consistency_bernoulli}, is bounded below by $N_s / M_s$, where $N_s = |\caln_s|$. Therefore, the condition that $d_{s,o} = o(\log M_s)$ imposes restrictions on the relative growth of the number of intervention and outcome units in the graph sequence that $N_s = o(M_s \log M_s)$. Despite this condition being required for consistency within our framework, it is not sufficient to substitute the condition on $d_{s,o}$, since the graph structure can impose highly imbalanced sizes for intervention sets which would violate positivity. Importantly, alternative treatment assignments, which are not as simplistic as a Bernoulli allocation, could bypass restrictions on relative sample size growth by incorporating $d_{s,o}$ in the experimental design to ensure that positivity is guaranteed.
\end{remark}

We compare our consistency results to others in the literature. Let $d_{s, \text{\tiny \it I}}$ denote the maximum size of outcome sets across intervention units in graph $\calg_s$, defined as 
$d_{s, \text{\tiny \it I}} = \max_{n \in \caln_s} |\calm_{s,n}|$.
The IPW estimator for the all-or-none effect under the simple Bernoulli design was recently studied in \cite{lu2025design}.
In their work, the size of the intervention sets is assumed to be bounded, $d_{s,o} = O(1)$, and the maximum size of an intervention unit's outcome set grows slower than the number of outcome units, 
$d_{s, \text{\tiny \it I}} = o(M_s)$. The conditions stated there for consistency are related but different to the ones provided here. Our work allows the size of the intervention sets to grow with the number of outcome units. Their condition on $d_{s, \text{\tiny \it I}}$ ensures that the sample includes units with sufficient variability in their intervention sets' treatment assignments. Therefore, it is related to our correlation condition, but one does not imply the other.

Furthermore, in the seminal work by \cite{Harshawetal2023}, the authors study the asymptotic behavior of their estimator which is developed under a linearity assumption on the potential outcomes. Their results are under a specific experimental design that assigns constant treatment in clusters of intervention units. Their main assumption for consistency is that $d_{s, \text{\tiny \it I}} d_{s,o}^3 = o(M_s)$ which limits the size of outcome and intervention sets simultaneously. If the size of intervention sets is bounded, then the condition becomes $d_{s, \text{\tiny \it I}} = o(M_s)$, as in the work by \cite{lu2025design}. However, the condition in \cite{Harshawetal2023} is more general, since it allows the size of intervention sets to grow, as in our work. The potentially strong linearity assumption on the potential outcomes allows the authors to use all outcome units for estimation of the all-or-none effect, and bypass certain positivity requirements for using IPW-based estimators.

\subsection{Estimation of estimands under stochastic interventions}
\label{subsec:stochastic-estimation}

Estimators for the stochastic intervention estimands in \cref{subsec:stochastic_effects} can be similarly acquired as weighted estimators of outcomes where the weights describe the relative likelihood of the observed treatment assignment under the stochastic intervention and the treatment assignment.
% First, we need to define what the stochastic intervention, $h_\alpha$, implies for the treatment assignment of the intervention set of outcome unit $m$. This follows similarly to the one in \cref{def:implied_trt_strategy}: For a set of intervention unit indices $\cals \subset \caln$ of cardinality $S$, let the $h_{\alpha, \cals}: \calw^S \rightarrow [0,1]$ denote the implied stochastic intervention on the interventional units in $\cals$, defined as
% \[
% h_{\alpha, \mathcal{S}}(\bm a) =
% \sum_{\substack{\bm w \in \mathcal{W}^N \\
% \bm w_{\mathcal{S}} = \bm a}} h_\alpha(\bm w).
% \]
For stochastic intervention, $h_\alpha$, define the estimator for the average potential outcome under this intervention in \cref{eq:stochastic_outcome} as
\[
\widehat Y_{h_\alpha} = \frac1M \sum_{m = 1}^M \frac{h_{m, \alpha}(\bm W_{\caln_m})}{\pi_{\caln_m}(\bm W_{\caln_m})} Y_m,
\]
and the estimator for a change in the intervention parameter from $\alpha$ to $\alpha'$ in \cref{eq:stochastic_effect} as
\[
\widehat \tau(\alpha, \alpha') = \widehat Y_{h_{\alpha'}} - \widehat Y_{h_\alpha}.
\]
For these estimators, we have the following result.

\begin{theorem}
Assume that $\pi$ is a treatment allocation strategy with $\pi_{\caln_m}(\bm a_{N_m}) > 0$ for all $(m, \bm a_{N_m})$ pairs where $\bm a \in \calw^{N_m}$ and $h_{m, \alpha}(\bm a_{N_m}) > 0$ or $h_{m, \alpha'}(\bm a_{N_m}) > 0$. Then,
\begin{enumerate}[leftmargin=*,label=(\alph*)]
\item $\E \Big( \widehat Y_{h_\alpha} \Big) = \overline Y_{h_\alpha}$, and as a result, $\E \big[ \widehat \tau(\alpha, \alpha') \big] = \tau(\alpha, \alpha')$.
\item 
\label{item:var_stoch}
$M^2 \Var \Big( \widehat Y_{h_\alpha} \Big)$ is equal to
\begin{align*}
& \sum_{m = 1}^M \left[ \sum_{\bm w_{\caln_m}} 
\pi_{\caln_m}(\bm w_{\caln_m}) \left( 1 - \pi_{\caln_m}(\bm w_{\caln_m}) \right)
 \left[ \frac{Y_m(\bm w_{\caln_m}) h_{m, \alpha}(\bm w_{\caln_m})}{\pi_{\caln_m}(\bm w_{\caln_m})} \right]^2 - \right. \\
 & \hspace{80pt} \left.
\sum_{\bm w_{\caln_m} \neq \bm w_{\caln_m}'} 
Y_m(\bm w_{\caln_m}) Y_m(\bm w_{\caln_m}')
h_{m, \alpha}(\bm w_{\caln_m}) h_{m, \alpha}(\bm w_{\caln_m}')
\right] + \\
+ &
\sum_{m \neq m'}
\sum_{\bm w_{\caln_m}, \bm w_{\caln_{m'}}} 
\left(
\frac{
\pi_{\caln_m, \caln_{m'}}(\bm w_{\caln_m}, \bm w_{\caln_{m'}})}
{
\pi_{\caln_m}(\bm w_{\caln_m}) \pi_{\caln_{m'}}(\bm w_{\caln_{m'}}) 
} 
-
1
\right)  
Y_m(\bm w_{\caln_m}) Y_{m'}(\bm w_{\caln_{m'}})
h_{m, \alpha}(\bm w_{\caln_m}) h_{m', \alpha}(\bm w_{\caln_{m'}}).
\end{align*}
% where 
% $\pi_{\caln_m, \caln_{m'}}(\bm w_{\caln_m}, \bm w_{\caln_{m'}}) =
% \pi(\bm W_{\caln_m} = \bm w_{\caln_m}, \bm W_{\caln_{m'}} = \bm w_{\caln_{m'}})$
% is the probability that the intervention set of units $m$ and $m'$ simultaneously have treatment levels $\bm w_{\caln_m}$ and $\bm w_{\caln_{m'}}$, respectively, under the treatment assignment.

\item $\Var\left[ \widehat \tau(\alpha, \alpha') \right]$ has the same form as $\Var \left( \widehat Y_{h_{\alpha}} \right)$ in part \ref{item:var_stoch}, except $h_{m,\alpha}(\cdot)$ is substituted by $h_{m,\alpha'}(\cdot) - h_{m,\alpha}(\cdot)$, and similarly for $h_{m',\alpha}(\cdot)$.

\end{enumerate}
\label{theorem:stochastic_unbiased}
\end{theorem}

% \begin{remark}
The result in \cref{theorem:stochastic_unbiased} is a general version of the result in \cref{theorem:unbiasedness}. To see this, note that if we set $h_{m, \alpha}(\bm w_{\caln_m}) = 1$ when $\bm w_{\caln_m} = \bm 0_{N_m}$, and $h_{m, \alpha}(\bm w) = 0$ otherwise, the estimator $\widehat Y_{h_\alpha}$ reverts to the estimator $\widehat Y(\bm 0)$, and similarly for $\widehat Y(\bm 1)$.
% \end{remark}

% \begin{remark}
It is interesting to investigate the terms in the variance of the estimator $\widehat Y_{h_\alpha}$. 
The first two lines correspond to terms for each outcome unit $m$. In the first line, the outcome unit appears only under a single treatment vector for its intervention set, $\bm w_{\caln_m}$. However, in the second line, it appears in the same term under two treatment vectors for its intervention set, $\bm w_{\caln_m}$ and $\bm w_{\caln_m}'$. Therefore, as long as the stochastic intervention assigns non-zero probability to at least two treatment vectors for $\caln_m$, this second line includes terms that correspond to treatment assignments that cannot occur simultaneously.
The last two lines in the variance of $\widehat Y_{h_\alpha}$ correspond to interactions between units $m$ and $m'$ under treatment vectors for their intervention sets $\bm w_{\caln_m}$ and $\bm w_{\caln_{m'}}$. These treatment vectors might be possible to observe simultaneously or not. For example, If the two units have overlapping intervention sets, $\caln_m \cap \caln_{m'}$ is not empty. Then, if $\bm w_{\caln_m}$ and $\bm w_{\caln_{m'}}$ specify different treatments to their overlapping intervention units, these terms include interactions of potential outcomes for units $m$ and $m'$ that are impossible to observe simultaneously.
% \end{remark}

%\begin{remark}
%For the estimators of the all-or-none quantities $\widehat Y(a)$, we saw in \cref{theorem:unbiasedness} that we can acquire unbiased estimators for the variance $\Var[ \widehat Y(\alpha)]$. 
For this reason, we cannot acquire unbiased estimators for the variance of $\widehat Y_{h_\alpha}$ for interventions $h_\alpha$ that are not point-mass distributions. % The reason for that is that the variance includes terms under treatment assignments that are impossible to be observed simultaneously. In contrast, under the intervention that sets all units to $a$, the stochastic intervention becomes a point mass, and only a single treatment vector is allowed for the intervention set of each outcome unit. Therefore, the ``problematic'' terms cancel out. 
Instead, we propose using $\widehat \Var \left( \widehat Y_{h_\alpha} \right)$, where $M^2 \widehat \Var \left( \widehat Y_{h_\alpha} \right)$ is equal to
\begin{align*}
& \sum_m \Bigg[
\left( 1 - \pi_{\caln_m}(\bm W_{\caln_m}) \right)
 \left( \frac{h_{m, \alpha}(\bm W_{\caln_m})}{\pi_{\caln_m}(\bm W_{\caln_m})} Y_m \right)^2 +
\frac{h_{m, \alpha}(\bm W_{\caln_m})}{\pi_{\caln_m}(\bm W_{\caln_m})}(1-h_{m, \alpha}(\bm W_{\caln_m})) Y_m^2
\Bigg] + \\
& \sum_{m \neq m'} \Bigg[
 \dfrac{h_{m, \alpha}(\bm W_{\caln_m}) h_{m', \alpha}(\bm W_{\caln_{m'}})}
{\pi_{\caln_m, \caln_{m'}}(\bm W_{\caln_m}, \bm W_{\caln_{m'}})}
\left(
\dfrac{
\pi_{\caln_m, \caln_{m'}}(\bm W_{\caln_m}, \bm W_{\caln_{m'}})}
{
\pi_{\caln_m}(\bm W_{\caln_m}) \pi_{\caln_{m'}}(\bm W_{\caln_{m'}}) 
} 
-
1
\right)  
Y_m Y_{m'} + \\
& \hspace{40pt}
+ \frac{h_{m, \alpha}(\bm W_{\caln_m})}{2 \pi_{\caln_m}(\bm W_{\caln_m})} \lambda_{m,m'}(\bm W_{\caln_m}) Y_m^2 +
\frac{h_{m', \alpha}(\bm W_{\caln_{m'}})}{2 \pi_{\caln_{m'}}(\bm W_{\caln_{m'}})} \lambda_{m',m}(\bm W_{\caln_{m'}}) Y_{m'}^2 \Bigg]
,
\end{align*}
for
\[
\lambda_{m,m'}(\bm w_{\caln_m}) = \ \ 
\smashoperator{\sum_{\substack{\bm w_{\caln_{m'}}: \\
\pi_{\caln_m, \caln_{m'}}(\bm w_{\caln_m}, \bm w_{\caln_{m'}}) = 0
}}} \ \
h_{m',\alpha}(\bm w_{\caln_{m'}}),
\]
and similarly for $\lambda_{m',m}(\bm w_{\caln_{m'}})$.
In the following proposition, we establish that this variance estimator is, in expectation, an upper bound for the theoretical variance of the stochastic intervention estimators.
%\end{remark}

\begin{proposition}[Variance estimator of the average potential outcome under stochastic interventions]
\label{prop:var_estimation_stoch}
Under the assumptions of \cref{theorem:stochastic_unbiased}, it holds that
$\E \left[ \widehat \Var \left( \widehat Y_{h_\alpha} \right) \right] 
\geq \Var \left( \widehat Y_{h_\alpha} \right) $.
\end{proposition}

The result on the estimation of the variance upper bound in \cref{prop:var_estimation_stoch} holds assuming only the stochastic positivity assumption in \cref{theorem:stochastic_unbiased}. If, additionally, all potential outcomes have the same sign (for example, when the outcome is binary), we can exclude the second term in the summation over $m$, and the second and third terms in the summation over $m$ and $m'$ in $\widehat \Var \left( \widehat Y_{h_\alpha} \right)$. Under this additional assumption, the new estimator will be a sharper upper bound of the variance, and should be preferred.

Next, we propose an estimator for the variance of the estimator of the contrast of average potential outcomes under stochastic interventions, and we show that it is, in expectation, at least as large as the true variance.
This estimator, denoted by $\widehat\Var\left[ \widehat \tau(\alpha, \alpha') \right]$, is similar to $\widehat\Var \left( \widehat Y_{h_\alpha} \right)$ with some additional book-keeping for the terms that cannot be estimated unbiasedly. Specifically, we set $M^2 \widehat \Var \left[ \widehat \tau(\alpha, \alpha') \right]$ equal to
\begin{align*}
    & \sum_{m = 1}^M \Bigg[
    \left( 1 - \pi_{\caln_m}(\bm W_{\caln_m}) \right)
    \left( \frac{h_{m, \alpha'}(\bm W_{\caln_m}) - h_{m, \alpha}(\bm W_{\caln_m})}{\pi_{\caln_m}(\bm W_{\caln_m})} Y_m \right)^2 \\
    & \hspace{40pt} +
    Y_m^2 \frac{\left| h_{m, \alpha'}(\bm W_{\caln_m}) - h_{m, \alpha}(\bm W_{\caln_m}) \right| }{\pi_{\caln_m}(\bm W_{\caln_m})}
\ \ \smashoperator{\sum_{\bm w_{\caln_m}': \bm w_{\caln_m}' \neq \bm W_{\caln_m}}} \ \ 
\left| h_{m, \alpha'}(\bm w_{\caln_m}') - h_{m, \alpha}(\bm w_{\caln_m}') \right| 
    \Bigg] + \\
&
\sum_{m \neq m'} \Bigg[
\frac{\left[ h_{m, \alpha'}(\bm W_{\caln_m}) - h_{m, \alpha}(\bm W_{\caln_m}) \right] \left[h_{m', \alpha'}(\bm W_{\caln_{m'}}) - h_{m', \alpha}(\bm W_{\caln_{m'}}) \right]}{\pi_{\caln_m, \caln_{m'}}(\bm W_{\caln_m}, \bm W_{\caln_{m'}})}
\left(
\frac{
\pi_{\caln_m, \caln_{m'}}(\bm W_{\caln_m}, \bm W_{\caln_{m'}})}
{
\pi_{\caln_m}(\bm W_{\caln_m}) \pi_{\caln_{m'}}(\bm W_{\caln_{m'}}) 
} 
-
1
\right)  
Y_m Y_{m'} \\
& \hspace{40pt} +
\frac{ \left| h_{m, \alpha'}(\bm W_{\caln_m}) - h_{m, \alpha}(\bm W_{\caln_m}) \right| }{2 \pi_{\caln_m}(\bm W_{\caln_m})}
\lambda^\tau_{m,m'}(\bm W_{\caln_m}) Y_m^2 \\
& \hspace{40pt}
+ \frac{\left| h_{m', \alpha'}(\bm W_{\caln_{m'}}) - h_{m', \alpha}(\bm W_{\caln_{m'}})  \right|}{2 \pi_{\caln_{m'}}(\bm W_{\caln_{m'}})} 
\lambda^\tau_{m',m}(\bm W_{\caln_{m'}}) Y_{m'}^2 \Bigg]
\end{align*}
where
\[
\lambda^\tau_{m,m'}(\bm w_{\caln_m}) = \smashoperator{\sum_{\substack{\bm w_{\caln_{m'}} \\
\pi_{\caln_m, \caln_{m'}}(\bm w_{\caln_m}, \bm w_{\caln_{m'}}) = 0 }}} \ \ 
\left| h_{m', \alpha'}(\bm w_{\caln_{m'}}) - h_{m', \alpha}(\bm w_{\caln_{m'}})  \right|,
\]
and similarly for  $\lambda^\tau_{m',m}(\bm w_{\caln_{m'}})$.
The following holds.

\begin{proposition}[Variance estimator of the contrast of average potential outcomes under stochastic interventions]
\label{prop:var_estimation_tau_stoch}
    Under the assumptions of \cref{theorem:stochastic_unbiased}, $\E \left\{ \widehat \Var \left[ \widehat \tau(\alpha, \alpha') \right] \right\} \geq \Var \left[ \widehat \tau (\alpha, \alpha') \right] $.
\end{proposition}

Importantly, all the terms in $\widehat \Var \left[ \widehat \tau(\alpha, \alpha') \right]$ are necessary for a valid upper bound of the true variance, {\it even when} all potential outcomes have the same sign. This is in contrast to the estimator of the average potential outcome under stochastic interventions, and unlike scenarios in the literature for estimating average treatment effects.

Next, we study the conditions under which the estimator for the stochastic intervention is consistent for the true estimand.

\begin{theorem}[Consistency of the IPW estimator of estimands under stochastic intervention]
Consider a sequence of bipartite graphs $\calg_s = \{\caln_s, \calm_s, A_s\}$, $s = 1, 2, \dots$ for which all potential outcomes are bounded and that the treatment assignment over the interventional units is $\pi_s$. Let 
$M_s = |\calm_s|$ be the number of outcome units under graph $\calg_s$, and
%
% --- Expectation wrt to intervention distn
% We use $\E_{h_m}$ to denote the expectation with respect to the stochastic intervention for the $m^{th}$ outcome unit, and  $\E_{h_m \otimes h_{m'}}$ with respect to independent stochastic interventions for the $m^{th}$ and $m'^{th}$ outcome units. 
% Let 
% $\Delta_{s, \alpha} = \max_{m \in \calm_S} \E_{h_m} \left[ {h_{m, \alpha}(\bm W_{\caln_m})} / {\pi_{s, \caln_m}(\bm W_{\caln_m})}  \right]$ be the largest expected weight across outcome units, 
%
% -- Expectation wrt to assignment
$\E_{\pi_s}$ denote the expectation with respect to the treatment assignment.
Let
\begin{itemize}[topsep=0pt,itemsep=0pt,leftmargin=*]
    \item $ \displaystyle \Delta_{s, \alpha} = \max_{m \in \calm_S} \E_{\pi_s} \left\{ \left[ \frac{h_{m, \alpha}(\bm W_{\caln_m})}{\pi_{s, \caln_m}(\bm W_{\caln_m})}  \right]^2 \right\}$ be the largest expected squared weight across outcome units, 
    \item $\mathcal{K}_s = \{ (m, m'): m \neq m' \text{ and }  \pi_{s, \caln_m, \caln_{m'}}(\bm w_{\caln_m}, \bm w_{\caln_{m'}}) \neq \pi_{s, \caln_m}(\bm w_{\caln_m}) \pi_{s, \caln_{m'}}(\bm w_{\caln_{m'}}) \text{ for some } \bm w_{\caln_m}$, $\bm w_{\caln_{m'}} \}$ be the set of pairs of outcome units whose intervention sets' treatment assignment is correlated, with cardinality $\kappa_s = |\mathcal{K}_s|$, and 
    \item $ \displaystyle \Gamma_{s, \alpha}^* = 
    \max_{(m, m') \in \mathcal{K}_s} \E_{\pi_s} \left\{ 
    \left| \frac{\pi_{s, \caln_m, \caln_{m'}}(\bm W_{\caln_m}, \bm W_{\caln_{m'}})}{\pi_{s, \caln_m}(\bm W_{\caln_m}) \pi_{s, \caln_{m'}}(\bm W_{\caln_{m'}})} - 1\right|
    \frac{h_{m, \alpha}(\bm W_{\caln_m}) h_{m', \alpha}(\bm W_{\caln_{m'}})}{\pi_{s, \caln_m}(\bm W_{\caln_m}) \pi_{s, \caln_{m'}}(\bm W_{\caln_{m'}})}
    \right\}$ be a measure of dependence in the treatment assignments of intervention sets corresponding to pairs of outcome units. 
\end{itemize}
If $M_s \rightarrow \infty$,
$\Delta_{s, \alpha} = o(M_s)$,
and
$\kappa_s \Gamma_{s, a} ^ * = o(M_s^2)$, 
then $\widehat Y_{h_\alpha}$ is consistent for $\overline Y_{h_\alpha}$. 
\label{theorem:consistency_stochastic}
\end{theorem}

As in \cref{theorem:consistency}, the two conditions can be interpreted as a positivity and a correlation condition. The positivity condition of $\Delta_{s, \alpha}$ is expressed based on the distance between the stochastic intervention and the propensity score, defined as the expected squared weight given to each outcome unit in the causal estimator. 
If the stochastic intervention is equal to a point-mass distribution for each outcome unit, $h_{m, \alpha}(\bm w_{\caln_m}) = I(\bm w_{\caln_m} = \bm a_{N_m})$, the condition reverts to the positivity condition on $\gamma_{s, a}$ for the all-or-none estimand in \cref{theorem:consistency}. 
On the other hand, if the stochastic intervention is specified to be equal to the propensity score, $h_{m, \alpha}(\bm w_{\caln_m}) = \pi_{s, \caln_m}(\bm w_{\caln_m})$ for all $\bm w_{\caln_m}$, then $\Delta_{s, \alpha} = 1$ and the positivity condition is trivially satisfied.
Of course, the latter case lacks scientific interest, but it provides an interesting insight on the plausibility of the positivity condition for estimands under stochastic interventions. When the stochastic intervention resembles the propensity score more closely, the positivity condition is more likely to hold, as the value of $\Delta_{s, \alpha}$ is smaller. Therefore, when estimands such as the all-or-none causal effect in \cref{theorem:consistency} cannot be estimated consistently due to positivity violations, alternative estimands under well-designed stochastic interventions might still be well-estimated.

As for the correlation condition, the one in \cref{theorem:consistency_stochastic} reverts back to the one in \cref{theorem:consistency} for point-mass distributions $h_{m, \alpha}$. In principle, in certain scenarios, carefully designed stochastic interventions that down-weigh ``problematic'' treatment vectors can render the correlation condition more plausible compared to point-mass distributions such as in the all-or-none effect. However, this might not be possible to achieve under general experimental designs $\pi_s$ and network structure that lead to a large correlation constant $\Gamma_{s, \alpha}^*$ irrespective of the stochastic intervention chosen. Investigating specific cases where this might be the case is outside the scope of this work.

% A Note I had on positivity before for the case between indicator and PS:
%
%The in-between cases: Note that $\gamma_{s, a}$ in \cref{theorem:consistency} is defined as $ \gamma_{s, a}  = \min_m \pi_{\caln_m}(a)$, the minimum across outcome units of the probability of the only possible assignment under the intervention. Here, our intervention could lead to many different treatment assignments, so this minimum is re-defined to be over all possible treatment vectors that can arise under the intervention. Set $\tilde \gamma_s = \min_m \min_{w_{\caln_m}: h_{m, \alpha}(w_{\caln_m}) > 0} \pi_{\caln_m}(w_{\caln_m})$. Then, $\Delta_{s, \alpha} \leq \tilde \gamma_s^{-1}$, and the condition of the theorem becomes $\tilde \gamma_s^{-1} = o(M_s)$, which is essentially the one in \cref{theorem:consistency}.
%
% And another note:
%
% In Bernoulli assignments, we might be able to show how the condition on positivity is weaker under the stochastic intervention because we can let the probability of the intervention go towards the probability of the realized assignment. That would allow for growing intervention sets (I believe), though this is restricted by the fact that these intervention sets have to be relatively non-overlapping to satisfy the dependence condition. Doing this would require some assumption on the outcomes, such as that their variance is constant across assignments.

\subsection{Estimation of the ``$+K$'' Estimand}

For simplicity, we consider experimental designs that pre-specify the number of treated intervention units. Specifically, under treatment assignment $\pi$, $N^t$ units in $\caln$ are assigned to treatment and the remaining $N^c = N - N^t$ units in $\caln$ are assigned to control. A completely randomized experiment on the set of intervention units satisfies this structure, though it is not the only assignment that is allowed here, and the assignment in \cref{sec:application} is another example.

For outcome unit $m \in \calm$, we use
$N_m^t = \sum_{n \in \caln_m} I(W_n = 1)$ to denote the number of units in its intervention set that are treated, and
$N_m^c = \sum_{n \in \caln_m} I(W_n = 0)$ for the number of units that are control.
The following definition of function $\rho_m$ specifies the weight that will be given to the outcome unit $m$ under each possible treatment vector on its intervention set. 

We define the function $\rho_m : \calw^{N_m} \mapsto \mathbb{R}$ as
\[
\rho_m(\bm w_{\caln_m}) = \frac1{\binom{N^c}{K} \pi_{\caln_m}(\bm w_{\caln_m})} \left[
\sum_{k = 0}^K \binom{N^c - N^c_m}{K - k}
\sum_{\bm w_{\caln_m}^{*k}} \pi_{\caln_m}(\bm w_{\caln_m}^{*k}) 
\right],
\]
where $\bm w_{\caln_m}^{*k}$ represents treatment vectors for the intervention set $\caln_m$ which are equal to $\bm w_{\caln_m}$ except with $k$ fewer units assigned to treatment. Specifically, the vector $\bm w_{\caln_m}^{*k}$ is of length $N_m$ and satisfies the following condition: there exists indices $k$ indices $i_1, i_2, \dots, i_k \in \{1,2, \dots, N_m\}$ such that
$[\bm w_{\caln_m}]_{i_l} = 1$, and $[\bm w_{\caln_m}^{*k}]_{i_l} = 0$ for $l = 1, 2, \dots, k$, and for all other indices they are equal:
$[\bm w_{\caln_m}]_j = [\bm w_{\caln_m}^{*k}]_j$ for all $j \in \{1, 2, \dots, N_m\} \setminus \{i_1, i_2, \dots, i_k\}$. 
For example, if $\caln_m$ includes three units and \(\bm w_{\caln_m} = (0, 1,1 )\), then \( \bm w_{\caln_m}^{*1} \in \{ (0, 1, 0), (0, 0, 1) \}, \)
and \( \bm w_{\caln_m}^{*2} = (0, 0, 0). \)
If the number of controls in the intervention set is smaller than $k$, then the corresponding terms in the summation are equal to 0.

\begin{theorem}
Assume that treatments are assigned on the interventional units $\caln$ following a design $\pi$ under which $N^t$ units are assigned to treatment and $N^c = N - N^t$ to control. Assume that for this experimental design, if $\pi_{\caln_m}(\bm w_{\caln_m}) > 0$ for some $\bm w_{\caln_m}$, then it also holds that 
$\pi_{\caln_m}(\bm w_{\caln_m}') > 0$ for all $\bm w_{\caln_m}'$ that are like $\bm w_{\caln_m}$ with any number of additional treated units, i.e., $\bm w'_{\caln_m} \in \{0, 1\}^{N_m}$ such that $[\bm w'_{\caln_m}]_j = 1$ if $[\bm w_{\caln_m}]_j = 1$. Then, the estimator
\[
\widehat Y^{+K} = \frac1M \sum_{m \in \calm} \rho_m(\bm W_{\caln_m}) Y_m
\]
is an unbiased estimator for $\overline Y^{+K}$.
\label{theorem:plusK_unbiasedness}
\end{theorem}

The positivity condition in \cref{theorem:plusK_unbiasedness} is relatively strong for general designs. It specifies that, if a treatment vector on $\caln_m$ is possible, all ``richer'' (in treatment) treatment vectors must also be possible. This condition is necessary to ensure that every unit could have been observed under its realized and its counterfactual treatment vector of assigning $K$ additional intervention units to treatment.

The form of the weights in function $\rho_m$ is complicated, and lacks obvious interpretation. We can acquire some intuition about the form of these weights by investigating their form in the special cases below.

\begin{specialcase}
For interventions that assign $K = 1$ additional units to treatment, the weights simplify to
\[
\rho_m(\bm W_{\caln_m}) = \frac1{N^c} \left[ N^c - N^c_m + \frac{\sum_{\bm w_{\caln_m}^{*1}}\pi_{\caln_m}(\bm W_{\caln_m}^{*1})}{\pi_{\caln_m}(\bm W_{\caln_m})} \right].
\]
The first term, $(N^c - N^c_m) / N^c$, describes how likely it is under the intervention that the control unit that receives the treatment is {\it not} connected to $m$. In that case, the observed outcome for $m$ is also the outcome under the intervention. Then, the second term describes how likely it is under the experimental design that one fewer units in $m$'s intervention set is treated than actually occurred. In that case, the observed treatment corresponds to what would have happened in that scenario the additional unit receives the treatment. 
\end{specialcase}

\begin{specialcase}
When $\pi$ specifies a completely randomized experiment on the set of interventional units $\caln$, the function of weights $\rho_m$ for the intervention that assign $K = 1$ additional units to treatment simplifies to
\[
\rho_m(\bm W_{\caln_m}) = \frac{1}{N^c} \left( N^c - N^c_m + N^t_m \frac{N^c - N^c_m}{N^t - N^t_m + 1} \right).
\]
In this special case, we can clearly see that the weight given to each outcome unit is intrinsically related to the number of interventional units it is connected to, and how many of these units are treated. Therefore, the structure of the graph is informative of the weight that each unit should be given. In that sense, the form of these weights are related to the work in \cite{doudchenko_causal_2020} where the authors discuss confounding by the graph structure.
\end{specialcase}

It is possible to extend estimation of $+K$ estimands to experimental designs $\pi$ that do not fix the number of treated units. By employing the law of iterative expectation in our results, we suspect that this extension would require little additional theoretical work. However, the form of the weights would likely be even harder to interpret. We leave such extension to future work.

%\commentG{[[Maybe: We can expand to experimental designs that do not fix the number of treated units using iterative expectation, but the weights become tedious. But in this case, we might have some issues with positivity. For general treatment assignments $\pi$, the positivity assumption would require that we would need a non-zero probability of any treatment on any intervention set. So the number of treated units $N^t$ according to $\pi$ has to be restricted according to the size of intervention sets. Or something like that?]]}

\section{Application}
\label{sec:application}

Since 2007, the network of high-speed rail (HSR) in China has developed rapidly with the construction of new lines; the goal of this rapid construction included supporting 
economic development by improving regional connectivity (\cite{lawrence2019china,ma2022localized}). 
Leveraging the recent construction of the HSR, \cite{borusyakhull2023} study the effect of market access growth on Chinese regional employment growth over the time period 2007–2016. 
% GP: I shortened the follwoing:
%
% In their work, the \emph{effective treatment} is defined as market access growth between 2007 and 2016, measured by combining data on the development of the HSR network and each area’s location and population, and predicting travel time between regions based on the operational speed of each HSR line as well as proxies of travel time by car and low-speed train.
% Their goal is to relate market access (MA) growth for a prefecture $m$, defined as $\log (\text{MA}^{2016}_m) - \log (\text{MA}^{2007}_m)$, to the corresponding growth in prefecture’s urban employment
Since market access growth may be endogenous,
% MA growth may be endogenous or may have unobserved determinants that are also associated with employment growth: employment may have grown in regions where more HSR construction is planned, irrespective of the fact that construction took place or not.
to resolve the potential Omitted Variable Bias (OVB) problem,
they propose an instrumental variable strategy by exploiting its exogenous determinants that are characterized by a (known) assignment process. 
Specifically, they assume that the timing of line completion is conditionally as good as random, and that 
% because built lines connect more regions, 
% Specifically, they assume that 
the 2016 completion status is completely randomized among lines with the same number of cross-prefecture ``links''. The number of links is the number of prefectures a line connects, that is, the number of prefectures that the line passes through at least one of its stations.  The main Beijing to Shanghai HSR line, which has the greatest number of links ($18$), is assumed to be completed with probability 1.
\cref{tab:links} reports the number of treated and control lines within strata defined by each line's number of links.
% The random completion status is thus formalized as a natural experiment that generates exogenous shocks that can be used as instruments for the treatment of interest in various strategies. 
% \cite{borusyakhull2023} generate a set of counterfactual HSR networks from the assumed assignment process and they compute MA growth for each draw to obtain the expected treatment.
% The observed MA growth is then recentered by substracting the expected MA growth to remove its systematic variation. They show that using the recentered treatment as an instrument for the realized treatment removes the omitted variable bias.

\subsection{The setup}

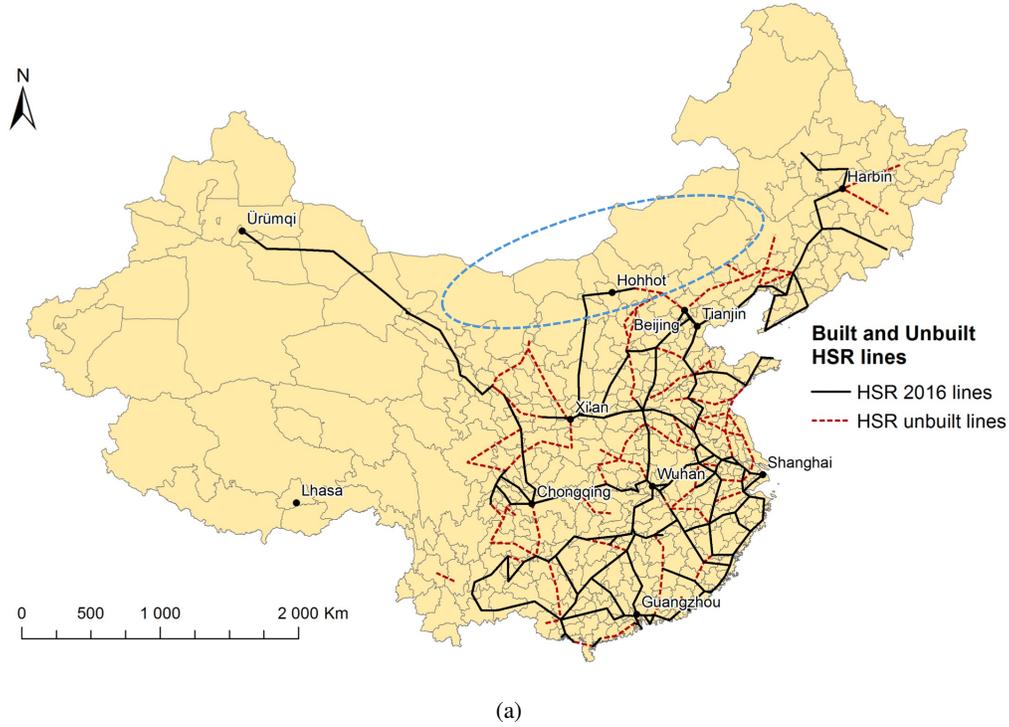
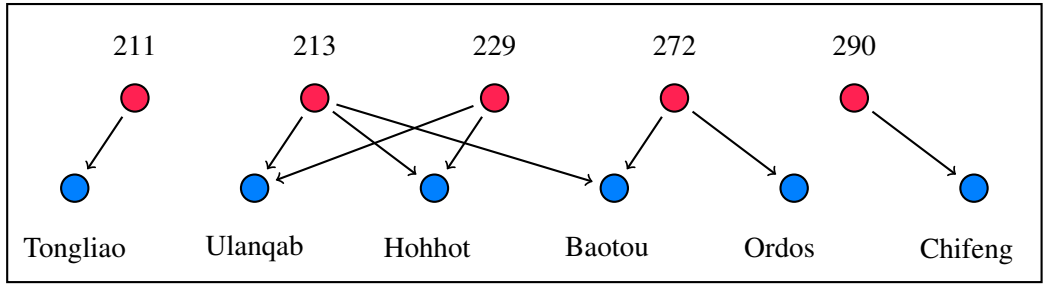
\begin{figure}[!t]
\centering
%\spacingset{1}

\subfloat[][]{
\includegraphics[width=0.8\linewidth]{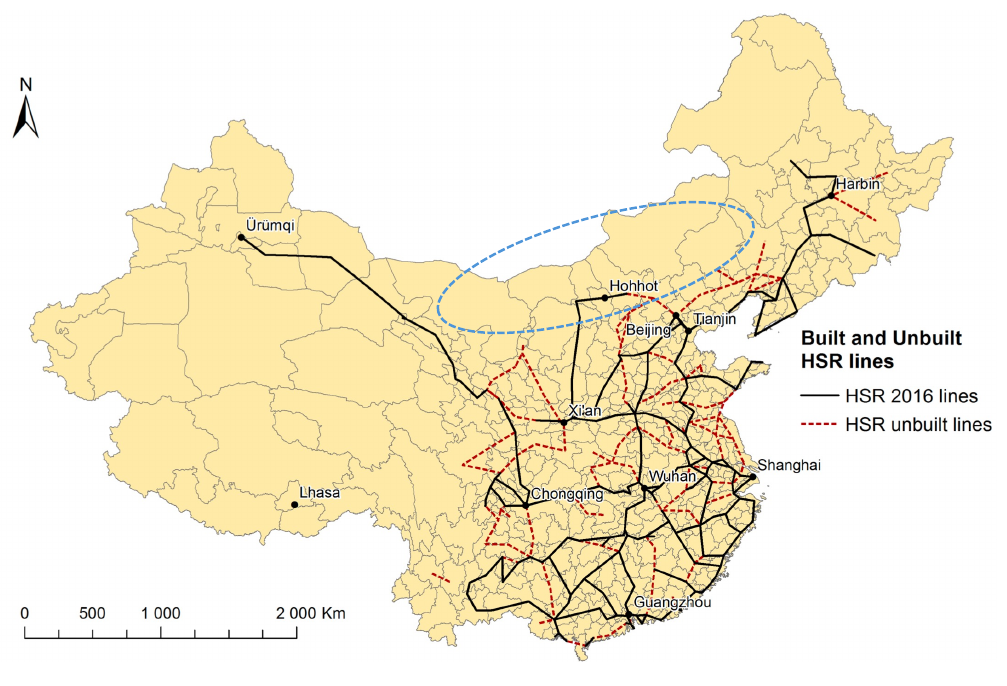}
\label{fig:application-lines}}

\vspace{10pt}

\subfloat[][]
{
\makebox[0.8\textwidth][c]{
\begin{tikzpicture}[thick,
  every node/.style={draw,circle},
  fsnode/.style={fill=myblue},
  ssnode/.style={fill=mygreen},
  -,shorten >= 3pt,shorten <= 3pt
]

% the vertices of U with custom names
\begin{scope}[start chain=going right,node distance=20mm]
\node[fsnode,on chain] (f1) [label=above: 211] {};
\node[fsnode,on chain] (f2) [label=above: 213] {};
\node[fsnode,on chain] (f3) [label=above: 229] {};
\node[fsnode,on chain] (f4) [label=above: 272] {};
\node[fsnode,on chain] (f5) [label=above: 290] {};
\end{scope}

% the vertices of V with custom names
% Reduce label distance for closer alignment
\begin{scope}[xshift=-0.8cm,yshift=-1.2cm,start chain=going right,node distance=20mm]
\node[ssnode,on chain] (s1) [label={[label distance=-2.5mm]below: Tongliao}] {};
\node[ssnode,on chain] (s2) [label={[label distance=-2.5mm]below: Ulanqab}] {};
\node[ssnode,on chain] (s3) [label={[label distance=-2.7mm]below: Hohhot\phantom{1}}] {};
\node[ssnode,on chain] (s4) [label={[label distance=-2.6mm]below: Baotou\phantom{1}}] {};
\node[ssnode,on chain] (s5) [label={[label distance=-2.6mm]below: Ordos\phantom{11}}] {};
\node[ssnode,on chain] (s6) [label={[label distance=-2.7mm]below: Chifeng\phantom{1}}] {};
\end{scope}

% the edges
\foreach \i in {1} \draw[->] (f1) edge (s\i);
\foreach \i in {2,3,4} \draw[->] (f2) edge (s\i);
\foreach \i in {2,3} \draw[->] (f3) edge (s\i);
\foreach \i in {4,5} \draw[->] (f4) edge (s\i);
\foreach \i in {6} \draw[->] (f5) edge (s\i);

% large rectangle covering all nodes
\node[draw, rectangle, fit margins={left=10pt, right=10pt, bottom=15pt, top=15pt}, 
fit=(f1) (f5) (s1) (s6)] (large_rect) {};

\end{tikzpicture}
}
\label{fig:appl_partial}
}
\caption{Case study: Evaluating the effect of HSR line completion on employment growth. (a) The illustration of the built and the unbuilt HSR lines with Inner Mongolia circled. (b) The sub-bipartite graph in Inner Mongolia province. The intervention units are the HSR lines identified with numbers, and the outcome units are prefectures in the province.}
\label{fig:application}
\end{figure}

We analyze this as a case study with bipartite interference. We study the impact of a line completion on employment growth.
The intervention units are the 150 HSR lines planned, completed or under construction as of April 2019. A line is considered treated if it was completed by 2016. There are 83 treated lines. The lines are illustrated in \cref{fig:application-lines} colored by completion status.
The outcome units are the 340 subprovince-level administrative divisions in mainland China, referred to as prefectures. 
The outcome is the growth in a prefecture’s urban employment from Chinese City Statistical Yearbooks\footnote{Employment is measured as “The
Average Number of Staff and Workers” (from the “People’s Living Conditions and Social
Security” chapter of the 2008–2017 China City Yearbooks (China
Statistics Press (2000–2017)). See \cite{borusyakhull2023} for details.}. Employment growth is measured as percentage change of employment between 2007 and 2016 and as difference in log 2016 employment and log 2007 employment.
%\commentG{Fabri, for effects we do percentage variation and difference in log outcomes. Can you say explicitly how the outcome is defined?}. 
We focus on the set of 237 prefectures with nonmissing outcome data and having at least one line passing through. See Data Appendix A in \cite{borusyakhull2023} for further details.
% We move away from  \cite{borusyakhull2023}, in that we do not specify any outcome model as a function of a specific effective treatment or exposure, and instead target 

\begin{table}[!b]
    \centering
    \begin{tabular}{lcccccccccc}
    \hline
    Number of links & 1 & 2 & 3 & 4 & 5 & 6 & 7 & 8 & 9 & 18 \\
    Number of treated & 25 & 23 & 9 & 5 & 7 & 6 & 2 & 1 & 4 & 1 \\
    Number of controls & 27 & 14 & 12 & 7 & 1 & 3 & 1 & 1 & 1 & 0 \\ \hline
    \end{tabular}
%     \caption{Caption}
%     \label{tab:my_label}
% \end{table}

% \begin{table}[!t]
%     \centering
%     \begin{tabular}{ccc}
%         nlink&treated & control \\
%          \hline
%         1 & 25&27\\
%         2 & 23&14\\
%         3&9&12\\
%         4&5&7\\
%         5&7&1\\
%         6&6&3\\
%         7&2&1\\
%         8&1&1\\
%         9&4&1\\
%         18&1&0\\
%         \hline
%     \end{tabular}
    \caption{Number of treated and control lines stratified by the number of links.}
    \label{tab:links}
\end{table}

%\commentFM{@Zhaoyan: It would be good to have for the bipartite graphs we consider (using 1st and 2nd degree) some summary statistics, average size of intervention sets and outcome sets, number of outcome units all under treatment or all under control. 
%You could also run randomization tests using the simple test statistics in Section 3.} 
%\guido{could we actually plot the bipartite graph?}
%\fabri{Zhaoyan has produced a map of China with completed and non completed (but planned) HSR lines, and a sub-bipartite graph for a region in China.}

In this study, the bipartite graph describes which lines' completion status is allowed to affect which prefecture's employment growth. We consider a first-degree bipartite graph
% , each associated with alternative assumptions on which intervention units may be influential for each outcome unit.
%The two bipartite graphs assume prefecture-line links based on first or second degree. In the first-degree network, 
where a prefecture is considered connected to an HSR line if the line passes through the stations in this prefecture. %, with the construction either completed or delayed by the year 2016. 
As an example, \cref{fig:appl_partial} illustrates the bipartite graph for the Inner Mongolia province.
We report descriptive statistics for the graph in \cref{tab:graph_description}. 
% 1.43\% of all intervention and outcome unit pairs are connected. The average number of prefectures connected to an HSR line is 3.81 and the average number of HSR lines connected to a prefecture is 2.14.
In \cref{supp_sec:application-second-degree}, we also consider an analysis that uses second-degree bipartite graph for illustration purposes.

% In order to describe these two bipartite graphs, we formulate two  bipartite matrices of size $150 \times 237$. 
\begin{table}[!t]
    \centering
    \begin{tabular}{lcc}
        \hline
        \% of graph entries equal to 1 & 1.43 % & 9.00 
        \\
        average size of intervention sets & 2.14 % & 13.50 
        \\
        average size of outcome sets & 3.81 % & 23.93 
        \\
        \% of HSR lines connected to only two prefectures & 35.33 % & 0.67 
        \\
        \% of prefecture connected to only one HSR line & 43.23 % & 0.75 
        \\
        maximum number of HSR lines connected to a prefecture & 9 % & 40 
        \\
        maximum number of prefectures connected to a HSR line & 19 % & 63 
        \\
        number of outcome units with all treated HSR lines &108 % &10
        \\
        number of outcome units with all control HSR lines &64 %&5
        \\
        \hline
    \end{tabular}
    \caption{Descriptive statistics of the first-degree bipartite graph between intervention units (HSR lines) and outcome units (prefectures).}
    \label{tab:graph_description}
\end{table}

% Within the bipartite framework, we target the estimands at the level of the outcome units which we introduced earlier, which correspond to specific interventions on the intervention units, that is, causal effects of constructing (or not) some of the planned HSR lines.  For that, we exploit the assumed random completion time in constructing the planned lines. 

For illustration purposes, we consider two different study designs with respect to the actual treatment assignment mechanism.
In one design, the completion status of an HSR line follows a completely randomized design (CRD) conditional on the line's number of links. That is, within strata of lines with the same number of links, a fixed number of lines are constructed at random. This design was adopted by \cite{borusyakhull2023}. \cref{tab:links} reports the number of treated and control lines, stratified by their number of links. As an example, within the stratum of the 52 HSR lines with one link, 25 are randomly chosen for construction.
In the second design, we assume that a line's completion status follows a Bernoulli distribution conditional on the number of links, with probability of being constructed equal to the observed proportion of completion.
For example, the probability of an HSR line to be completed in the stratum with one link is equal to $p=25/(25+27)$. 
% The two designs differ with respect to the correlation of treatment assignment across intervention units: in the CRD design, units with the same number of links have negatively correlated treatments, whereas in the Bernoulli design, the treatment assignment is independent across all intervention units. 
For both designs, we perform the randomization tests discussed in \cref{sec:randomization_tests}.
We consider both designs to illustrate how the design and the choice of estimand play a role in the plausibility of the positivity assumption, which
%The correlation of treatment across intervention units can lead to violation of the positivity assumption for certain estimands, as 
we discuss in 
\cref{subsec:application-positivity}.

We target the following estimands: All-or-None, All-vs-Status Quo, Status Quo-vs-None, and the $+1$ estimand.
We also target three different estimands under stochastic interventions.
The first one is aimed at comparing with the +1 estimand: we consider a hypothetical assignment $h_\alpha^{(1)}$, where within each stratum, the number of treated units is increased by 1, and the number of control units is reduced by 1. For example, in the stratum with one link, 26 units are treated and 26 are not in the CRD design, and units are assigned to treatment with probability $p=26/52$ in the Bernoulli design. The stratum with 18 links has a single treated unit that is always assigned treatment.
% , that will remain so through the randomization since no control units can be added to the treated group. 
The second stochastic intervention we consider, $h_\alpha^{(2)}$, assigns treatment to half of the control units within each stratum (in addition to the units actually treated). For strata with an odd number of control units, the smaller integer is considered. Since the strata with 5, 7, 8, 9 and 18 links have none or one control unit, the number of treated units for these strata does not change under the intervention. 
For the first two stochastic interventions, the corresponding estimand compares the average outcome under the intervention against the observed outcome, defined as
\[
\tau^{(j)}_\text{stoch} = \overline Y_{h_\alpha^{(j)}} - \overline Y,
\]
for $j = 1, 2.$
The third estimand with stochastic interventions we consider is aimed at evaluating the impact on the outcome from treating  95\% versus 5\% of the HSR lines. These estimands are conceptually and effectively very close to all-or-none estimands, but their estimation suffers less from positivity violations. Specifically, if $h_p$ is the distribution of a Bernoulli allocation with probability $p$ on the treated units, the estimand is equal to
\[
\tau^{(3)}_\text{stoch} = \overline Y_{h_{0.95}} - \overline Y_{h_{0.05}}.
\]

% To summarize, we consider two designs for the actual treatment allocation, two versions of the bipartite graph describing intervention-outcome unit dependencies, and seven estimands in each scenario, amounting to $2 \times 2 \times 7 = 28$ estimated quantities.

\subsection{The Positivity Assumption}
\label{subsec:application-positivity}

Before presenting and commenting on results, we discuss potential challenges with regards to the positivity assumption on the outcome units' intervention sets that arise when estimating different estimands in the various settings we consider. For the all-or-none effect, the assumption requires that all units could have intervention sets that are fully treated or control, and similarly for the status-quo effects. For the $+1$ estimand, if a treatment vector is possible under the design, all other treatment vectors with at least the same treated units need also be possible. For the estimands under stochastic interventions, treatment vectors that are possible under the stochastic intervention need also be possible under the design. Therefore, different estimands have different requirements for the corresponding positivity assumption to hold.

In our analysis, if an outcome unit's intervention set does not satisfy positivity, then it is excluded from the data set.
Table \ref{tab:posviolations} reports the number of outcome units that must be excluded in the estimation of each estimand because of positivity violations under the different designs.
For example,  24 out of 237 outcome units are excluded for estimation of the all-or-none effect because of positivity violation in CRD, that is, their intervention set cannot be all treated or all control under this design. Among the remaining units, 94 prefectures have all treated lines, 57 prefectures have all connected lines under control, and these are the outcomes units that will be effectively used in the estimation of the all-or-none estimand. 
The positivity assumption is violated for a fewer number of outcome units under a Bernoulli design compared to a CRD. That is because correlation of the treatment assignment across intervention units implies that it is more common for the treatment vectors considered in the estimand to be impossible under the design. 
The positivity violation under the Bernoulli design arise from the 18 outcome units that are connected with the longer HSR line that is treated with probability $1$.
% , and can never be observed with all connected intervention units under control.

Complete randomization designs provide interesting insights on the positivity assumption and the design of stochastic interventions. In CRDs, if an outcome unit's intervention set can be all treated, then the positivity assumption for the $+K$ estimand also holds. This implies that $+K$ estimands have weaker positivity conditions than all-or-none estimands under CRD. Furthermore, it implies that, for outcome units that can be all-treated, we can design stochastic interventions with at least the same treated units as observed or more, and these would immediately satisfy the stochastic intervention positivity.
Lastly, if an outcome unit satisfies the positivity condition for the all-or-none estimand in that its intervention set can be fully control and fully treated, then it can experience all intermediate treatment vectors as well. Therefore, it also satisfies the positivity condition for any stochastic intervention.

%\commentG{Zhaoyan mentioned that under a CR design, if (1, 1, ..., 1) satisfies positivity, then the +K positivity assumption should also hold. I agree with her. This would imply that +K estimands have weaker positivity conditions than all or nothing in CR settings. But we need to think about it.}

%\commentG{Another nice point by Zhaoyan (derived from the previous observation): If all or nothing can be estimated, it means that all units can have (0, 0, ..., 0) and (1, 1, ..., 1), which would imply that they can have all intermediate treatment vectors as well. Which as a result would imply that we can estimate causal effects under any stochastic intervention.}

%\commentG{Based on the above, if $\pi(1, 1, ... 1) > 0$ for everyone, we can design stochastic interventions that are CR designs with \# of treated higher than \# treated in observed, and then these would satisfy the stochastic intervention positivity.}

\begin{table}[!t]
    \centering
    \begin{tabular}{c|ccccccc}
        \makecell{Design} & All-or-None & \makecell{All-vs-\\Status Quo} & \makecell{Status Quo-vs-\\None} & +1 & $\tau^{(1)}_\text{stoch}$ & $\tau^{(2)}_\text{stoch}$ & $\tau^{(3)}_\text{stoch}$ \\
        \hline
        CRD & 24 & 0 & 24 & 0 & 0 & 0 & 0\\
%        CRD (2) & 127 & 7 & 127 & 7 & 7 & 7& 0\\
%        \hline
        Bernoulli & 18 & 0 & 18 & 0 & 0 & 0& 0\\
%        Bernoulli (2) & 52 & 0 & 52 & 0 & 0 & 0& 0\\
    \end{tabular}
    \caption{Number of outcomes units violating positivity for the various estimands under different experimental designs and the first-degree bipartite graph.}
    \label{tab:posviolations}
\end{table}

\subsection{Results from Randomization Testing}

We begin by using the randomization testing framework of \cref{sec:randomization_tests} to test for the presence of treatment effects, that is, deviations from the sharp null hypothesis. We employ test statistics defined at both the outcome unit and the intervention unit level, as described in Equations (\ref{eq:random-outcome-level}) and (\ref{eq:y_teststatistics_n}).

\begin{table}[!b]
    \centering
    \begin{tabular}{ccccc}
    &&&\multicolumn{2}{c}{P-values}\\
      Design & Unit level & \makecell{Test\\statistic} &\makecell{outcome: \\percentage\\ (\%) variation}  & \makecell{outcome: \\ log\\ difference} \\
         
        \hline
       & Outcome & $W^{tot}$ &0.067 &0.052\\
      CRD & Outcome & $ \overline W$ &0.112&0.110\\
        & Intervention & Difference &0.006&0.004\\
        \hline
       & Outcome & $W^{tot}$ &0.070&0.058\\
      Bernoulli & Outcome &  $\overline W$ &0.130&0.124\\
       & Intervention & Difference &0.021&0.018\\
        \hline
    \end{tabular}
    \caption{P-values for randomization tests for Fisher's sharp null. Results correspond to the two definitions of the outcome along the columns, and the two designs (CRD, Bernoulli) and test statistics along the rows.}
    \label{tab:tests1degree}
\end{table}

Results for both hypothesized designs are reported in Table \ref{tab:tests1degree}. P-values under CRDs are smaller across all test statistics and levels of analysis, indicating that there is stronger evidence of treatment effects under this design. This is expected because of the improved precision in completely randomized designs compared to Bernoulli trials that translates  into more powerful hypothesis testing \citep{imbens_causal_2015}. The CR design achieves better inferential performance through the fixed number of treated units: this eliminates uncertainty about the groups' size, which reduces variance of test statistics  and leads to more powerful tests; by design, completely randomized experiments ensure more balanced treatment and control groups;  the distribution of test statistics under the null hypothesis is narrower, allowing  sharper distinctions between null and alternative hypotheses.

We find some deviations from the sharp null hypothesis in both designs, suggesting the presence of nontrivial effects. All p-values are below the nominal 0.05 level for the test statistics at the intervention unit level.
At the outcome unit level, p-values for the test statistic that employs $W^{tot}$ are smaller compared to those that employ $\overline W$. This likely reflects that what matters for employment growth is the number of completed lines, not merely the proportion of completed lines. 

In this setting, test statistics at the intervention level appear to have more power to detect causal effects than test statistics at the outcome unit level. As discussed in \cref{sec:randomization_tests}, there is no reason to expect that test statistics at one level are uniformly more powerful than those at the outcome level. In our setting, the intervention-level statistics more clearly reveal deviations from the null: specifically, they suggest that the completion of high-speed rail lines has an impact on employment growth in certain prefectures.

\subsection{Estimates of different causal estimands}

In \cref{tab:application-estimates}, we report estimates and standard errors obtained by omitting, for each estimand separately, the outcome units that violate the positivity condition, as described in \cref{subsec:application-positivity} and listed in \cref{tab:posviolations}.
For comparability, in \cref{tab:application-estimates-union} we present estimates employing the same set of units for all estimands, by excluding the outcome units that violate positivity for at least one estimand. Reported standard errors for the all-or-none estimand and the estimands employing stochastic interventions are based on the variance upper bounds derived in \cref{sec:estimation}.
% Specifically, in Tables \ref{tab:est2log diffunion} and \ref{tab:est2percvarunion} units are used that violate none of the positivity conditions for all estimands (see Table 3).

%\commentG{Fabri, I made the tables prettier -- can you do the same in the supplement?}

\begin{table}[!b]
    \centering
    \begin{tabular}{cccccccc}
         \makecell{Design} & All-or-None & \makecell{All-vs-\\Status Quo} & \makecell{Status Quo-vs-\\None} & +1 & $\tau^{(1)}_\text{stoch}$ & $\tau^{(2)}_\text{stoch}$ & $\tau^{(3)}_\text{stoch}$\\
         \hline
         \\
         \multicolumn{8}{c}{Outcome: Difference of log employment between 2007 and 2016} \\
        \cmidrule{2-7}  
         CRD  &0.106&  -0.003& 0.091& -0.008&-0.014&0.005&0.094\\
         & (0.102) & (0.049) & (0.076) &&(0.047)&(0.125)&(0.101)\\
         \hline
         Bernoulli  & 0.105&-0.005&0.105& -0.008&-0.013 &  0.007 &0.095\\
         & (0.094) & (0.059) & (0.073) &&(0.046)&(0.116)&(0.097)\\
          \hline \\
         \multicolumn{8}{c}{Outcome: Percent variation of employment between 2007 and 2016} \\
        \cmidrule{2-7} 
          CRD &13.744&  -0.818& 11.956&  -0.969&-2.282&1.034&12.292\\
          & (13.146)  &  (6.438)  &  (9.598)& &(6.210)&(17.547)&(13.324)\\
         \hline
         Bernoulli & 13.482 & -1.102 & 14.112 &  -0.965 & -2.149 &   1.237 &12.414\\
         & (12.037)     & (7.634)  &  (9.318)&&(6.087)&(16.102)&(12.748)\\
        \end{tabular}
    \caption{Estimates and standard errors (in parentheses) for the effects (columns) of HSR line completion on employment. The outcome is measured as difference in log employment, and as percent variation in employment between 2007 and 2016 (big rows). Results are reported under the different experimental designs (rows). Estimates are obtained removing units that violate positivity for each of the estimands separately.}
    
    \label{tab:application-estimates}
\end{table}

The all-or-none effect appears to be driven largely by lines that have already been completed. This is evident from the sizable status quo-vs-none effect. It is also evident from the negligible effect estimates for all estimands that describe treating additional units, such as the all-vs-status quo effect that describes the additional effect of completing all remaining lines, the $+1$ estimand that describes the effect of completing one more line randomly, $\tau^{(1)}_\text{stoch}$ which resembles the $+1$ estimand but stochastically, and $\tau^{(2)}_\text{stoch}$ which describes treating half of the remaining lines within each stratum.
% Interestingly, the +1 effect and the all-vs-status quo effect have similar signs and magnitudes. This suggests that completing one additional line or all remaining lines leads to the same modest negative impact, reinforcing the notion that most of the effect has already been realized.

The estimate of $\tau^{(3)}_\text{stoch}$, which was designed to resemble the all-or-none effect, is in fact of similar magnitude, with limited gains in variance, if any. Perhaps, this can be explained by the need to bound more terms in the variance formula for effects under stochastic interventions, which might render the employed upper bound more conservative for stochastic interventions rather than for the all-or-none effect.
Similarly, the estimate of $\tau^{(1)}_\text{stoch}$, which was designed to resemble the $+1$ effect, is comparable in size and sign within their uncertainty limits. %, suggesting consistency of results across local perturbations in treatment.

\begin{table}[!t]
    \centering
    \begin{tabular}{cccccccc}
         \makecell{Design} & All-or-None & \makecell{All-vs-\\Status Quo} & \makecell{Status Quo-vs-\\None} & +1 & $\tau^{(1)}_\text{stoch}$ & $\tau^{(2)}_\text{stoch}$ & $\tau^{(3)}_\text{stoch}$\\
         \hline
         \\
         \multicolumn{8}{c}{Outcome: Difference of log employment between 2007 and 2016} \\
        \cmidrule{2-7}  
        CRD  &0.106&  0.016& 0.091& -0.007&-0.006&0.010&0.094\\
         & (0.102) & (0.055) & (0.076) &&(0.047)&(0.129)&(0.096)\\
         \hline
         Bernoulli  & 0.105&0.000&0.105& -0.007&-0.018 &  0.013&0.095 \\
         & (0.094) & (0.062) & (0.073) &&(0.046)&(0.123)&(0.096)\\
          \hline \\
         \multicolumn{8}{c}{Outcome: Percent variation of employment between 2007 and 2016} \\
        \cmidrule{2-7} 
          CRD  &13.744&  1.789& 11.956&  -0.880&-1.034&1.652 &12.292 \\
          & (13.146)  &  (7.114)  &  (9.598) &&(6.304)&(18.046)&(12.508)\\
         \hline
         Bernoulli & 13.482 & -0.630 & 14.112 &  -0.917 & -2.842 &   2.022 &12.414\\
         & (12.037)     & (8.017)  &  (9.318)&&(6.171) &(17.162)&(12.506)\\
        \end{tabular}
    \caption{Estimates and standard errors (in parentheses) for the effects (columns) of HSR line completion on employment. The outcome is measured as difference in log employment, and as percent variation in employment between 2007 and 2016 (big rows). Results are reported under the different experimental designs (rows). Estimates are obtained removing units that violate positivity for any estimand, resulting in 213 units under CRD and 219 units under Bernoulli design.}
    
    \label{tab:application-estimates-union}
\end{table}

\section{Discussion and directions for future research}

This paper studies methods for causal inference in settings characterized by bipartite interference, where intervention and outcome units are drawn from distinct populations, linked by a known bipartite graph. While interference in causal inference has received growing attention in recent years, bipartite interference moves beyond the traditional unipartite or partially interfering structures. 

In a general setting, we introduce policy-relevant estimands and corresponding design-based estimators that accommodate general forms of interference without restrictive assumptions on potential outcomes or exposure mappings.
A central innovation lies in our design-based estimation framework, which yields unbiased estimators under arbitrary designs. 
%This approach allows for consistent inference in large samples, under reasonable and verifiable assumptions on the bipartite graph structure.
We highlight how different estimands, ranging from all-or-none contrasts to stochastic and incremental interventions, vary in their statistical efficiency and interpretability, and we provide conditions under which they can be estimated reliably.
Notably, we investigate the graph and sample size conditions under which design-based estimators of all-or-none and stochastic estimands are consistent, under general designs.
Our application to high-speed rail construction in China illustrates both the richness and the practical challenges of implementing these methods in policy-relevant domains.

For practitioners, our framework offers a flexible yet principled toolkit for evaluating interventions in settings such as environmental policy, education, infrastructure planning, and public health, that often exhibit a bipartite structure. Importantly, the estimands we define, especially those based on incremental treatment expansions or stochastic interventions, can often better align with feasible policy levers, making our methods directly relevant for policy learning and optimization in experimental as well as quasi-experimental settings.

There are a number of open questions that persist for future research. First,
% we assume knowledge of the bipartite graph structure. It would be interesting to incorporate uncertainty in the bipartite graph in the estimation framework. Second, 
although our estimands allow for general forms of treatment effect heterogeneity, we do not explicitly model or estimate heterogeneous effects with respect to characteristics of intervention or outcome units, an extension that could be particularly valuable for policy. 
Furthermore, a promising line of work would investigate how features of the graph such as centrality, clustering, or degree distribution mediate causal effects, with implications for optimal treatment allocation and network-based policy design (e.g., \cite{kim2024optimalenvironmentalpoliciespolicy} and \cite{Viviano2022}).
% Third, our design-based estimators rely on positivity conditions that may be violated in sparse or highly unbalanced graphs, suggesting the need for robust estimation or smoothing techniques when full support is lacking.
% Furthermore, extending these methods to observational settings requires further method development.
% such as graph-aware propensity score methods.
%potentially drawing on recent advances in network and spatial causal inference. 
% Fourth, a promising line of work would also investigate how features of the graph such as centrality, clustering, or degree distribution mediate causal effects, with implications for optimal treatment allocation and network-based policy design.
Lastly, an interesting line of future research would provide a general framework for finding optimal experimental designs for targeting specific estimands.
We hope this paper offers a foundation for future methodological and empirical work in this emerging area, and provides researchers and practitioners with tools to address the complex causal questions in bipartite settings.

\section*{Acknowledgments}
We are grateful to Cory Zigler and Fredrik S\"{a}vje for the insightful discussions.
We would like to thank Miguel Nunes for excellent research assistance. This work was supported by EUI Research Council funding (2023 and 2024), by a generous grant from the Office for Naval Research under grant numbers N00014-17-1-2131 and N00014-19-1-2468 and by a gift from Amazon.

\bibliographystyle{abbrvnat}
\bibliography{refs.bib}

\begin{thebibliography}{68}
\providecommand{\natexlab}[1]{#1}
\providecommand{\url}[1]{\texttt{#1}}
\expandafter\ifx\csname urlstyle\endcsname\relax
  \providecommand{\doi}[1]{doi: #1}\else
  \providecommand{\doi}{doi: \begingroup \urlstyle{rm}\Url}\fi

\bibitem[Aronow(2012)]{aronow2012}
P.~Aronow.
\newblock A general method for detecting interference between units in
  randomized experiments.
\newblock \emph{Sociological Methods \& Research}, 41\penalty0 (1):\penalty0
  3--16, 2012.

\bibitem[Aronow and Samii(2013)]{aronow2013conservative}
P.~M. Aronow and C.~Samii.
\newblock Conservative variance estimation for sampling designs with zero
  pairwise inclusion probabilities.
\newblock \emph{Survey Methodology}, 39\penalty0 (1):\penalty0 231--241, 2013.

\bibitem[Aronow and Samii(2017)]{aronow2017estimating}
P.~M. Aronow and C.~Samii.
\newblock Estimating average causal effects under general interference, with
  application to a social network experiment.
\newblock \emph{The Annals of Applied Statistics}, 11\penalty0 (4):\penalty0
  1912--1947, 2017.

\bibitem[Athey et~al.(2018)Athey, Eckles, and Imbens]{athey2018exact}
S.~Athey, D.~Eckles, and G.~W. Imbens.
\newblock Exact p-values for network interference.
\newblock \emph{Journal of the American Statistical Association}, 113\penalty0
  (521):\penalty0 230--240, 2018.

\bibitem[Baird et~al.(2018)Baird, Bohren, McIntosh, and
  {\"O}zler]{baird2018optimal}
S.~Baird, J.~A. Bohren, C.~McIntosh, and B.~{\"O}zler.
\newblock Optimal design of experiments in the presence of interference.
\newblock \emph{Review of Economics and Statistics}, 100\penalty0 (5):\penalty0
  844--860, 2018.

\bibitem[Bajari et~al.(2023)Bajari, Burdick, Imbens, Masoero, McQueen,
  Richardson, and Rosen]{bajari2023experimental}
P.~Bajari, B.~Burdick, G.~W. Imbens, L.~Masoero, J.~McQueen, T.~S. Richardson,
  and I.~M. Rosen.
\newblock Experimental design in marketplaces.
\newblock \emph{Statistical Science}, 38\penalty0 (3):\penalty0 458--476, 2023.

\bibitem[Basse and Feller(2018)]{basse2018analyzing}
G.~Basse and A.~Feller.
\newblock Analyzing two-stage experiments in the presence of interference.
\newblock \emph{Journal of the American Statistical Association}, 113\penalty0
  (521):\penalty0 41--55, 2018.

\bibitem[Basse et~al.(2019{\natexlab{a}})Basse, Feller, and
  Toulis]{basse2019randomization}
G.~Basse, A.~Feller, and P.~Toulis.
\newblock Randomization tests of causal effects under interference.
\newblock \emph{Biometrika}, 2019{\natexlab{a}}.

\bibitem[Basse and Airoldi(2018)]{BasseAiroldi2018}
G.~W. Basse and E.~M. Airoldi.
\newblock Limitations of design-based causal inference and a/b testing under
  arbitrary and network interference.
\newblock \emph{Sociological Methodology}, 48:\penalty0 136--151, 2018.

\bibitem[Basse et~al.(2019{\natexlab{b}})Basse, Feller, and
  Toulis]{bassefellertoulis}
G.~W. Basse, A.~Feller, and P.~Toulis.
\newblock {Randomization tests of causal effects under interference}.
\newblock \emph{Biometrika}, 106\penalty0 (2):\penalty0 487--494, 02
  2019{\natexlab{b}}.
\newblock ISSN 0006-3444.
\newblock \doi{10.1093/biomet/asy072}.

\bibitem[Borusyak and Hull(2023)]{borusyakhull2023}
K.~Borusyak and P.~Hull.
\newblock Non-random exposure to exogenous shocks.
\newblock \emph{Econometrica}, 91\penalty0 (6):\penalty0 2155--2185, 2023.

\bibitem[Borusyak et~al.(2025{\natexlab{a}})Borusyak, Hull, and
  Jaravel]{borusyak2025design}
K.~Borusyak, P.~Hull, and X.~Jaravel.
\newblock Design-based identification with formula instruments: A review.
\newblock \emph{The Econometrics Journal}, 28\penalty0 (1):\penalty0 83--108,
  2025{\natexlab{a}}.

\bibitem[Borusyak et~al.(2025{\natexlab{b}})Borusyak, Hull, and
  Jaravel]{borusyak2025practical}
K.~Borusyak, P.~Hull, and X.~Jaravel.
\newblock A practical guide to shift-share instruments.
\newblock \emph{Journal of Economic Perspectives}, 39\penalty0 (1):\penalty0
  181--204, 2025{\natexlab{b}}.

\bibitem[Bowers et~al.(2013)Bowers, Fredrickson, and
  Panagopoulos]{bowers2013reasoning}
J.~Bowers, M.~M. Fredrickson, and C.~Panagopoulos.
\newblock Reasoning about interference between units: A general framework.
\newblock \emph{Political Analysis}, 21\penalty0 (1):\penalty0 97--124, 2013.

\bibitem[Brennan et~al.(2022)Brennan, Mirrokni, and
  Pouget-Abadie]{Brennanetal2022}
J.~Brennan, V.~Mirrokni, and J.~Pouget-Abadie.
\newblock Cluster randomized designs for one-sided bipartite experiments.
\newblock \emph{Advances in Neural Information Processing Systems},
  35:\penalty0 37962--37974, 2022.

\bibitem[Chattopadhyay et~al.(2023)Chattopadhyay, Imai, and
  Zubizarreta]{chattopadhyay2023design}
A.~Chattopadhyay, K.~Imai, and J.~R. Zubizarreta.
\newblock Design-based inference for generalized network experiments with
  stochastic interventions.
\newblock \emph{arXiv preprint arXiv:2312.03268}, 2023.

\bibitem[Chen et~al.(2024)Chen, Bargagli-Stoffi, Kim, Henneman, and
  Nethery]{chen2024differenceindifferencesbipartitenetworkinterference}
K.~L. Chen, F.~J. Bargagli-Stoffi, R.~C. Kim, L.~R.~F. Henneman, and R.~C.
  Nethery.
\newblock Difference-in-differences under bipartite network interference: A
  framework for quasi-experimental assessment of the effects of environmental
  policies on health.
\newblock 2024.

\bibitem[Choi(2017)]{choi2017networkexp}
D.~Choi.
\newblock Estimation of monotone treatment effects in network experiments.
\newblock \emph{Journal of the American Statistical Association}, 112\penalty0
  (519):\penalty0 1147--1155, 2017.

\bibitem[Crema(2022)]{Crema2022}
A.~Crema.
\newblock School competition and classroom segregation.
\newblock \emph{Unpublished Manuscript}, 2022.

\bibitem[D{\'\i}az~Mu{\~n}oz and Van Der~Laan(2012)]{diaz2012population}
I.~D{\'\i}az~Mu{\~n}oz and M.~Van Der~Laan.
\newblock Population intervention causal effects based on stochastic
  interventions.
\newblock \emph{Biometrics}, 68\penalty0 (2):\penalty0 541--549, 2012.

\bibitem[Doudchenko et~al.(2020)Doudchenko, Zhang, Drynkin, Airoldi, Mirrokni,
  and Pouget-Abadie]{doudchenko_causal_2020}
N.~Doudchenko, M.~Zhang, E.~Drynkin, E.~M. Airoldi, V.~Mirrokni, and
  J.~Pouget-Abadie.
\newblock Causal {Inference} with {Bipartite} {Designs}.
\newblock {SSRN} {Scholarly} {Paper} 3757188, Social Science Research Network,
  Rochester, NY, Nov. 2020.

\bibitem[Eckles et~al.(2017)Eckles, Karrer, and
  Ugander]{EcklesKarrerUgander+2017}
D.~Eckles, B.~Karrer, and J.~Ugander.
\newblock Design and analysis of experiments in networks: Reducing bias from
  interference.
\newblock \emph{Journal of Causal Inference}, 5\penalty0 (1):\penalty0
  20150021, 2017.
\newblock \doi{doi:10.1515/jci-2015-0021}.

\bibitem[Fisher(1935)]{Fisher1935}
R.~Fisher.
\newblock \emph{The Design of Experiments}.
\newblock Oliver and Boyd, London, first edition, 1935.

\bibitem[Forastiere et~al.(2021)Forastiere, Airoldi, and
  Mealli]{forastiere_identification_2021}
L.~Forastiere, E.~M. Airoldi, and F.~Mealli.
\newblock Identification and {Estimation} of {Treatment} and {Interference}
  {Effects} in {Observational} {Studies} on {Networks}.
\newblock \emph{Journal of the American Statistical Association}, 116\penalty0
  (534):\penalty0 901--918, Apr. 2021.
\newblock ISSN 0162-1459.
\newblock \doi{10.1080/01621459.2020.1768100}.
\newblock Publisher: Taylor \& Francis.

\bibitem[Forastiere et~al.(2022)Forastiere, Mealli, Wu, and
  Airoldi]{forastiere_estimating_2022}
L.~Forastiere, F.~Mealli, A.~Wu, and E.~Airoldi.
\newblock Estimating {Causal} {Effects} under {Network} {Interference} with
  {Bayesian} {Generalized} {Propensity} {Scores}.
\newblock \emph{Jounral of Machine Learning Research}, 23\penalty0
  (289):\penalty0 1--61, 2022.
\newblock arXiv: 1807.11038.

\bibitem[Goldsmith-Pinkham et~al.(2020)Goldsmith-Pinkham, Sorkin, and
  Swift]{goldsmith2020bartik}
P.~Goldsmith-Pinkham, I.~Sorkin, and H.~Swift.
\newblock Bartik instruments: What, when, why, and how.
\newblock \emph{American Economic Review}, 110\penalty0 (8):\penalty0
  2586--2624, 2020.

\bibitem[Grossi et~al.(2020)Grossi, Lattarulo, Mariani, Mattei, and
  Oner]{grossi2020synthetic}
G.~Grossi, P.~Lattarulo, M.~Mariani, A.~Mattei, and O.~Oner.
\newblock Synthetic control group methods in the presence of interference: The
  direct and spillover effects of light rail on neighborhood retail activity.
\newblock \emph{arXiv preprint arXiv:2004.05027}, 2020.

\bibitem[Grossi et~al.(2025)Grossi, Mattei, and
  Papadogeorgou]{grossi2025spatial}
G.~Grossi, A.~Mattei, and G.~Papadogeorgou.
\newblock Spatial vertical regression for spatial panel data: Evaluating the
  effect of the florentine tramway's first line on commercial vitality.
\newblock \emph{arXiv preprint arXiv:2505.00450}, 2025.

\bibitem[Halloran and Struchiner(1995)]{halloran_causal_1995}
M.~E. Halloran and C.~J. Struchiner.
\newblock Causal inference in infectious diseases.
\newblock \emph{Epidemiology}, 6\penalty0 (2):\penalty0 142, 1995.
\newblock ISSN 1044-3983.

\bibitem[Harshaw et~al.(2023)Harshaw, S{\"a}vje, Eisenstat, Mirrokni, and
  Pouget-Abadie]{Harshawetal2023}
C.~Harshaw, F.~S{\"a}vje, D.~Eisenstat, V.~Mirrokni, and J.~Pouget-Abadie.
\newblock {Design and analysis of bipartite experiments under a linear
  exposure-response model}.
\newblock \emph{Electronic Journal of Statistics}, 17\penalty0 (1):\penalty0
  464 -- 518, 2023.
\newblock \doi{10.1214/23-EJS2111}.

\bibitem[Henneman et~al.(2019{\natexlab{a}})Henneman, Choirat, Ivey, Cummiskey,
  and Zigler]{henneman_characterizing_2019}
L.~R.~F. Henneman, C.~Choirat, C.~Ivey, K.~Cummiskey, and C.~M. Zigler.
\newblock Characterizing population exposure to coal emissions sources in the
  {United} {States} using the {HyADS} model.
\newblock \emph{Atmospheric Environment}, 203:\penalty0 271--280, Apr.
  2019{\natexlab{a}}.
\newblock ISSN 1352-2310.
\newblock \doi{10.1016/j.atmosenv.2019.01.043}.

\bibitem[Henneman et~al.(2019{\natexlab{b}})Henneman, Choirat, and
  Zigler]{henneman_accountability_2019}
L.~R.~F. Henneman, C.~Choirat, and C.~M. Zigler.
\newblock Accountability {Assessment} of {Health} {Improvements} in the
  {United} {States} {Associated} with {Reduced} {Coal} {Emissions} {Between}
  2005 and 2012.
\newblock \emph{Epidemiology}, 30\penalty0 (4):\penalty0 477, July
  2019{\natexlab{b}}.
\newblock ISSN 1044-3983.
\newblock \doi{10.1097/EDE.0000000000001024}.

\bibitem[Hudgens and Halloran(2008)]{hudgens_toward_2008}
M.~Hudgens and E.~Halloran.
\newblock Toward causal inference with interference.
\newblock \emph{Journal of the American Statistical Association}, 103\penalty0
  (482):\penalty0 832--842, June 2008.

\bibitem[Imbens and Rubin(2015{\natexlab{a}})]{imbens2015causal}
G.~W. Imbens and D.~B. Rubin.
\newblock \emph{Causal Inference in Statistics, Social, and Biomedical
  Sciences}.
\newblock Cambridge University Press, 2015{\natexlab{a}}.

\bibitem[Imbens and Rubin(2015{\natexlab{b}})]{imbens_causal_2015}
G.~W. Imbens and D.~B. Rubin.
\newblock \emph{Causal {Inference} for {Statistics}, {Social}, and {Biomedical}
  {Sciences}: {An} {Introduction}}.
\newblock Cambridge University Press, Cambridge, 2015{\natexlab{b}}.
\newblock ISBN 978-0-521-88588-1.
\newblock \doi{10.1017/CBO9781139025751}.

\bibitem[Jayachandran et~al.(2017)Jayachandran, de~Laat, Lambin, Stanton, Audy,
  and Thomas]{Jayachandranetal2017}
S.~Jayachandran, J.~de~Laat, E.~Lambin, C.~Y. Stanton, R.~Audy, and N.~E.
  Thomas.
\newblock Cash for carbon: A randomized trial of payments for ecosystem
  services to reduce deforestation.
\newblock \emph{Science}, 357\penalty0 (6348):\penalty0 267--273, 2017.

\bibitem[Kelli et~al.(2019)Kelli, Kim, Samman~Tahhan, Liu, Ko, Hammadah,
  Sullivan, Sandesara, Alkhoder, Choudhary, et~al.]{kelli2019living}
H.~M. Kelli, J.~H. Kim, A.~Samman~Tahhan, C.~Liu, Y.-A. Ko, M.~Hammadah,
  S.~Sullivan, P.~Sandesara, A.~A. Alkhoder, F.~K. Choudhary, et~al.
\newblock Living in food deserts and adverse cardiovascular outcomes in
  patients with cardiovascular disease.
\newblock \emph{Journal of the American Heart Association}, 8\penalty0
  (4):\penalty0 e010694, 2019.

\bibitem[Kim et~al.(2024)Kim, Bargagli-Stoffi, Chen, and
  Nethery]{kim2024optimalenvironmentalpoliciespolicy}
R.~C. Kim, F.~J. Bargagli-Stoffi, K.~L. Chen, and R.~C. Nethery.
\newblock Towards optimal environmental policies: Policy learning under
  arbitrary bipartite network interference.
\newblock 2024.

\bibitem[Lawrence et~al.(2019)Lawrence, Bullock, and Liu]{lawrence2019china}
M.~Lawrence, R.~Bullock, and Z.~Liu.
\newblock \emph{China's high-speed rail development}.
\newblock World Bank Publications, 2019.

\bibitem[Leung(2022)]{Leung2022}
M.~P. Leung.
\newblock Causal inference under approximate neighborhood interference.
\newblock \emph{Econometrica}, 90\penalty0 (1):\penalty0 267--293, 2022.
\newblock \doi{https://doi.org/10.3982/ECTA17841}.

\bibitem[Li and Wager(2022)]{LiWager2022}
S.~Li and S.~Wager.
\newblock Random graph asymptotics for treatment effect estimation under
  network interference.
\newblock \emph{The Annals of Statistics}, 50\penalty0 (4):\penalty0
  2334--2358, 2022.

\bibitem[Liu et~al.(2016)Liu, Hudgens, and Becker-Dreps]{liu_inverse_2016}
L.~Liu, M.~G. Hudgens, and S.~Becker-Dreps.
\newblock On inverse probability-weighted estimators in the presence of
  interference.
\newblock \emph{Biometrika}, 103\penalty0 (4):\penalty0 829--842, Dec. 2016.
\newblock ISSN 0006-3444.
\newblock \doi{10.1093/biomet/asw047}.

\bibitem[Lo et~al.(2023)Lo, Olivella, and Imai]{LoOlivellaImai2023}
A.~Lo, S.~Olivella, and K.~Imai.
\newblock A statistical model of bipartite networks: Application to
  cosponsorship in the united states senate.
\newblock \emph{ArXiv}, 2305.05833v1, 2023.

\bibitem[Lu et~al.(2025)Lu, Shi, Fang, Zhang, and Ding]{lu2025design}
S.~Lu, L.~Shi, Y.~Fang, W.~Zhang, and P.~Ding.
\newblock Design-based causal inference in bipartite experiments.
\newblock \emph{arXiv preprint arXiv:2501.09844}, 2025.

\bibitem[Luo et~al.(2012)Luo, Small, Li, and Rosenbaum]{luo_inference_2012}
X.~Luo, D.~S. Small, C.-S.~R. Li, and P.~R. Rosenbaum.
\newblock Inference {With} {Interference} {Between} {Units} in an {fMRI}
  {Experiment} of {Motor} {Inhibition}.
\newblock \emph{Journal of the American Statistical Association}, 107\penalty0
  (498):\penalty0 530--541, June 2012.
\newblock ISSN 0162-1459.
\newblock \doi{10.1080/01621459.2012.655954}.

\bibitem[Ma(2022)]{ma2022localized}
X.~Ma.
\newblock \emph{Localized bargaining: The political economy of China's
  high-Speed Railway program}.
\newblock Oxford University Press, 2022.

\bibitem[Manski(2013)]{manski2013identification}
C.~F. Manski.
\newblock Identification of treatment response with social interactions.
\newblock \emph{The Econometrics Journal}, 16\penalty0 (1):\penalty0 S1--S23,
  2013.

\bibitem[Ogburn et~al.(2022)Ogburn, Sofrygin, Díaz, and van~der
  Laan]{ogburn_causal_2022}
E.~L. Ogburn, O.~Sofrygin, I.~Díaz, and M.~J. van~der Laan.
\newblock Causal {Inference} for {Social} {Network} {Data}.
\newblock \emph{Journal of the American Statistical Association}, 0\penalty0
  (0):\penalty0 1--15, 2022.
\newblock ISSN 0162-1459.
\newblock \doi{10.1080/01621459.2022.2131557}.
\newblock Publisher: Taylor \& Francis \_eprint:
  https://doi.org/10.1080/01621459.2022.2131557.

\bibitem[Papadogeorgou et~al.(2019)Papadogeorgou, Mealli, and
  Zigler]{papadogeorgou2019causal}
G.~Papadogeorgou, F.~Mealli, and C.~M. Zigler.
\newblock Causal inference with interfering units for cluster and population
  level treatment allocation programs.
\newblock \emph{Biometrics}, 75\penalty0 (3):\penalty0 778--787, 2019.

\bibitem[Papadogeorgou et~al.(2022)Papadogeorgou, Imai, Lyall, and
  Li]{papadogeorgou2022causal}
G.~Papadogeorgou, K.~Imai, J.~Lyall, and F.~Li.
\newblock Causal inference with spatio-temporal data: estimating the effects of
  airstrikes on insurgent violence in iraq.
\newblock \emph{Journal of the Royal Statistical Society Series B: Statistical
  Methodology}, 84\penalty0 (5):\penalty0 1969--1999, 2022.

\bibitem[Pouget-Abadie et~al.(2019)Pouget-Abadie, Aydin, Schudy, Brodersen, and
  Mirrokni]{pouget-abadie_variance_2019}
J.~Pouget-Abadie, K.~Aydin, W.~Schudy, K.~Brodersen, and V.~Mirrokni.
\newblock Variance {Reduction} in {Bipartite} {Experiments} through
  {Correlation} {Clustering}.
\newblock In \emph{Advances in {Neural} {Information} {Processing} {Systems}},
  volume~32. Curran Associates, Inc., 2019.

\bibitem[Puelz et~al.(2022)Puelz, Basse, Feller, and
  Toulis]{puelzbassefellertoulis2022}
D.~Puelz, G.~Basse, A.~Feller, and P.~Toulis.
\newblock A graph-theoretic approach to randomization tests of causal effects
  under interferenc.
\newblock \emph{Journal of the Royal Statistical Society, Series B},
  84\penalty0 (1):\penalty0 174--204, 2022.

\bibitem[Rosenbaum(2007)]{Rosenbaum2007}
P.~R. Rosenbaum.
\newblock Interference between units in randomized experiments.
\newblock \emph{Journal of the American Statistical Association}, 102\penalty0
  (477):\penalty0 191--200, 2007.

\bibitem[Rubin(1980)]{rubin1980randomization}
D.~B. Rubin.
\newblock Comment on: Randomization analysis of experimental data: The fisher
  randomization test.
\newblock \emph{Journal of the American Statistical Association}, 75\penalty0
  (371):\penalty0 591--593, 1980.

\bibitem[S{\"a}vje(2024)]{Savje2024}
F.~S{\"a}vje.
\newblock {Causal inference with misspecified exposure mappings: separating
  definitions and assumptions}.
\newblock \emph{Biometrika}, 111\penalty0 (1):\penalty0 1 -- 15, 2024.

\bibitem[S{\"a}vje et~al.(2021)S{\"a}vje, Aronow, and Hudgens]{Savje2021}
F.~S{\"a}vje, P.~M. Aronow, and M.~G. Hudgens.
\newblock Average treatment effects in the presence of unknown interference.
\newblock \emph{Annals of Statistics}, 49\penalty0 (2):\penalty0 673--701,
  2021.

\bibitem[Schnell and Papadogeorgou(2020)]{Schnell2020mitigating}
P.~M. Schnell and G.~Papadogeorgou.
\newblock {Mitigating unobserved spatial confounding when estimating the effect
  of supermarket access on cardiovascular disease deaths}.
\newblock \emph{The Annals of Applied Statistics}, 14\penalty0 (4):\penalty0
  2069 -- 2095, 2020.
\newblock \doi{10.1214/20-AOAS1377}.

\bibitem[Sekhon and Shem-Tov(2020)]{sekhon2020inference}
J.~S. Sekhon and Y.~Shem-Tov.
\newblock Inference on a new class of sample average treatment effects.
\newblock \emph{Journal of the American Statistical Association}, pages 1--7,
  2020.

\bibitem[Sobel(2006)]{sobel2006randomized}
M.~E. Sobel.
\newblock What do randomized studies of housing mobility demonstrate? causal
  inference in the face of interference.
\newblock \emph{Journal of the American Statistical Association}, 101\penalty0
  (476):\penalty0 1398--1407, 2006.

\bibitem[Song and Papadogeorgou(2024)]{song2024bipartite}
Z.~Song and G.~Papadogeorgou.
\newblock Bipartite causal inference with interference, time series data, and a
  random network.
\newblock \emph{arXiv}, 2024.

\bibitem[Stock(1989)]{stock1989nonparametric}
J.~H. Stock.
\newblock Nonparametric policy analysis.
\newblock \emph{Journal of the American Statistical Association}, 84\penalty0
  (406):\penalty0 567--575, 1989.

\bibitem[Stock(1991)]{stock1989waste}
J.~H. Stock.
\newblock Nonparametric policy analysis:an application to estimating hazardous
  waste cleanup benefits.
\newblock \emph{Nonparametric and Semiparametric Methods in Econometrics and
  Statistics: Proceedings of the Fifth International Symposium in Economic
  Theory and Econometrics. Copertina anteriore William A. Barnett, James
  Powell, George E. Tauchen (Eds.)}, Chapter 3:\penalty0 77--98, 1991.

\bibitem[Tchetgen~Tchetgen and VanderWeele(2012)]{tchetgen_causal_2012}
E.~J. Tchetgen~Tchetgen and T.~J. VanderWeele.
\newblock On causal inference in the presence of interference.
\newblock \emph{Statistical Methods in Medical Research}, 21\penalty0
  (1):\penalty0 55--75, Feb. 2012.
\newblock ISSN 0962-2802, 1477-0334.
\newblock \doi{10.1177/0962280210386779}.

\bibitem[Viviano(2025)]{Viviano2022}
D.~Viviano.
\newblock Experimental design under network interference.
\newblock \emph{The Review of Economic Studies}, 92\penalty0 (2):\penalty0
  1257–1292, 2025.

\bibitem[Wang et~al.(2023)Wang, Samii, Chang, and Aronow]{wang2023design}
Y.~Wang, C.~Samii, H.~Chang, and P.~Aronow.
\newblock Design-based inference for spatial experiments with unknown
  interference.
\newblock \emph{arXiv preprint arXiv:2010.13599v4}, 2023.

\bibitem[Wooldridge(2015)]{wooldridge2015control}
J.~M. Wooldridge.
\newblock Control function methods in applied econometrics.
\newblock \emph{Journal of Human Resources}, 50\penalty0 (2):\penalty0
  420--445, 2015.

\bibitem[Zigler and Papadogeorgou(2021)]{zigler2021bipartite}
C.~M. Zigler and G.~Papadogeorgou.
\newblock Bipartite causal inference with interference.
\newblock \emph{Statistical science: a review journal of the Institute of
  Mathematical Statistics}, 36\penalty0 (1):\penalty0 109, 2021.

\bibitem[Zigler et~al.(2025)Zigler, Liu, Mealli, and Forastiere]{zigler2023air}
C.~M. Zigler, V.~Liu, F.~Mealli, and L.~Forastiere.
\newblock Bipartite interference and air pollution transport: Estimating health
  effects of power plant interventions.
\newblock \emph{Biostatistics}, 26\penalty0 (1), 2025.

\end{thebibliography}
\clearpage

\appendix
\setstretch{1.4}

\setcounter{page}{1}
\setcounter{section}{0}    
\renewcommand{\thetheorem}{S.\arabic{theorem}}
\renewcommand{\thedefinition}{S.\arabic{definition}}
\renewcommand{\theproposition}{S.\arabic{proposition}}
\renewcommand{\theassumption}{S.\arabic{assumption}}
\renewcommand{\theequation}{S.\arabic{equation}}
\renewcommand{\thetable}{S.\arabic{table}}
\renewcommand{\thefigure}{S.\arabic{figure}}

\begin{center}
    \Large Supplementary material
\end{center}

\section{Proofs of theoretical results}
\label{supp_sec:proofs}

\subsection{All-or-none and status-quo estimators}

\begin{proof}[Proof of \cref{theorem:unbiasedness}]
We use $\caln_m^c$ to denote the complement of $\caln_m$. Let $\bm a = \bm 0_N, \bm 1_N$ and $\bm a_{N_m} = \bm 0_{N_m}, \bm 1_{N_m}$:
\begin{enumerate}[label = (\alph*)]
\item 
\begin{align*}
\E\Big[\widehat Y(\bm a) \Big]
&=
\sum_{\bm w \in \{0, 1\}^N} \left[ 
\frac1M \sum_{m = 1}^M
\frac{I(\bm w_{\caln_m} = \bm a_{N_m})}
{\pi_{\caln_m}(\bm w_{\caln_m})} 
Y_m
\right] \pi(\bm w) \\
&= \frac1M \sum_{m = 1}^M
\sum_{\bm w_{\caln_m}} \sum_{\bm w_{\caln_m^c}}
\frac{I(\bm w_{\caln_m} = \bm a_{N_m})}
{\pi_{\caln_m}(\bm w_{\caln_m})} 
Y_m(\bm a_{N_m})
\pi(\bm w) \\
&=
\frac1M \sum_{m = 1}^M \sum_{\bm w_{\caln_m}}
\frac{I(\bm w_{\caln_m} =\bm a_{N_m})}
{\pi_{\caln_m}(\bm w_{\caln_m})} 
Y_m(\bm a_{N_m})
\sum_{\bm w_{\caln_m^c}} \pi(\bm w) \\
&=
\frac1M \sum_{m = 1}^M \sum_{\bm w_{\caln_m}}
I(\bm w_{\caln_m} =\bm a_{N_m})
Y_m(\bm a_{N_m})
\\
&= \overline Y( \bm a)
\end{align*}

\item
\begin{align*}
M^2 & \Var \Big[ \widehat Y(\bm a) \Big] 
= M^2 \E \left( \widehat Y(\bm a) - \overline Y(\bm a) \right) ^2 \\
&=
\E \left[ \sum_{m = 1}^M \left( \frac{I(\bm W_{\caln_m} = \bm a_{N_m})}{\pi_{\caln_m} (\bm W_{\caln_m})} Y_m - Y_m(\bm a_{N_m}) \right) \right]^2 \\
&=
\E \left[ \sum_{m = 1}^M \left( \frac{I(\bm W_{\caln_m} =\bm a_{N_m})}{\pi_{\caln_m} (\bm a_{N_m})} Y_m(\bm a_{N_m}) - Y_m(\bm a_{N_m}) \right) \right]^2 \\
&=
\E \left[ \sum_{m = 1}^M [ I(\bm W_{\caln_m} = \bm a_{N_m}) - \pi_{\caln_m} (\bm a_{N_m}) ] \frac{Y_m(\bm a_{N_m})}{\pi_{\caln_m} (\bm a_{N_m})} \right]^2 \\
&=
\E \Bigg[ \sum_{m = 1}^M
[ I(\bm W_{\caln_m} = \bm a_{N_m}) - \pi_{\caln_m} (\bm a_{N_m})]^2 \frac{ Y_m(\bm a_{N_m})^2}{ \pi_{\caln_m}^2 (\bm a_{N_m})^2} + \\
& \hspace{10pt}
 \sum_{m \neq m'}
[I(\bm W_{\caln_m} = \bm a_{N_m}) - \pi_{\caln_m} (\bm a_{N_m})] [I(\bm W_{\caln_{m'}} = \bm a_{N_{m'}}) - \pi_{\caln_{m'}} (\bm a_{N_{m'}}) ]
\frac{Y_m(\bm a_{N_m}) Y_{m'}(\bm a_{N_{m'}})}{\pi_{\caln_m} (\bm a_{N_m}) \pi_{\caln_{m'}} (\bm a_{N_{m'}})} \Bigg] \\
&= \sum_{m = 1}^M
\pi_{\caln_m}(\bm a_{N_m}) \big( 1 - \pi_{\caln_m}(\bm a_{N_m}) \big)
\frac{ Y_m(\bm a_{N_m})^2}{ \pi_{\caln_m}^2 (\bm a_{N_m})^2} \ + \\
& \hspace{30pt}
\sum_{m \neq m'}
\left( \pi_{\caln_m \cup \caln_{m'}}(\bm a_{|\caln_m \cup \caln_{m'}|}) - \pi_{\caln_m}(\bm a_{N_m}) \pi_{\caln_{m'}}(\bm a_{N_{m'}}) \right) 
\frac{Y_m(\bm a_{N_m}) Y_{m'}(\bm a_{N_{m'}})}{\pi_{\caln_m} (\bm a_{N_m}) \pi_{\caln_{m'}} (\bm a_{N_{m'}})},
\end{align*}
where the last line holds because $\E \big[ I(\bm W_{\caln_m} = \bm a_{N_m}) - \pi_{\caln_m} (\bm a_{N_m}) \big]^2$ is the variance of the indicator, and
$\E \big[ (I(\bm W_{\caln_m} = \bm a_{\caln_m}) - \pi_{\caln_m} (\bm a_{N_m})) (I(\bm W_{\caln_{m'}} = \bm a_{N_{m'}}) - \pi_{\caln_{m'}} (\bm a_{N_{m'}})) \big] $ is their covariance.

\item For the covariance term we have
\begin{align*}
    \mathrm{Cov}\Big[ \widehat Y(\bm 0_N), \widehat Y(\bm 1_N) \Big] &= 
    \frac1{M^2} \sum_{m, m'} \frac{Y_m(\bm 1_{N_m}) Y_{m'}(\bm 0_{N_{m'}})}{\pi_{\caln_m}(\bm 1_{N_m}) \pi_{\caln_{m'}}(\bm 0_{N_{m'}})} \text{Cov} \left( I(\bm W_{\caln_m} = \bm 1_{N_{m}}), I(\bm W_{\caln_{m'}} = \bm 0_{N_{m'}}) \right) \\
    = \frac1{M^2} \sum_{m, m'} &\frac{Y_m(\bm 1_{N_m}) Y_{m'}(\bm 0_{N_{m'}})}{\pi_{\caln_m}(\bm 1_{N_m}) \pi_{\caln_{m'}}(\bm 0_{N_{m'}})} \left[ \pi(\bm W_{\caln_m} = \bm 1_{N_{m}}, \bm W_{\caln_{m'}} = \bm 0_{N_{m'}}) - \pi_{\caln_m}(\bm 1_{N_{m}}) \pi_{\caln_{m'}}(\bm 0_{N_{m'}}) \right].
\end{align*}
To get the result, we split the sum in pairs $(m, m')$ for which $ \pi(\bm W_{\caln_m} = \bm 1_{N_{m}}, \bm W_{\caln_{m'}} = \bm 0_{N_{m'}}) \neq 0$ and those with $ \pi(\bm W_{\caln_m} = \bm 1_{N_{m}}, \bm W_{\caln_{m}} = \bm 0_{N_{m'}}) = 0$ (such as those with $\caln_m \cap \caln_{m'} \neq \varnothing$).
\end{enumerate}
\end{proof}

\begin{proof}[Proof of \cref{prop:var_estimation}]
Write the true variance of the estimator for the average potential outcome, $\Var \Big[ \widehat Y(\bm a) \Big]$, as
\begin{align*}
& \smashoperator{\sum_m}
c_{m,m}
\frac{Y_m(\bm a_{N_m})^2} {\pi_{\caln_m}(\bm a_{N_m}) ^2} \\
& + \smashoperator{\sum_{\substack{m \neq m': \\ \pi_{\caln_m \cup \caln_{m'}}(\bm a_{|\caln_m \cup \caln_{m'}|}) \neq 0}}}
c_{m,m'}
\frac{Y_m(\bm a_{N_m})Y_{m'}(\bm a_{N_{m'}})} {\pi_{\caln_m}(\bm a_{N_m}) \pi_{\caln_{m'}}(\bm a_{N_{m'}})} \\
& + \smashoperator{\sum_{\substack{m \neq m': \\ \pi_{\caln_m \cup \caln_{m'}}(\bm a_{|\caln_m \cup \caln_{m'}|}) = 0}}}
c_{m,m'}
\frac{Y_m(\bm a_{N_m})Y_{m'}(\bm a_{N_{m'}})} {\pi_{\caln_m}(\bm a_{N_m}) \pi_{\caln_{m'}}(\bm a_{N_{m'}})} \\
=& 
\smashoperator{\sum_m}
\pi_{\caln_m}(\bm a_{N_m}) \left( 1 - \pi_{\caln_m}(\bm a_{N_m}) \right)
\frac{Y_m(\bm a_{N_m})^2} {\pi_{\caln_m}(\bm a_{N_m}) ^2} \\
& + \ \ \smashoperator{\sum_{\substack{m \neq m': \\ \pi_{\caln_m \cup \caln_{m'}}(\bm a_{|\caln_m \cup \caln_{m'}|}) \neq 0}}} \ \ 
\left[
\pi_{\caln_m \cup \caln_{m'}}(\bm a_{|\caln_m \cup \caln_{m'}|}) - \pi_{\caln_m}(\bm a_{N_m}) \pi_{\caln_{m'}}(\bm a_{N_{m'}})
\right]
\frac{Y_m(\bm a_{N_m})Y_{m'}(\bm a_{N_{m'}})} {\pi_{\caln_m}(\bm a_{N_m}) \pi_{\caln_{m'}}(\bm a_{N_{m'}})} \\
& - \ \ \smashoperator{\sum_{\substack{m \neq m': \\ \pi_{\caln_m \cup \caln_{m'}}(\bm a_{|\caln_m \cup \caln_{m'}|}) = 0}}} \ \ 
Y_m(\bm a_{N_m})Y_{m'}(\bm a_{N_{m'}})
\end{align*}
The first two terms of $\widehat \Var \left[ \widehat Y(\bm a) \right]$ are unbiased for the first two terms of $\Var \Big[ \widehat Y(\bm a) \Big]$. The proof for this follows steps that are identical to ones in the proof of unbiasedness for $\widehat Y(\bm a)$ seen in the proof of part \ref{item:bias_yhat} of \cref{theorem:unbiasedness}, which allows us to substitute $I(\bm W_{\caln_m} = \bm a_{N_m})$ with $\pi_{\caln_m}(\bm a_{N_m})$ and $I(\bm W_{\caln_m \cup \caln_{m'}} = \bm a_{|\caln_m \cup \caln_{m'}|})$ with $\pi_{\caln_m \cup\caln_{m'}}(\bm a_{|\caln_m \cup\caln_{m'}|})$.
For the last term, notice that 
\[
- Y_m(\bm a_{N_m}) Y_{m'}(\bm a_{N_{m'}}) \leq \frac12 \left( 
Y_m(\bm a_{N_m})^2 + Y_{m'}(\bm a_{N_{m'}})
\right),
\]
since for any two real numbers $a, b \in \mathbb{R}$, $(a + b)^2 \geq 0$. Then, following identical steps we can show that the third term of 
$\widehat \Var \left[ \widehat Y(\bm a) \right]$ is greater or equal in expectation than the third term of
$\Var \left[ \widehat Y(\bm a) \right]$, which completes the result. 

If the outcome has the same sign for all units, $Y_m(\bm a_{N_m}) Y_{m'}(\bm a_{N_{m'}}) \geq 0$, the third term in $\widehat \Var \left[ \widehat Y(\bm a) \right]$ is upper-bounded by 0. Then, this term need not be used to guarantee that
$\E \left[ \widehat \Var \left[ \widehat Y(\bm a) \right] \right] \geq \Var \left[ \widehat Y(\bm a) \right]$, and therefore should be excluded.

Following identical steps, we can show that the first term of $\widehat{\text{Cov}}^\text{LB} \left[ \widehat Y(\bm 0_N), \widehat Y(\bm 1_N) \right]$ which is equal to
\begin{align*}
\frac1{M^2}
\smashoperator{\sum_{\substack{m, m': \\ \pi_{\caln_m, \caln_{m'}}(\bm 0_{N_m}, \bm 1_{N_{m'}}) \neq 0}}}
\frac
{
I(\bm W_{\caln_m} = \bm 0_{N_m}) I(\bm W_{\caln_{m'}} = \bm 1_{N_{m'}})
}
{
\pi_{\caln_m}(\bm 0_{N_m}) \pi_{\caln_{m'}}(\bm 1_{N_{m'}})
}
\left[ 1 - \frac{\pi_{\caln_m}(\bm 0_{N_m}) \pi_{\caln_{m'}}(\bm 1_{N_{m'}})}{\pi_{\caln_m, \caln_{m'}}(\bm 0_{N_m}, \bm 1_{N_{m'}})} \right] 
    Y_m Y_{m'} 
\end{align*}
is unbiased for the first term of the true covariance $\text{Cov}^\text{LB} \left[ \widehat Y(\bm 0_N), \widehat Y(\bm 1_N) \right]$ in part \ref{item:var_aon} of the theorem, which is equal to
\begin{align*}
\frac1{M^2}
\smashoperator{\sum_{\substack{m, m': \\ \pi_{\caln_m, \caln_{m'}}(\bm 0_{N_m}, \bm 1_{N_{m'}}) \neq 0}}} 
    \left[ \pi_{\caln_m, \caln_{m'}}(\bm 0_{N_m}, \bm 1_{N_{m'}}) - \pi_{\caln_m}(\bm 0_{N_m}) \pi_{\caln_{m'}}(\bm 1_{N_{m'}}) \right] \frac{Y_m(\bm 0_{N_m}) Y_{m'}(\bm 1_{N_{m'}})}{\pi_{\caln_m}(\bm 0_{N_m}) \pi_{\caln_{m'}}(\bm 1_{N_{m'}})}
.
\end{align*}
Then, for the second term of $\widehat{\text{Cov}}^\text{LB} \left[ \widehat Y(\bm 0_N), \widehat Y(\bm 1_N) \right]$ we have that
\begin{align*}
\E \Bigg[
    &
\frac1{2M^2}
\smashoperator{\sum_{\substack{m, m': \\ \pi_{\caln_m, \caln_{m'}}(\bm 0_{N_m}, \bm 1_{N_{m'}}) = 0}}}
\left( 
\frac{I(\bm W_{\caln_m} = \bm 0_{N_m}) Y_m^2}{\pi_{\caln_m}(\bm 0_{N_m})} +
\frac{I(\bm W_{\caln_{m'}} = \bm 1_{N_{m'}}) Y_{m'}^2}{\pi_{\caln_{m'}}(\bm 1_{N_{m'}})} 
\right)
\Bigg] \\
&=
\frac1{2M^2}
\smashoperator{\sum_{\substack{m, m': \\ \pi_{\caln_m, \caln_{m'}}(\bm 0_{N_m}, \bm 1_{N_{m'}}) = 0}}}
\left( 
Y_m(\bm 0_{N_m})^2 +
Y_{m'}(\bm 1_{N_m'})^2 
\right) \\
& \geq \frac1{M^2}
\smashoperator{\sum_{\substack{m, m': \\ \pi_{\caln_m, \caln_{m'}}(\bm 0_{N_m}, \bm 1_{N_{m'}}) = 0}}}
Y_m(\bm 0_{N_m})
Y_{m'}(\bm 1_{N_m'}),
\end{align*}
which is the second term in $\text{Cov}^\text{LB} \left[ \widehat Y(\bm 0_N), \widehat Y(\bm 1_N) \right]$,
where the last inequality holds because for any $a, b \in \mathbb{R}$, $(a-b)^2 \geq 0$.
From this we have that
$\E \left\{ \widehat{\text{Cov}}^\text{LB} \left[ \widehat Y(\bm 0_N), \widehat Y(\bm 1_N)  \right] \right\} \leq
\text{Cov}^\text{LB} \left[ \widehat Y(\bm 0_N), \widehat Y(\bm 1_N) \right] $ which implies the result.

\end{proof}

% \begin{assumption}
%     All potential outcomes for all outcome units are bounded, \iteg, there exists $B \in \mathbb{R}^+$ such that $|Y_m (\bm w)| \leq B$ for all $m$ and $\bm w \in \calw^N$. 
%     \label{ass:po_bounded}
% \end{assumption}

\begin{proof}[Proof of \cref{theorem:consistency}]
% We are working under the assumption that all potential outcomes are non-negative and bounded in \cref{ass:po_bounded}.
Since the estimator is unbiased, it suffices to show that the variance converges to 0. From \cref{theorem:unbiasedness}, we have the form of $\Var[\widehat Y(a)]$, which we can write as
\begin{align*}
    % \Var[\widehat Y(\bm a)] =
    &  \frac1{M_s^2} \left[ 
    \sum_{m = 1}^{M_s} \left( \frac1{\pi_{s, \caln_m}(\bm a_{N_{m}})} - 1 \right) Y_m(\bm a_{N_{m}})^2 +
    \sum_{m \neq m'} \left( \frac{\pi_{s,\caln_m \cup \caln_{m'}}(\bm a_{|\caln_m \cup \caln_{m'}|})}{\pi_{s,\caln_m}(\bm a_{N_m}) \pi_{s,\caln_{m'}}(\bm a_{N_{m'}})} - 1 \right) Y_m(\bm a_{N_{m}}) Y_{m'}(\bm a_{N_{m'}}) 
    \right] \\
&\leq \frac1{M_s^2} \left[ 
    \sum_{m = 1}^{M_s} \left( \frac1{\pi_{s, \caln_m}(\bm a_{N_{m}})} - 1 \right) Y_m(\bm a_{\caln_m})^2 +
    \sum_{m \neq m'} \left| \frac{\pi_{s,\caln_m \cup \caln_{m'}}(\bm a_{|\caln_m \cup \caln_{m'}|})}{\pi_{s,\caln_m}(\bm a_{N_m}) \pi_{s,\caln_{m'}}(\bm a_{N_{m'}})} - 1 \right| |Y_m(\bm a_{N_{m}})| |Y_{m'}(\bm a_{N_{m'}})| 
    \right] 
    %\tag{Using that the terms in the first summation are always positive.}
    \\
    & \leq \frac{B^2}{M_s^2} \left[ M_s (\gamma_{s, a}^{-1} - 1) +
    \kappa_s \gamma_{s, a}^* \right],
\end{align*}
where in the last equality we have used that $\pi_{s,\caln_m \cup \caln_{m'}}(\bm a_{|\caln_m \cup \caln_{m'}|}) =\pi_{s,\caln_m}(\bm a_{N_m}) \pi_{s,\caln_{m'}}(\bm a_{N_{m'}})$ for pairs $(m, m') \not \in \mathcal K_s$.
Therefore, if $\gamma_{s, a}^{-1} = o(M_s)$, and $\kappa_s \gamma_{s, a} ^ * = o(M_s^2)$, then the estimator is consistent.

\end{proof}

\begin{proof}[Proof of \cref{cor:consistency_bernoulli}]
Under a Bernoulli design, $\pi_{\caln_m}(\bm a) = p^{N_m}$ where $p = P(W_n = a) \in (0, 1)$. Then,
\begin{itemize}[leftmargin=*, label=-]
    \item $\gamma_{s, a} = \min_m \pi_{\caln_m}(\bm a_{N_m}) = p^{d_{s, o}}$, where $d_{s, o}$ is the maximum size of outcome units' intervention set, 
    \item $\mathcal K_s$ is the set of outcome unit pairs with overlapping intervention sets, and
    \item $\displaystyle \gamma_{s, a}^* = \max_{(m, m') \in \mathcal{K}_s} \left\{ \frac{\pi_{s,\caln_m \cup \caln_{m'}}(\bm a_{|\caln_m \cup \caln_{m'}|})}{\pi_{s,\caln_m}(\bm a_{N_m}) \pi_{s,\caln_{m'}}(\bm a_{N_{m'}})} - 1 \right\} = p^{- d_{s, \kappa}} - 1 \leq p^{-d_{s, \kappa}}$.
\end{itemize}
Then, if $d_{s, o} = o(\log(M_s))$, and $d_{s, \kappa} = o(\log(M_s^2 / \kappa_s))$, the conditions of \cref{theorem:consistency} are satisfied, and the estimator is consistent.

% \begin{align*}
%     \Var[\widehat Y(a)]
%     &= \frac1{M_s^2} \left[ 
%     \sum_{m = 1}^{M_s} \frac{p^{N_{m,s}}(1 - p^{N_{m,s}}) Y_m(a)^2}{p^{2N_{m,s}}} 
%     +
%     \sum_{m \neq m'} \frac{(p^{N_{m,m',s}} - p^{N_{m,s}} p^{N_{m',s}}) Y_m(a) Y_{m'}(a)}{p^{N_{m,s}} p^{N_{m',s}}}
%     \right] \\
%     %
%     &= 
%     \frac1{M_s^2} \left[ 
%     \sum_{m = 1}^{M_s} \left(\frac1{p^{N_{m,s}}} - 1 \right) Y_m(a)^2 
%     +
%     \sum_{m \neq m'} \left( \frac1{p^{N_{m,s} + N_{m',s} - N_{m,m',s}}} - 1 \right) Y_m(a) Y_{m'}(a)
%     \right] \\
%     %
% \intertext{where $N_{m,m',s} = |\caln_{s, m} \cup \caln_{s, m'}|$ is the number of interventional units in the union of the intervention sets for outcome units $m$ and $m'$ under graph $\calg_s$, and therefore
% $N_{m,m',s} \leq N_{m,s} + N_{m',s}$}
%     %
%     &\leq 
%     \frac{B^2}{M_s^2} \left[ 
%     \sum_{m = 1}^{M_s} \left(\frac1{p^{N_{m,s}}} - 1 \right)
%     +
%     \sum_{m \neq m'} \left( \frac1{p^{N_{m,s} + N_{m',s} - N_{m,m',s}}} - 1 \right)
%     \right] \\
%     %
%     &\leq 
%     \frac{B^2}{M_s^2} \left[ M_s \left(\frac1{p^{d_{s, o}}} - 1 \right)
%     +
%     \kappa_s \left( \frac1{p^{d_{s, \kappa}}} - 1 \right)
%     \right], \\
%     %
% \intertext{where in the last equation we have used that the pairs $(m,m')$ without overlapping intervention sets do not contribute to the second sum,}
%     %
%     &= \frac{B^2}{M_s} \left(\frac1{p^{d_{s, o}}} - 1 \right) +
%     \frac{B^2}{M_s^2} \kappa_s \left( \frac1{p^{d_{s, \kappa}}} - 1 \right).
% \end{align*}

\end{proof}

\subsection{Estimators for stochastic interventions}

\begin{proof}[Proof of \cref{theorem:stochastic_unbiased}]
We prove the unbiasedness of the estimator under stochastic interventions and derive the form of its variance.
\begin{enumerate}[label = (\alph*)]
\item 
\begin{align*}
\E ( \widehat Y_{h_\alpha} ) &= \sum_{\bm w \in \calw^N} \left\{ 
\frac1M \sum_{m = 1}^M \frac{h_{m, \alpha}(\bm w_{\caln_m})}{\pi_{\caln_m}(\bm w_{\caln_m})} Y_m \right\} \pi(\bm w) \\
&= \frac1M \sum_{m = 1}^M \sum_{\bm w_{\caln_m}} \sum_{\bm w_{\caln \setminus \caln_m}} \left\{ \frac{h_{m, \alpha}(\bm w_{\caln_m})}{\pi_{\caln_m}(\bm w_{\caln_m})} Y_m(\bm w_{\caln_m}) \pi(\bm w) \right\} \\
&= \frac1M \sum_{m = 1}^M \sum_{\bm w_{\caln_m}} \left\{ \frac{h_{m, \alpha}(\bm w_{\caln_m})}{\pi_{\caln_m}(\bm w_{\caln_m})} Y_m(\bm w_{\caln_m})  \sum_{\bm w_{\caln \setminus \caln_m}} \pi(\bm w) \right\}  \\
&= \frac1M \sum_{m = 1}^M \sum_{\bm w_{\caln_m}}  h_{m, \alpha}(\bm w_{\caln_m}) Y_m(\bm w_{\caln_m})   \\
&= \overline Y_{h_\alpha}
\end{align*}

\item For the variance estimator, we have that
\begin{align*}
M^2 \Var (\widehat Y {h_\alpha} )  &= 
\E \left\{ 
\left[
\sum_{m = 1}^M \frac{h_{m, \alpha}(\bm W_{\caln_m})}{\pi_{\caln_m}(\bm W_{\caln_m})} Y_m -
\sum_{m = 1}^M \sum_{\bm w_{\caln_m}} h_{m, \alpha}(\bm w_{\caln_m}) Y_m(\bm w_{\caln_m})
\right]^2
\right\} \\
&= \E \left\{ 
\left\{
\sum_{m = 1}^M \left[ \frac{h_{m, \alpha}(\bm W_{\caln_m})}{\pi_{\caln_m}(\bm W_{\caln_m})} Y_m -
\sum_{\bm w_{\caln_m}} h_{m, \alpha}(\bm w_{\caln_m}) Y_m(\bm w_{\caln_m})
\right] \right\}^2
\right\} \\
&= \E \left\{ \sum_{m = 1}^M
T_{1,m}(\bm W_{\caln_m}) +  \sum_{m \neq m'} T_{2,m,m'}(\bm W_{\caln_m}, \bm W_{\caln_{m'}})
\right\},
\end{align*}
where
\[
T_{1,m}(\bm W_{\caln_m}) = \left[
 \frac{h_{m, \alpha}(\bm W_{\caln_m})}{\pi_{\caln_m}(\bm W_{\caln_m})} Y_m -
\sum_{\bm w_{\caln_m}} h_{m, \alpha}(\bm w_{\caln_m}) Y_m(\bm w_{\caln_m})
\right]^2
\]
and 
\begin{align*}
T_{2,m,m'}(\bm W_{\caln_m}, \bm W_{\caln_{m'}}) &=
\left[ \frac{h_{m, \alpha}(\bm W_{\caln_m})}{\pi_{\caln_m}(\bm W_{\caln_m})} Y_m -
\sum_{\bm w_{\caln_m}} h_{m, \alpha}(\bm w_{\caln_m}) Y_m(\bm w_{\caln_m}) \right] \times \\
& \hspace{40pt} \times
\left[  \frac{h_{m', \alpha}(\bm W_{\caln_{m'}})}{\pi_{\caln_{m'}}(\bm W_{\caln_{m'}})} Y_{m'} -
\sum_{\bm w_{\caln_{m'}}} h_{m', \alpha}(\bm w_{\caln_{m'}}) Y_{m'}(\bm w_{\caln_{m'}}) \right].
\end{align*}
Then, we write
\begin{align*}
& T_{1,m}(\bm W_{\caln_m}) = \\
&= \left[
\sum_{\bm w_{\caln_m}} \frac{h_{m, \alpha}(\bm w_{\caln_m})}{\pi_{\caln_m}(\bm w_{\caln_m})} Y_m(\bm w_{\caln_m}) I(\bm W_{\caln_m} = \bm w_{\caln_m}) - h_{m, \alpha}(\bm w_{\caln_m}) Y_m(\bm w_{\caln_m})
\right]^2 \\
&= \left[
\sum_{\bm w_{\caln_m}} \left[ I(\bm W_{\caln_m} = \bm w_{\caln_m}) - \pi_{\caln_m}(\bm w_{\caln_m}) \right]
\frac{h_{m, \alpha}(\bm w_{\caln_m})}{\pi_{\caln_m}(\bm w_{\caln_m})} Y_m(\bm w_{\caln_m})
\right]^2 \\
&=
\sum_{\bm w_{\caln_m}} \left[ I(\bm W_{\caln_m} = \bm w_{\caln_m}) - \pi_{\caln_m}(\bm w_{\caln_m}) \right]^2 \left[
\frac{h_{m, \alpha}(\bm w_{\caln_m})}{\pi_{\caln_m}(\bm w_{\caln_m})} Y_m(\bm w_{\caln_m})
\right]^2 + \\
& \hspace{40pt} +
\sum_{\bm w_{\caln_m}, \bm w_{\caln_m}'} 
\left[ I(\bm W_{\caln_m} = \bm w_{\caln_m}) - \pi_{\caln_m}(\bm w_{\caln_m}) \right]
\left[ I(\bm W_{\caln_m} = \bm w_{\caln_m}') - \pi_{\caln_m}(\bm w_{\caln_m}') \right] \times \\
& \hspace{100pt} \times
\frac{h_{m, \alpha}(\bm w_{\caln_m})}{\pi_{\caln_m}(\bm w_{\caln_m})} Y_m(\bm w_{\caln_m}) \frac{h_{m, \alpha}(\bm w_{\caln_m}')}{\pi_{\caln_m}(\bm w_{\caln_m}')} Y_m(\bm w_{\caln_m}')
\end{align*}
Note that the only randomness in the expression above comes through $I(\bm W_{\caln_m} = \cdot)$. Note also that using the formula for the variance and covariance of binary indicators we have that
\[
\E \left\{ \left[ I(\bm W_{\caln_m} = \bm w_{\caln_m}) - \pi_{\caln_m}(\bm w_{\caln_m}) \right]^2 \right\} = 
\pi_{\caln_m}(\bm w_{\caln_m}) \left[ 1 - \pi_{\caln_m}(\bm w_{\caln_m}) \right]
\]
and
\begin{align*}
\E \big[ I(\bm W_{\caln_m} = \bm w_{\caln_m}) - \pi_{\caln_m}(\bm w_{\caln_m}) \big]
\big[ I(\bm W_{\caln_m} = \bm w_{\caln_m}') - \pi_{\caln_m}(\bm w_{\caln_m}') \big] =
- \pi_{\caln_m}(\bm w_{\caln_m}) \pi_{\caln_m}(\bm w_{\caln_m}').
\end{align*}
From these we can acquire that
\begin{align*}
\E \Bigg[ \sum_{m = 1}^M & T_{1,m}(\bm W_{\caln_m}) \Bigg] = \\
&= 
\sum_{m = 1}^M \left\{ \sum_{\bm w_{\caln_m}}
\pi_{\caln_m}(\bm w_{\caln_m}) \left[ 1 - \pi_{\caln_m}(\bm w_{\caln_m}) \right]
\left[
\frac{h_{m, \alpha}(\bm w_{\caln_m})}{\pi_{\caln_m}(\bm w_{\caln_m})} Y_m(\bm w_{\caln_m})
\right]^2 + \right. \\
& \left. \hspace{40pt} 
+ \sum_{\bm w_{\caln_m}, \bm w_{\caln_m}'} \left[ - \pi_{\caln_m}(\bm w_{\caln_m}) \pi_{\caln_m}(\bm w_{\caln_m}') \right]
\frac{h_{m, \alpha}(\bm w_{\caln_m}) h_{m, \alpha}(\bm w_{\caln_m}')
Y_m(\bm w_{\caln_m})  Y_m(\bm w_{\caln_m}')}
{\pi_{\caln_m}(\bm w_{\caln_m}) \pi_{\caln_m}(\bm w_{\caln_m}')}
\right\} \\
&= 
\sum_{m = 1}^M \left\{ \sum_{\bm w_{\caln_m}}
\pi_{\caln_m}(\bm w_{\caln_m}) \left[ 1 - \pi_{\caln_m}(\bm w_{\caln_m}) \right]
\left[
\frac{h_{m, \alpha}(\bm w_{\caln_m})}{\pi_{\caln_m}(\bm w_{\caln_m})} Y_m(\bm w_{\caln_m})
\right]^2 - \right. \\
& \left. \hspace{40pt} 
- \sum_{\bm w_{\caln_m}, \bm w_{\caln_m}'}
h_{m, \alpha}(\bm w_{\caln_m}) h_{m, \alpha}(\bm w_{\caln_m}')
Y_m(\bm w_{\caln_m})  Y_m(\bm w_{\caln_m}')
\right\}.
\end{align*}
Now we turn our attention to $T_{2,m,m'}$. We write:
\begin{align*}
    & T_{2,m,m'} (\bm W_{\caln_m}, \bm W_{\caln_{m'}}) = \\
    &=
    \left[ \sum_{\bm w_{\caln_m}} \frac{h_{m, \alpha}(\bm w_{\caln_m})}{\pi_{\caln_m}(\bm w_{\caln_m})} Y_m(\bm w_{\caln_m}) I(\bm W_{\caln_m} = \bm w_{\caln_m}) -
    \sum_{\bm w_{\caln_m}} h_{m, \alpha}(\bm w_{\caln_m}) Y_m(\bm w_{\caln_m}) \right] \times \\
    & \hspace{20pt} \times
    \left[ \sum_{\bm w_{\caln_{m'}}} \frac{h_{m', \alpha}(\bm w_{\caln_{m'}})}{\pi_{\caln_{m'}}(\bm w_{\caln_{m'}})} Y_{m'}(\bm w_{\caln_{m'}}) I(\bm W_{\caln_{m'}} = \bm w_{\caln_{m'}}) -
    \sum_{\bm w_{\caln_{m'}}} h_{m', \alpha}(\bm w_{\caln_{m'}}) Y_{m'}(\bm w_{\caln_{m'}}) \right] \\
    &=
    \left[
    \sum_{\bm w_{\caln_m}} \frac{h_{m, \alpha}(\bm w_{\caln_m})}{\pi_{\caln_m}(\bm w_{\caln_m})} Y_m(\bm w_{\caln_m}) \left[ I(\bm W_{\caln_m} = \bm w_{\caln_m}) - \pi_{\caln_m}(\bm w_{\caln_m}) \right]
    \right] \times \\
    & \hspace{20pt} \times
    \left[ \sum_{\bm w_{\caln_{m'}}} \frac{h_{m', \alpha}(\bm w_{\caln_{m'}})}{\pi_{\caln_{m'}}(\bm w_{\caln_{m'}})} Y_{m'}(\bm w_{\caln_{m'}}) \left[ I(\bm W_{\caln_{m'}} = \bm w_{\caln_{m'}}) -
    \pi_{\caln_{m'}}(\bm w_{\caln_{m'}}) \right] \right] \\
    &= \sum_{\bm w_{\caln_m}}  \sum_{\bm w_{\caln_{m'}}} 
    \frac{h_{m, \alpha}(\bm w_{\caln_m})}{\pi_{\caln_m}(\bm w_{\caln_m})}
    \frac{h_{m', \alpha}(\bm w_{\caln_{m'}})}{\pi_{\caln_{m'}}(\bm w_{\caln_{m'}})}
    Y_m(\bm w_{\caln_m}) Y_{m'}(\bm w_{\caln_{m'}}) \times \\
    & \hspace{40pt} \times
    \left[ I(\bm W_{\caln_m} = \bm w_{\caln_m}) - \pi_{\caln_m}(\bm w_{\caln_m}) \right] 
    \left[ I(\bm W_{\caln_{m'}} = \bm w_{\caln_{m'}}) - \pi_{\caln_{m'}}(\bm w_{\caln_{m'}})
    \right].
\end{align*}
Again, we can use the correlation of Bernoulli indicators for calculating the expected value of product of centered indicators, since
\begin{align*}
& \E \left\{ \left[ I(\bm W_{\caln_m} = \bm w_{\caln_m}) - \pi_{\caln_m}(\bm w_{\caln_m}) \right] 
    \left[ I(\bm W_{\caln_{m'}} = \bm w_{\caln_{m'}}) - \pi_{\caln_{m'}}(\bm W_{\caln_{m'}})
    \right] \right\} = \\
&= \pi_{\caln_m, \caln_{m'}}(\bm w_{\caln_m}, \bm w_{\caln_{m'}}) -
\pi_{\caln_m}(\bm w_{\caln_m}) \pi_{\caln_{m'}}(\bm w_{\caln_{m'}}),
\end{align*}
where 
$\pi_{\caln_m, \caln_{m'}(\bm w_{\caln_m}, \bm w_{\caln_{m'}}}) =
\pi(\bm W_{\caln_m} = \bm w_{\caln_m}, \bm W_{\caln_{m'} = \bm w_{\caln_{m'}}})$.

From this, we have that
\begin{align*}
    & \E \left\{  \sum_{m \neq m'} T_{2,m,m'}(\bm W_{\caln_m}, \bm W_{\caln_{m'}}) \right\} =\\
    & =
\sum_{m \neq m'} \left[
\sum_{\bm w_{\caln_m}, \bm w_{\caln_{m'}}} 
\left(
\pi_{\caln_m, \caln_{m'}}(\bm w_{\caln_m}, \bm w_{\caln_{m'}}) -
\pi_{\caln_m}(\bm w_{\caln_m}) \pi_{\caln_{m'}}(\bm w_{\caln_{m'}}) 
\right) \times \right. \\
& \hspace{200pt} \left.    
\frac{
Y_m(\bm w_{\caln_m}) Y_{m'}(\bm w_{\caln_{m'}})
h_{m, \alpha}(\bm w_{\caln_m}) h_{m', \alpha}(\bm w_{\caln_{m'}})}
{\pi_{\caln_m}(\bm w_{\caln_m}) \pi_{\caln_{m'}}(\bm w_{\caln_{m'}})}
\right] \\
&=
\sum_{m \neq m'}
\sum_{\bm w_{\caln_m}, \bm w_{\caln_{m'}}} 
\left(
\frac{
\pi_{\caln_m, \caln_{m'}}(\bm w_{\caln_m}, \bm w_{\caln_{m'}})}{
\pi_{\caln_m}(\bm w_{\caln_m}) \pi_{\caln_{m'}}(\bm w_{\caln_{m'}})} - 1 
\right)   
Y_m(\bm w_{\caln_m}) Y_{m'}(\bm w_{\caln_{m'}})
h_{m, \alpha}(\bm w_{\caln_m}) h_{m', \alpha}(\bm w_{\caln_{m'}}),
\end{align*}
which completes the proof.

\end{enumerate}
\end{proof}

\begin{proof}[Proof of \cref{prop:var_estimation_stoch}]
$M^2 \Var \left( \widehat Y_{h_\alpha} \right)$ is equal to
\begin{align*}
& \sum_{m = 1}^M \left[ \sum_{\bm w_{\caln_m}} 
\pi_{\caln_m}(\bm w_{\caln_m}) \left( 1 - \pi_{\caln_m}(\bm w_{\caln_m}) \right)
 \left[ \frac{Y_m(\bm w_{\caln_m}) h_{m, \alpha}(\bm w_{\caln_m})}{\pi_{\caln_m}(\bm w_{\caln_m})} \right]^2 - \right. \\
 & \hspace{80pt} \left.
\sum_{\bm w_{\caln_m} \neq \bm w_{\caln_{m'}}} 
Y_m(\bm w_{\caln_m}) Y_m(\bm w_{\caln_{m'}})
h_{m, \alpha}(\bm w_{\caln_m}) h_{m, \alpha}(\bm w_{\caln_{m'}})
\right] + \\
+ &
\sum_{m \neq m'}
\sum_{\bm w_{\caln_m}, \bm w_{\caln_{m'}}} 
\left(
\frac{
\pi_{\caln_m, \caln_{m'}}(\bm w_{\caln_m}, \bm w_{\caln_{m'}})}
{
\pi_{\caln_m}(\bm w_{\caln_m}) \pi_{\caln_{m'}}(\bm w_{\caln_{m'}}) 
} 
-
1
\right)  
Y_m(\bm w_{\caln_m}) Y_{m'}(\bm w_{\caln_{m'}})
h_{m, \alpha}(\bm w_{\caln_m}) h_{m', \alpha}(\bm w_{\caln_{m'}}).
\end{align*}

The first term of the true variance can be estimated unbiasedly using 
\begin{align*}
\sum_{m = 1}^M  
 \left( 1 - \pi_{\caln_m}(\bm W_{\caln_m}) \right)
 \left[ \frac{h_{m, \alpha}(\bm W_{\caln_m})}{\pi_{\caln_m}(\bm W_{\caln_m})} Y_m\right]^2
\end{align*}

The proof for this goes as follows:

\begin{align*}
& \E\left[\sum_{m = 1}^M  
 \left( 1 - \pi_{\caln_m}(\bm W_{\caln_m}) \right)
 \left[ \frac{h_{m, \alpha}(\bm W_{\caln_m})}{\pi_{\caln_m}(\bm W_{\caln_m})} Y_m\right]^2\right] = \\
& \sum_{\bm w} \left[\sum_{m = 1}^M  
 \left( 1 - \pi_{\caln_m}(\bm W_{\caln_m}) \right)
 \left[ \frac{h_{m, \alpha}(\bm W_{\caln_m})}{\pi_{\caln_m}(\bm W_{\caln_m})} Y_m\right]^2 \right]\pi(\bm w) = \\
 & \sum_{m = 1}^M \sum_{\bm w_{\caln_m}} \sum_{\bm w_{\caln_m^c}}  
 \left( 1 - \pi_{\caln_m}(\bm w_{\caln_m}) \right)
 \left[ \frac{h_{m, \alpha}(\bm w_{\caln_m})}{\pi_{\caln_m}(\bm w_{\caln_m})} Y_m(\bm w_{\caln _m})\right]^2 \pi(\bm w) = \\
 & \sum_{m = 1}^M  \sum_{\bm w_{\caln_m}}   
 \left( 1 - \pi_{\caln_m}(\bm w_{\caln_m}) \right)
 \left[ \frac{h_{m, \alpha}(\bm w_{\caln_m})}{\pi_{\caln_m}(\bm w_{\caln_m})} Y_m(\bm w_{\caln _m})\right]^2 \sum_{\bm w_{\caln_m^c}} \pi(\bm w) = \\
 & \sum_{m = 1}^M \left[ \sum_{\bm w_{\caln_m}} 
\pi_{\caln_m}(\bm w_{\caln_m}) \left( 1 - \pi_{\caln_m}(\bm w_{\caln_m}) \right)
 \left[ \frac{Y_m(\bm w_{\caln_m}) h_{m, \alpha}(\bm w_{\caln_m})}{\pi_{\caln_m}(\bm w_{\caln_m})} \right]^2 \right]
\end{align*}
where the last equality comes from $\sum_{\bm w_{\caln_m^c}} \pi(\bm w)=\pi(\bm w_{\caln_m})$. 

For the second term of the true variance consider: 
\begin{align*}
    &
- \sum_{m = 1}^M \sum_{\bm w_{\caln_m} \neq \bm w_{\caln_{m'}}} 
Y_m(\bm w_{\caln_m}) Y_m(\bm w_{\caln_{m'}})
h_{ \alpha}(\bm w_{\caln_m}) h_{m, \alpha}(\bm w_{\caln_{m'}})
 \\
\leq \ &
\frac1{2} \sum_{m = 1}^M \sum_{\bm w_{\caln_m} \neq \bm w_{\caln_{m'}}}
\left( Y_m(\bm w_{\caln_m})^2 + Y_m(\bm w_{\caln_{m'}})^2  \right)h_{m, \alpha}(\bm w_{\caln_m}) h_{m, \alpha}(\bm w_{\caln_{m'}}) \\
= & \frac1{2} \sum_{m = 1}^M \left[ \sum_{\bm w_{\caln_m}}
 Y_m(\bm w_{\caln_m})^2 h_{m, \alpha}(\bm w_{\caln_m})\sum_{\bm w_{\caln_{m'}} \neq \bm w_{\caln_m}}  h_{m, \alpha}(\bm w_{\caln_{m'}})    \right. \\
& \hspace{40pt} + \left. \sum_{\bm w_{\caln_{m'}}}
 Y_m(\bm w_{\caln_{m'}})^2  h_{m, \alpha}(\bm w_{\caln_{m'}}) \sum_{\bm w_{\caln_m} \neq \bm w_{\caln_{m'}}}  h_{m, \alpha}(\bm w_{\caln_m}) \right] \\
 = & \sum_{m = 1}^M \sum_{\bm w_{\caln_m}} Y_m(\bm w_{\caln_m})^2  h_{m, \alpha}(\bm w_{\caln_m}) (1- h_{m, \alpha}(\bm w_{\caln_m}))
\end{align*}
where the first inequality holds because for any $a, b \in \mathbb{R}$, $(a+b)^2 \geq 0 \iff -2ab\leq a^2+b^2$.  An unbiased estimator for this quantity is:
\begin{align*}
\sum_{m = 1}^M  
 \left( 1 - h_{m, \alpha}(\bm W_{\caln_m}) \right)
 \frac{h_{m, \alpha}(\bm W_{\caln_m})}{\pi_{\caln_m}(\bm W_{\caln_m})} Y_m^2
\end{align*}
The proof of unbiasedness for this estimator of the upper bound of the second term follows steps that are identical to ones in the proof of unbiasedness of the first term, and it is omitted.

The last term of the true variance can be divided into pairs of $\bm w_{\caln_m}$ and $\bm w_{\caln_{m'}}$ that can be observed simultaneously or not:
\begin{align*}
& \sum_{m \neq m'} \left[ \ \ \ \ \ \ \ \ \ \ \ \
\smashoperator{
\sum_{\substack{\bm w_{\caln_m}, \bm w_{\caln_{m'}}\\ \pi_{\caln_m, \caln_{m'}}(\bm w_{\caln_m}, \bm w_{\caln_{m'}}) \neq 0}}}  \ \ \ \
\left(
\frac{
\pi_{\caln_m, \caln_{m'}}(\bm w_{\caln_m}, \bm w_{\caln_{m'}})}
{
\pi_{\caln_m}(\bm w_{\caln_m}) \pi_{\caln_{m'}}(\bm w_{\caln_{m'}}) 
} 
-
1
\right)  
Y_m(\bm w_{\caln_m}) Y_{m'}(\bm w_{\caln_{m'}})
h_{m, \alpha}(\bm w_{\caln_m}) h_{m', \alpha}(\bm w_{\caln_{m'}}) \right.\\
& \hspace{40pt} + \left. \hspace{30pt}
\smashoperator{\sum_{\substack{\bm w_{\caln_m}, \bm w_{\caln_{m'}}\\ \pi_{\caln_m, \caln_{m'}}(\bm w_{\caln_m}, \bm w_{\caln_{m'}}) = 0}} } \ \ \ \ 
\left(
\frac{
\pi_{\caln_m, \caln_{m'}}(\bm w_{\caln_m}, \bm w_{\caln_{m'}})}
{
\pi_{\caln_m}(\bm w_{\caln_m}) \pi_{\caln_{m'}}(\bm w_{\caln_{m'}}) 
} 
-
1
\right)  
Y_m(\bm w_{\caln_m}) Y_{m'}(\bm w_{\caln_{m'}})
h_{m, \alpha}(\bm w_{\caln_m}) h_{m', \alpha}(\bm w_{\caln_{m'}}) \right]
\end{align*}

The first term can be estimated unbiasedly, while the second will be bounded. The estimator for the first term is
\begin{align*}
    \sum_{m \neq m'}
 \dfrac{h_{m, \alpha}(\bm W_{\caln_m}) h_{m', \alpha}(\bm W_{\caln_{m'}})}
{\pi_{\caln_m, \caln_{m'}}(\bm W_{\caln_m}, \bm W_{\caln_{m'}})}
\left(
\dfrac{
\pi_{\caln_m, \caln_{m'}}(\bm W_{\caln_m}, \bm W_{\caln_{m'}})}
{
\pi_{\caln_m}(\bm W_{\caln_m}) \pi_{\caln_{m'}}(\bm W_{\caln_{m'}}) 
} 
-
1
\right)  
Y_m Y_{m'}
\end{align*}
Then, it holds that
\begin{align*}
& \E\left[\sum_{m \neq m'}
 \dfrac{h_{m, \alpha}(\bm W_{\caln_m}) h_{m', \alpha}(\bm W_{\caln_{m'}})}
{\pi_{\caln_m, \caln_{m'}}(\bm W_{\caln_m}, \bm W_{\caln_{m'}})}
\left(
\dfrac{
\pi_{\caln_m, \caln_{m'}}(\bm W_{\caln_m}, \bm W_{\caln_{m'}})}
{
\pi_{\caln_m}(\bm W_{\caln_m}) \pi_{\caln_{m'}}(\bm W_{\caln_{m'}}) 
} 
-
1
\right)  
Y_m Y_{m'} \right] = \\
& \sum_{m \neq m'} \ \ \ \ \ \ \ \ 
\smashoperator{\sum_{\substack{\bm w_{\caln_m}, \bm w_{\caln_{m'}}\\ \pi_{\caln_m, \caln_{m'}}(\bm w_{\caln_m}, \bm w_{\caln_{m'}}) \neq 0}} } \ \ \ \ \ 
\left(
\frac{
\pi_{\caln_m, \caln_{m'}}(\bm w_{\caln_m}, \bm w_{\caln_{m'}})}
{
\pi_{\caln_m}(\bm w_{\caln_m}) \pi_{\caln_{m'}}(\bm w_{\caln_{m'}}) 
} 
-
1
\right)  
Y_m(\bm w_{\caln_m}) Y_{m'}(\bm w_{\caln_{m'}})
h_{m, \alpha}(\bm w_{\caln_m}) h_{m', \alpha}(\bm w_{\caln_{m'}}).
\end{align*}
The proof for this follows steps that are identical to ones in the proof of unbiasedness of the first term of the true variance. 

For the last term, we start by deriving the upper bound. Because $\pi_{\caln_m, \caln_{m'}}(\bm w_{\caln_m}, \bm w_{\caln_{m'}}) = 0$, it equals:
\begin{align*}
& \sum_{m \neq m'} \ \ \ \ \ \ \ \ 
\smashoperator{\sum_{\substack{\bm w_{\caln_m}, \bm w_{\caln_{m'}}\\ \pi_{\caln_m, \caln_{m'}}(\bm w_{\caln_m}, \bm w_{\caln_{m'}}) = 0}} } \ \ \ \ \ \
\left(
-
1
\right)  
Y_m(\bm w_{\caln_m}) Y_{m'}(\bm w_{\caln_{m'}})
h_{m, \alpha}(\bm w_{\caln_m}) h_{m', \alpha}(\bm w_{\caln_{m'}}) \\
& \leq \frac1{2}
\sum_{m \neq m'} \ \ \ \ \ \ \ \ 
\smashoperator{\sum_{\substack{\bm w_{\caln_m}, \bm w_{\caln_{m'}}\\ \pi_{\caln_m, \caln_{m'}}(\bm w_{\caln_m}, \bm w_{\caln_{m'}}) = 0}} } \ \ \ \ \ \
\left( Y_m(\bm w_{\caln_m})^2 + Y_{m'}(\bm w_{\caln_{m'}} )^2 \right) h_{m, \alpha}(\bm w_{\caln_m}) h_{m', \alpha}(\bm w_{\caln_{m'}}) \\
& = \frac1{2} \sum_{m \neq m'}^M \left[ \sum_{\bm w_{\caln_m}}
 Y_m(\bm w_{\caln_m})^2 h_{m, \alpha}(\bm w_{\caln_m})  \ \ \ \smashoperator{\sum_{\substack{\bm w_{\caln_m'}:\\ \pi_{\caln_m, \caln_{m'}}(\bm w_{\caln_m}, \bm w_{\caln_{m'}}) = 0}}} \   h_{m, \alpha}(\bm w_{\caln_{m'}})    \right. \\
& \hspace{60pt} + \left. \sum_{\bm w_{\caln_{m'}}}
 Y_m(\bm w_{\caln_{m'}})^2  h_{m, \alpha}(\bm w_{\caln_{m'}}) \ \ \ \smashoperator{\sum_{\substack{\bm w_{\caln_m}:\\ \pi_{\caln_m, \caln_{m'}}(\bm w_{\caln_m}, \bm w_{\caln_{m'}}) = 0}}} \  h_{m, \alpha}(\bm w_{\caln_m}) \right] \\
 &= 
 \sum_{m \neq m'}^M \left[ \frac1{2} \sum_{\bm w_{\caln_m}}
 Y_m(\bm w_{\caln_m})^2 h_{m, \alpha}(\bm w_{\caln_m})    \lambda_{m,m'}(\bm w_{\caln_m}) \right. \\
& \hspace{60pt} + \left. \frac1{2} \sum_{\bm w_{\caln_{m'}}}
 Y_m(\bm w_{\caln_{m'}})^2  h_{m, \alpha}(\bm w_{\caln_{m'}}) \lambda_{m',m}(\bm w_{\caln_{m'}}) \right]
\end{align*}
for
\[
\lambda_{m,m'}(\bm w_{\caln_m}) = \ \ 
\smashoperator{\sum_{\substack{\bm w_{\caln_{m'}}: \\
\pi_{\caln_m, \caln_{m'}}(\bm w_{\caln_m}, \bm w_{\caln_{m'}}) = 0
}}} \ \
h_{m',\alpha}(\bm w_{\caln_{m'}}).
\]

This quantity can be estimated using 
\begin{align*}
\sum_{m \neq m'} \Bigg[\frac{h_{m, \alpha}(\bm W_{\caln_m})}{2 \pi_{\caln_m}(\bm W_{\caln_m})} \lambda_{m,m'}(\bm W_{\caln_m}) Y_m^2 +
\frac{h_{m', \alpha}(\bm W_{\caln_{m'}})}{2 \pi_{\caln_{m'}}(\bm W_{\caln_{m'}})} \lambda_{m',m}(\bm W_{\caln_{m'}}) Y_{m'}^2 \Bigg]
\end{align*}

Proof of unbiasedness for this quantity is identical of the proof of the first term of the variance estimator.

\end{proof}

\begin{proof}[Proof of \cref{prop:var_estimation_tau_stoch}]    
First, for the true variance we have that 
$M^2 \Var \left[ \widehat \tau(\alpha, \alpha') \right]$ is equal to
\begin{align*}
& \sum_{m = 1}^M \left\{ \sum_{\bm w_{\caln_m}} 
\pi_{\caln_m}(\bm w_{\caln_m}) \left( 1 - \pi_{\caln_m}(\bm w_{\caln_m}) \right)
 \left\{ \frac{Y_m(\bm w_{\caln_m}) \left[ h_{m, \alpha'}(\bm w_{\caln_m}) -  h_{m, \alpha}(\bm w_{\caln_m})\right]}{\pi_{\caln_m}(\bm w_{\caln_m})} \right\} ^2 - \right. \\
 & \hspace{30pt} \left.
\sum_{\bm w_{\caln_m} \neq \bm w_{\caln_m}'} 
Y_m(\bm w_{\caln_m}) Y_m(\bm w_{\caln_m}')
\left[ h_{m, \alpha'}(\bm w_{\caln_m}) - h_{m, \alpha}(\bm w_{\caln_m}) \right] 
\left[ h_{m, \alpha'}(\bm w_{\caln_m}') - h_{m, \alpha}(\bm w_{\caln_m}') \right]
\right\} + \\
+ &
\sum_{m \neq m'}
\sum_{\bm w_{\caln_m}, \bm w_{\caln_{m'}}} 
\Bigg\{ \left(
\frac{
\pi_{\caln_m, \caln_{m'}}(\bm w_{\caln_m}, \bm w_{\caln_{m'}})}
{
\pi_{\caln_m}(\bm w_{\caln_m}) \pi_{\caln_{m'}}(\bm w_{\caln_{m'}}) 
} 
-
1
\right)  
Y_m(\bm w_{\caln_m}) Y_{m'}(\bm w_{\caln_{m'}}) \times \\
&\hspace{100pt} \times
\left[ h_{m, \alpha'}(\bm w_{\caln_m}) - h_{m, \alpha}(\bm w_{\caln_m}) \right] \left[h_{m', \alpha'}(\bm w_{\caln_{m'}}) - h_{m', \alpha}(\bm w_{\caln_{m'}}) \right]
\Bigg\} .
\end{align*}

The first term of $\widehat \Var \left[ \widehat \tau(\alpha, \alpha') \right]$ is unbiased for the first term in the summation over $m$ of the true variance $\Var \left[ \widehat \tau(\alpha, \alpha') \right]$. The proof for this follows steps identical to the corresponding term in \cref{prop:var_estimation_stoch}.

The second term in the summation over $m$ for $\Var \left[ \widehat \tau(\alpha, \alpha') \right]$ is slightly more complicated than the corresponding term in $\Var \left( \widehat Y_{h_\alpha} \right)$ in \cref{prop:var_estimation_stoch} because the product of differences of the stochastic interventions need not be positive. For an outcome unit $m$, we write:
\begin{align*}
& - \smashoperator{ \sum_{\bm w_{\caln_m} \neq \bm w_{\caln_m}'}}
Y_m(\bm w_{\caln_m}) Y_m(\bm w_{\caln_m}') 
\left[ h_{m, \alpha'}(\bm w_{\caln_m}) - h_{m, \alpha}(\bm w_{\caln_m}) \right] 
\left[ h_{m, \alpha'}(\bm w_{\caln_m}') - h_{m, \alpha}(\bm w_{\caln_m}') \right]  \\
&= \ \  - \smashoperator{\sum_{\substack{\bm w_{\caln_m} \neq \bm w_{\caln_m}' \\ \text{pd}(\bm w_{\caln_m}, \bm w_{\caln_m}') \geq 0}}} \ \ 
Y_m(\bm w_{\caln_m}) Y_m(\bm w_{\caln_m}')
\left[ h_{m, \alpha'}(\bm w_{\caln_m}) - h_{m, \alpha}(\bm w_{\caln_m}) \right] 
\left[ h_{m, \alpha'}(\bm w_{\caln_m}') - h_{m, \alpha}(\bm w_{\caln_m}') \right] + \\
& \hspace{20pt}
 \ \ - \smashoperator{ \sum_{\substack{\bm w_{\caln_m} \neq \bm w_{\caln_m}' \\ \text{pd}(\bm w_{\caln_m}, \bm w_{\caln_m}') < 0}}} \ \ 
Y_m(\bm w_{\caln_m}) Y_m(\bm w_{\caln_m}')
\left[ h_{m, \alpha'}(\bm w_{\caln_m}) - h_{m, \alpha}(\bm w_{\caln_m}) \right] 
\left[ h_{m, \alpha'}(\bm w_{\caln_m}') - h_{m, \alpha}(\bm w_{\caln_m}') \right],
\intertext{where $\text{pd}(\bm w_{\caln_m}, \bm w_{\caln_m}') = \left[ h_{m, \alpha'}(\bm w_{\caln_m}) - h_{m, \alpha}(\bm w_{\caln_m}) \right] 
\left[ h_{m, \alpha'}(\bm w_{\caln_m}') - h_{m, \alpha}(\bm w_{\caln_m}') \right] $ represents the product of differences of the stochastic interventions at the two treatment vectors,}
& \leq \frac12 \ \ 
\smashoperator{\sum_{\substack{\bm w_{\caln_m} \neq \bm w_{\caln_m}' \\ \text{pd}(\bm w_{\caln_m}, \bm w_{\caln_m}') \geq 0}}} \ \ 
\left[ Y_m(\bm w_{\caln_m})^2 + Y_m(\bm w_{\caln_m}')^2 \right]
\left[ h_{m, \alpha'}(\bm w_{\caln_m}) - h_{m, \alpha}(\bm w_{\caln_m}) \right] 
\left[ h_{m, \alpha'}(\bm w_{\caln_m}') - h_{m, \alpha}(\bm w_{\caln_m}') \right] \\
& \hspace{15pt}
- \frac12 \ \  \smashoperator{ \sum_{\substack{\bm w_{\caln_m} \neq \bm w_{\caln_m}' \\ \text{pd}(\bm w_{\caln_m}, \bm w_{\caln_m}') < 0}}} \ \ 
\left[ Y_m(\bm w_{\caln_m}) ^ 2 + Y_m(\bm w_{\caln_m}') ^ 2 \right]
\left[ h_{m, \alpha'}(\bm w_{\caln_m}) - h_{m, \alpha}(\bm w_{\caln_m}) \right] 
\left[ h_{m, \alpha'}(\bm w_{\caln_m}') - h_{m, \alpha}(\bm w_{\caln_m}') \right],
\intertext{where we have used that for $a, b \in \mathbb{R}$ and $c > 0$ it holds that $-2abc \leq (a^2 + b^2)c$, whereas for $c < 0$ it holds that $-2abc \leq - (a^2 + b^2) c$ (these are derived from $(a\pm b)^2 \geq 0).$ With steps similar to those in the proof of \cref{prop:var_estimation_stoch}:}
& = \sum_{\bm w_{\caln_m}} \Bigg\{ Y_m(\bm w_{\caln_m})^2
\left[ h_{m, \alpha'}(\bm w_{\caln_m}) - h_{m, \alpha}(\bm w_{\caln_m}) \right] 
\ \ \smashoperator{\sum_{\substack{\bm w_{\caln_m}': \bm w_{\caln_m}' \neq \bm w_{\caln_m} \\ \text{pd}(\bm w_{\caln_m}, \bm w_{\caln_m}') \geq 0}}} \ \ 
\left[ h_{m, \alpha'}(\bm w_{\caln_m}') - h_{m, \alpha}(\bm w_{\caln_m}') \right] \Bigg\} \\
& \hspace{20pt} - 
\sum_{\bm w_{\caln_m}} \Bigg\{ Y_m(\bm w_{\caln_m})^2
\left[ h_{m, \alpha'}(\bm w_{\caln_m}) - h_{m, \alpha}(\bm w_{\caln_m}) \right] 
\ \ \smashoperator{\sum_{\substack{\bm w_{\caln_m}': \bm w_{\caln_m}' \neq \bm w_{\caln_m} \\ \text{pd}(\bm w_{\caln_m}, \bm w_{\caln_m}') < 0}}} \ \ 
\left[ h_{m, \alpha'}(\bm w_{\caln_m}') - h_{m, \alpha}(\bm w_{\caln_m}') \right] \Bigg\} \\
& =
\sum_{\bm w_{\caln_m}} \Bigg\{ Y_m(\bm w_{\caln_m})^2 \left| h_{m, \alpha'}(\bm w_{\caln_m}) - h_{m, \alpha}(\bm w_{\caln_m}) \right|
\ \ \smashoperator{\sum_{\bm w_{\caln_m}': \bm w_{\caln_m}' \neq \bm w_{\caln_m}}} \ \ 
\left| h_{m, \alpha'}(\bm w_{\caln_m}') - h_{m, \alpha}(\bm w_{\caln_m}') \right| \Bigg\} .
\end{align*}
The fact that the second term in $\widehat \Var \left[ \widehat \tau(\alpha, \alpha') \right]$ estimates this upper bound unbiasedly follows identically to the other terms, and the proof is therefore omitted.

For the third term in $\Var \left[ \widehat \tau(\alpha, \alpha') \right]$ which involves the summation over pairs of distinct units $(m, m')$ with their treatment vectors $\bm w_{\caln_m}$ and $\bm w_{\caln_{m'}}$, we split the summation $\sum_{\bm w_{\caln_m}, \bm w_{\caln_{m'}}}$ into two sums: one over pairs of treatment vectors that can be observed simultaneously ($\pi_{\caln_m, \caln_{m'}}(\bm w_{\caln_m}, \bm w_{\caln_{m'}}) \neq 0$), and one over pairs of treatment vectors that cannot be observed simultaneously ($\pi_{\caln_m, \caln_{m'}}(\bm w_{\caln_m}, \bm w_{\caln_{m'}}) = 0$). The fact that the third term in $\widehat \Var \left[ \widehat \tau(\alpha, \alpha') \right]$ estimates the term for the former unbiasedly follows with identical steps to the previous terms, and the proof is therefore omitted. We focus on the summation over treatment vectors that can\textit{not} be observed simultaneously which, since $\pi_{\caln_m, \caln_{m'}}(\bm w_{\caln_m}, \bm w_{\caln_{m'}}) = 0$, can be written as
\begin{align*}
& - \ \ \smashoperator{\sum_{\substack{\bm w_{\caln_m}, \bm w_{\caln_{m'}} \\
\pi_{\caln_m, \caln_{m'}}(\bm w_{\caln_m}, \bm w_{\caln_{m'}}) = 0}}} \ \
Y_m(\bm w_{\caln_m}) Y_{m'}(\bm w_{\caln_{m'}})
\left[ h_{m, \alpha'}(\bm w_{\caln_m}) - h_{m, \alpha}(\bm w_{\caln_m}) \right] \left[h_{m', \alpha'}(\bm w_{\caln_{m'}}) - h_{m', \alpha}(\bm w_{\caln_{m'}}) \right].
\intertext{Following steps similar to above, we have that}
\leq & \ \ \frac12 \ \ \ \ 
\smashoperator{\sum_{\substack{\bm w_{\caln_m}, \bm w_{\caln_{m'}} \\
\pi_{\caln_m, \caln_{m'}}(\bm w_{\caln_m}, \bm w_{\caln_{m'}}) = 0 \\
\text{pd}(\bm w_{\caln_m}, \bm w_{\caln_{m'}}) \geq 0}}} \ \ 
\left[ Y_m(\bm w_{\caln_m})^2 + Y_{m'}(\bm w_{\caln_{m'}})^2 \right]
\left[ h_{m, \alpha'}(\bm w_{\caln_m}) - h_{m, \alpha}(\bm w_{\caln_m}) \right] \left[h_{m', \alpha'}(\bm w_{\caln_{m'}}) - h_{m', \alpha}(\bm w_{\caln_{m'}}) \right] 
\\
& \hspace{10pt}
- \frac12 \ \ \ \  \smashoperator{\sum_{\substack{\bm w_{\caln_m}, \bm w_{\caln_{m'}} \\
\pi_{\caln_m, \caln_{m'}}(\bm w_{\caln_m}, \bm w_{\caln_{m'}}) = 0 \\
\text{pd}(\bm w_{\caln_m}, \bm w_{\caln_{m'}}) < 0}}} \ \ 
\left[ Y_m(\bm w_{\caln_m}) ^ 2 + Y_{m'}(\bm w_{\caln_{m'}}) ^ 2 \right]
\left[ h_{m, \alpha'}(\bm w_{\caln_m}) - h_{m, \alpha}(\bm w_{\caln_m}) \right] \left[h_{m', \alpha'}(\bm w_{\caln_{m'}}) - h_{m', \alpha}(\bm w_{\caln_{m'}})  \right], \\
\intertext{where here we define $\text{pd}(\bm w_{\caln_m}, \bm w_{\caln_{m'}}) =
\left[ h_{m, \alpha'}(\bm w_{\caln_m}) - h_{m, \alpha}(\bm w_{\caln_m}) \right] \left[h_{m', \alpha'}(\bm w_{\caln_{m'}}) - h_{m', \alpha}(\bm w_{\caln_{m'}})  \right]$,}
= &
\frac12 \sum_{\bm w_{\caln_m}} Y_m(\bm w_{\caln_m})^2 \left| h_{m, \alpha'}(\bm w_{\caln_m}) - h_{m, \alpha}(\bm w_{\caln_m}) \right|
\ \ \smashoperator{\sum_{\substack{\bm w_{\caln_{m'}} \\
\pi_{\caln_m, \caln_{m'}}(\bm w_{\caln_m}, \bm w_{\caln_{m'}}) = 0 }}} \ \ 
\left| h_{m', \alpha'}(\bm w_{\caln_{m'}}) - h_{m', \alpha}(\bm w_{\caln_{m'}})  \right| \\
& \hspace{10pt}
+ \frac12 \sum_{\bm w_{\caln_{m'}}}  Y_{m'}(\bm w_{\caln_{m'}})^2
\left| h_{m', \alpha'}(\bm w_{\caln_{m'}}) - h_{m', \alpha}(\bm w_{\caln_{m'}})  \right|
\ \ \smashoperator{\sum_{\substack{\bm w_{\caln_m} \\
\pi_{\caln_m, \caln_{m'}}(\bm w_{\caln_m}, \bm w_{\caln_{m'}}) = 0 }}} \ \ 
\left[ h_{m, \alpha'}(\bm w_{\caln_m}) - h_{m, \alpha}(\bm w_{\caln_m}) \right],
\intertext{which for $\lambda^\tau_{m,m'}(\bm w_{\caln_m}) = \smashoperator{\sum_{\substack{\bm w_{\caln_{m'}} \\
\pi_{\caln_m, \caln_{m'}}(\bm w_{\caln_m}, \bm w_{\caln_{m'}}) = 0 }}} \ \ 
\left| h_{m', \alpha'}(\bm w_{\caln_{m'}}) - h_{m', \alpha}(\bm w_{\caln_{m'}})  \right| $ can be written as}
&= \frac12 \sum_{\bm w_{\caln_m}} Y_m(\bm w_{\caln_m})^2 \left| h_{m, \alpha'}(\bm w_{\caln_m}) - h_{m, \alpha}(\bm w_{\caln_m}) \right|
\lambda^\tau_{m,m'}(\bm w_{\caln_m}) \\
& \hspace{10pt}
+ \frac12 \sum_{\bm w_{\caln_{m'}}}  Y_{m'}(\bm w_{\caln_{m'}})^2
\left| h_{m', \alpha'}(\bm w_{\caln_{m'}}) - h_{m', \alpha}(\bm w_{\caln_{m'}})  \right|
\lambda^\tau_{m', m}(\bm w_{\caln_{m'}}).
\end{align*}
Therefore, with similar steps to the other terms, we have that the last two terms in $\widehat \Var \left[ \widehat \tau(\alpha, \alpha') \right]$ estimate unbiasedly these two terms, which in turn form an upper bound for the quantity of interest in the true variance.
\end{proof}

\begin{proof}[Proof of \cref{theorem:consistency_stochastic}]
From \cref{theorem:stochastic_unbiased}, we know the estimator is unbiased and that for its variance we have that
\begin{align*}
\Var(\widehat Y_{h_\alpha}) \leq & \frac1{M_s^2}
\Bigg\{ \sum_{m = 1}^{M_s} \left[ \sum_{\bm w_{\caln_m}} 
\pi_{s, \caln_m}(\bm w_{\caln_m}) \left( 1 - \pi_{s, \caln_m}(\bm w_{\caln_m}) \right)
\frac{ | Y_m(\bm w_{\caln_m}) | ^2 h_{m, \alpha}(\bm w_{\caln_m})^2 }{\pi_{s, \caln_m}(\bm w_{\caln_m})^2} \ + \right. \\
 & \hspace{80pt} \left.
\sum_{\bm w_{\caln_m} \neq \bm w_{\caln_m}'} 
\left|  Y_m(\bm w_{\caln_m}) Y_m(\bm w_{\caln_m}') \right| \ 
h_{m, \alpha}(\bm w_{\caln_m}) h_{m, \alpha}(\bm w_{\caln_m}')
\right] + \\
+ &
\sum_{m \neq m'}
\sum_{\bm w_{\caln_m}, \bm w_{\caln_{m'}}} 
\left|
\frac{
\pi_{s, \caln_m, \caln_{m'}}(\bm w_{\caln_m}, \bm w_{\caln_{m'}})}
{
\pi_{s, \caln_m}(\bm w_{\caln_m}) \pi_{s, \caln_{m'}}(\bm w_{\caln_{m'}}) 
} 
-
1
\right|
\left| Y_m(\bm w_{\caln_m}) Y_{m'}(\bm w_{\caln_{m'}}) \right|
h_{m, \alpha}(\bm w_{\caln_m}) h_{m', \alpha}(\bm w_{\caln_{m'}})
\Bigg\} \\
\leq & 
\frac{B^2}{M_s^2} \sum_{m = 1}^{M_s} \left\{ \sum_{\bm w_{\caln_m}} 
\pi_{s, \caln_m}(\bm w_{\caln_m}) \left( 1 - \pi_{s, \caln_m}(\bm w_{\caln_m}) \right)
 \left[ \frac{h_{m, \alpha}(\bm w_{\caln_m})}{\pi_{s, \caln_m}(\bm w_{\caln_m})} \right]^2 + \right. \\
 & \hspace{80pt} \left.
\sum_{\bm w_{\caln_m} \neq \bm w_{\caln_{m'}}} 
h_{m, \alpha}(\bm w_{\caln_m}) h_{m, \alpha}(\bm w_{\caln_m}') \right\} + \\
& + \frac{B^2}{M_s^2}
\sum_{m \neq m'} 
\sum_{\bm w_{\caln_m}, \bm w_{\caln_{m'}}} 
\left|
\frac{
\pi_{s, \caln_m, \caln_{m'}}(\bm w_{\caln_m}, \bm w_{\caln_{m'}})}
{
\pi_{s, \caln_m}(\bm w_{\caln_m}) \pi_{s, \caln_{m'}}(\bm w_{\caln_{m'}}) 
} 
-
1
\right| 
h_{m, \alpha}(\bm w_{\caln_m}) h_{m', \alpha}(\bm w_{\caln_{m'}})
 \\
= &
\frac{B^2}{M_s^2} \sum_{m = 1}^M \sum_{\bm w_{\caln_m}} 
\pi_{s, \caln_m}(\bm w_{\caln_m}) \left( 1 - \pi_{s, \caln_m}(\bm w_{\caln_m}) \right)
\left[
\frac{h_{m, \alpha}(\bm w_{\caln_m})}{\pi_{s, \caln_m}(\bm w_{\caln_m})} \right]^2  + \\
& + \frac{B^2}{M_s^2} \sum_{m = 1}^{M_s}
 \sum_{\bm w_{\caln_m} \neq \bm w_{\caln_m}'}
h_{m, \alpha}(\bm w_{\caln_m}) h_{m, \alpha}(\bm w_{\caln_m}') + \\
 & + \frac{B^2}{M_s^2}
\sum_{(m, m') \in \mathcal K_s} \left\{
\sum_{\bm w_{\caln_m}, \bm w_{\caln_{m'}}} 
\left|
\frac{\pi_{s, \caln_m, \caln_{m'}}(\bm w_{\caln_m}, \bm w_{\caln_{m'}})}{\pi_{s, \caln_m}(\bm w_{\caln_m}) \pi_{s, \caln_{m'}}(\bm w_{\caln_{m'}})} - 1 
\right|
h_{m, \alpha}(\bm w_{\caln_m}) h_{m', \alpha}(\bm w_{\caln_{m'}})
\right\} \\
\leq &
\frac{B^2}{M_s^2} \left\{  \sum_{m = 1}^{M_s} \E_{\pi_s} \left\{ \left[
\frac{h_{m, \alpha}(\bm W_{\caln_m})}{\pi_{s, \caln_m}(\bm W_{\caln_m})}  \right]^2 \right\} +
\sum_{m = 1}^{M_s}
 \sum_{\bm w_{\caln_m}}
h_{m, \alpha}(\bm w_{\caln_m}) \left[ 1 - h_{m, \alpha}(\bm w_{\caln_m}) \right] \right.+ \\
& \hspace{40pt} \left.
\sum_{(m, m') \in \mathcal K_s} \E_{\pi_s} \left\{ \left| 
\frac
{\pi_{s, \caln_m, \caln_{m'}}(\bm W_{\caln_m}, \bm W_{\caln_{m'}})}
{\pi_{s, \caln_m}(\bm W_{\caln_m}) \pi_{s, \caln_{m'}}(\bm W_{\caln_{m'}})}
-1
\right| 
\frac{h_{m, \alpha}(\bm W_{\caln_m})}{\pi_{s, \caln_m}(\bm W_{\caln_m})}
\frac{h_{m', \alpha}(\bm W_{\caln_{m'}})}{\pi_{s, \caln_{m'}}(\bm W_{\caln_{m'}})}
\right\} \right\} \\
\leq & B^2 \left( \frac{\Delta_{s, \alpha}}{M_s} + \frac1{M_s} + \frac{\kappa_s \Gamma_{s, \alpha^*}}{{M_s^2}} \right).
\end{align*}
Therefore, if $M_s \rightarrow \infty$, $\Delta_{s, \alpha} = o(M_s)$ and $\kappa_s \Gamma_{s, \alpha}^* = o(M_s^2)$, then the estimator is consistent.

\end{proof}

\subsection{Estimators of $+K$ estimands}

\begin{proof}[Proof of \cref{theorem:plusK_unbiasedness}]
To build intuition, we first include the proof for $K = 1$, and then we generalize it to any value $K \geq 1$.

\paragraph{For $K = 1$:}
    We study $\E \left[ \overline Y^\textrm{+1} - \frac1M \sum_{m \in \calm} \rho_m(\bm W_{\caln_m}) Y_m \right] $, where the expectation is with respect to the treatment assignment), and show that the weights defined in the theorem satisfy that this expectation is equal to 0. 
    We use $\caln \setminus \caln_m$ to denote the interventional units that are not part of outcome unit $m$'s intervention set, and $\caln_m^c$ for the set of control units in outcome unit $m$'s intervention set.
    \begin{align*}
        \E & \left[ \overline Y^\textrm{+1} - \frac1M \sum_{m \in \calm} \rho_m(\bm W_{\caln_m}) Y_m \right] = \\
        &= \sum_{\bm w} \pi(\bm w) \left[
        \frac{1}{N^c}\sum_{n\in \caln^c} \frac1M \sum_{m\in{\cal M}} Y_m(\bm W + e_{\bm n}) -
        \frac1M \sum_{m \in \calm} \rho_m(\bm W_{\caln_m}) Y_m
        \right] \\
        &= \frac1M \sum_{m \in \calm} \sum_{\bm w} \pi(\bm w) \left\{
        \frac1{N^c} \sum_{n \in \caln^c} Y_m(\bm W + e_{\bm n}) - \rho_m(\bm w_{\caln_m}) Y_m(\bm W_{\caln_m}) 
        \right\} \\
        &= \frac1M \sum_{m \in \calm} \sum_{\bm w_{\caln \setminus \caln_m}} \sum_{\bm w_{\caln_m}} \pi(\bm w) \left\{
        \frac1{N^c} \sum_{n \in \caln_m^c} Y_m(\bm W_{\caln_m} + e_{\bm n}) +
        \left[ \frac{N^c - N^c_m}{N^c}
        - \rho_m(\bm W_{\caln_m}) \right] Y_m(\bm W_{\caln_m})
        \right\} \\
         &= \frac1M \sum_{m \in \calm} \sum_{\bm w_{\caln_m}} \pi_{\caln_m}(\bm w_{\caln_m}) \left\{
        \frac1{N^c} \sum_{n \in \caln_m^c} Y_m(\bm W_{\caln_m} +e_{\bm n}) +
        \left[ \frac{N^c - N^c_m}{N^c}
        - \rho_m(\bm W_{\caln_m}) \right] Y_m(\bm W_{\caln_m})
        \right\}. \numberthis \label{supp_eq:positivity_plusK}
\end{align*}
Since we do not impose any condition across outcomes, to make this expectation equal to 0, the term corresponding to each outcome unit has to be equal to 0. So we drop the summation over $m \in \calm$ from the following calculations. 
We split the sum over $\bm w_{\caln_m}$ to three cases, all control units in $\caln_m$, at least one control unit (but not all controls), and all treated units. Then, this quantity is equal to:
%We also use $\pi_{\caln \setminus \caln_m \mid \bm W_{\caln_m} = \bm w_{\caln_m}}(\bm w_{\caln \setminus \caln_m})$ to denote the conditional distribution on $\caln \setminus \caln_m$ when outcome unit $m$'s intervention set treatment is equal to $\bm w_{\caln_m}.$
\begin{align*}
        &
        \frac{\pi_{\caln_m}(\bm 0)}{N^c}  \sum_{n \in \caln_m^c} Y_m(\bm 0 + e_n) +  \pi_{\caln_m}(\bm 0) \left[ \frac{N^c - N^c_m}{N^c}
        - \rho_m(\bm 0) \right] Y_m(\bm 0) + \\
        &
        \sum_{\substack{\bm w_{\caln_m}:\\ \text{at least one control}\\ \text{not all controls}}} 
        \Bigg\{
        \frac{\pi_{\caln_m}(\bm w_{\caln_m})}{N^c} \sum_{n \in \caln_m^c} Y_m(\bm w_{\caln_m} + e_n) + 
        \pi_{\caln_m}(\bm w_{\caln_m})
        \left[ \frac{N^c - N^c_m}{N^c} - \rho_m(\bm w_{\caln_m}) \right] 
        Y_m(\bm w_{\caln_m})
        \Bigg\} - \\
        & \hspace{20pt}
        \pi_{\caln_m}(\bm 1) \rho_m(\bm 1) Y_m(\bm 1)
\end{align*}
From the above, it becomes evident which potential outcomes show up in which terms. For example, the potential outcome under all controls $Y_m(\bm 0)$ only appears once in the summation. We re-write this expression by studying ``how often'' each of the potential outcomes appears across terms, as
\begin{align*}
        & \hspace{57pt} \Bigg[ \sum_{\substack{\bm w_{\caln_m}^*: \\ \text{all treated} \\ \text{except one}}}
          \frac{\pi_{\caln_m}(\bm w_{\caln_m}^*)}{N^c} - \pi_{\caln_m}(\bm 1) \rho_m(\bm 1) \Bigg]
        Y_m(\bm 1) + \\
        &  \sum_{\substack{\bm w_{\caln_m}: \\ \text{at least one} \\  \text{control, }\\ \text{not all control}}} 
        \Bigg\{
        \sum_{\substack{\bm w_{\caln_m}^{*1}: \\
        \text{equal to $\bm w_{\caln_m}$}\\\text{with one fewer} \\ \text{treated unit}}}
        \frac{\pi_{\caln_m}(\bm w_{\caln_m}^{*1})}{N^c} +
        \pi_{\caln_m}(\bm w_{\caln_m})
        \left[ \frac{N^c - N^c_m}{N^c} - \rho_m(\bm w_{\caln_m}) \right] 
        \Bigg\} Y_m(\bm w_{\caln_m}) + \\
        & \hspace{57pt} \Bigg\{
        \pi_{\caln_m}(\bm 0) \left[ \frac{N^c - N^c_m}{N^c}
        - \rho_m(\bm 0) \right]
        \Bigg\} Y_m (\bm 0) \\
        %%%%%
        &= 
        \sum_{\bm w_{\caln_m}} 
        \Bigg\{
        \sum_{\substack{\bm w_{\caln_m}^{*1}: \\
        \text{equal to $\bm w_{\caln_m}$}\\\text{with one fewer} \\ \text{treated unit}}}
        \frac{\pi_{\caln_m}(\bm w_{\caln_m}^{*1})}{N^c} +
        \pi_{\caln_m}(\bm w_{\caln_m})
        \left[ \frac{N^c - N^c_m}{N^c} - \rho_m(\bm W_{\caln_m}) \right] 
        \Bigg\} Y_m(\bm W_{\caln_m}).
    \end{align*}
We solve the equation that the coefficient of each $Y_m(\bm w_{\caln_m})$ is equal to 0:
\begin{align*}
    \sum_{\substack{\bm w_{\caln_m}^*: \\
        \text{equal to $\bm w_{\caln_m}$}\\\text{with one fewer} \\ \text{treated unit}}}
        \frac{\pi_{\caln_m}(\bm w_{\caln_m}^*)}{N^c} =&
        \pi_{\caln_m}(\bm w_{\caln_m})
        \left[ \rho_m(\bm w_{\caln_m}) - \frac{N^c - N^c_m}{N^c} \right] \\
      \iff  
        \rho_m(\bm W_{\caln_m}) &= 
        \frac1{N^c} \left[ N^c - N^c_m + \dfrac{\sum_{\bm w_{\caln_m}^*}
       \pi_{\caln_m}(\bm w_{\caln_m}^*)}{\pi_{\caln_m}(\bm w_{\caln_m})} \right]
\end{align*}
So the specified weights indeed lead to unbiased estimation of $\overline Y^{+1}$.

\vspace{5pt}
\noindent
\underline{Positivity assumption and \cref{supp_eq:positivity_plusK}:} We see that, if $\bm w_{\caln_m}$ is a possible treatment vector for outcome unit $m$'s intervention set under the treatment assignment $\pi$, then the bias will include terms that involve $Y_m(\bm w_{\caln_m} + e_n)$. Therefore, for the bias to be zero, we need to require that $\bm w_{\caln_m}' = \bm w_{\caln_m} + e_n$ is also possible under $\pi$. Iteratively, we would require that, if $\bm w_{\caln_m}$ is such that $\pi_{\caln_m}(\bm w_{\caln_m}) > 0$, then $\pi_{\caln_m}(\bm w_{\caln_m}') > 0$ for all vectors $\bm w_{\caln_m}'$ such that $[\bm w_{\caln_m}']_j = 1$ if $[\bm w_{\caln_m}]_j = 1$. This condition is essentially a positivity constraint: any treatment vector that is possible under the intervention is also required to be possible under the realized treatment assignment.

\paragraph{For a general $K \geq 1$:} The proof is very similar here, except we have to be careful to keep track of all terms that include potential outcomes corresponding to the same treatment. The set $\caln_m^c$ denotes the control units that are connected to $m$, so $\caln^c \setminus \caln_m^c$ denotes the control units that are not connected to $m$. First, note that we can write
\begin{align*}
    \E\left[ \overline Y^{+K} \right] &=
    \E \left[ \frac{1}{\binom{N^c}{K}}\sum_{\bm n} \frac1M \sum_{m\in \calm} Y_m(\bm W+e_{\bm n}) \right] \\
&= \frac1M \sum_{m \in \calm}  \E \Bigg[ \frac{1}{\binom{N^c}{K}} \sum_{k = 0}^K \
\sum_{\substack{\bm n_m:\phantom{'} \\ k \text{ units} \\ \text{in } \caln_m^c}}  \ 
\sum_{\substack{\bm n_m': \\ K - k \text{ units} \\ \text{in } \caln^c \setminus \caln_m^c }}
Y_m(\bm W+e_{\bm n})  \\
&=
\frac1M \sum_{m \in \calm}  \E \Bigg[ \frac{1}{\binom{N^c}{K}} \sum_{k = 0}^K 
\binom{N^c - N^c_m}{K - k}
\sum_{\substack{\bm n_m: \\ k \text{ units} \\ \text{in } \caln_m^c}} 
Y_m(\bm W_{\caln_m} +e_{\bm n_m})
\Bigg],
\end{align*}
where we have split the sum over $K$ additional treated units in $\bm n$ in summing over vectors of units such that $l = 0, 1, \dots, K$ of them are connected to $m$ and the rest are not, and noted that which of the units in $\caln^c \setminus \caln_m^c$ are treated does not affect the outcome for unit $m$. Based on this, we write
\begin{align*}
    \E & \left[ \overline Y^\textrm{+K} - \frac1M \sum_{m \in \calm} \rho_m(\bm W_{\caln_m}) Y_m \right] = \\
    &= \frac1M \sum_{m \in \calm} \E \Bigg[ \frac{1}{\binom{N^c}{K}} \sum_{k = 0}^K  \binom{N^c - N^c_m}{K - k}
    \sum_{\substack{\bm n_m: \\ k \text{ units} \\ \text{in } \caln_m^c}} 
    Y_m(\bm W_{\caln_m} +e_{\bm n_m}) - \rho_m(\bm W_{\caln_m}) Y_m(\bm W_{\caln_m}) \Bigg]
\end{align*}
Since we do not impose any restrictions on the potential outcomes, for this quantity to be zero, the term corresponding to each $m \in \calm$ has to be equal to 0. Then, we re-write the expectation as
\begin{align*}
    & \sum_{\bm w_{\caln \setminus \caln_m}} \sum_{\bm w_{\caln_m}} \pi(\bm w) \Bigg[ \frac{1}{\binom{N^c}{K}} \sum_{k = 0}^K  \binom{N^c - N^c_m}{K - k}
    \sum_{\substack{\bm n_m: \\ k \text{ units} \\ \text{in } \caln_m^c}} 
    Y_m(\bm W_{\caln_m} +e_{\bm n_m}) - \rho_m(\bm W_{\caln_m}) Y_m(\bm W_{\caln_m}) \Bigg] \\
    &= \sum_{\bm w_{\caln_m}} \pi_{\caln_m}(\bm w_{\caln_m}) \Bigg[ \frac{1}{\binom{N^c}{K}} \sum_{k = 0}^K  \binom{N^c - N^c_m}{K - k}
    \sum_{\substack{\bm n_m: \\ k \text{ units} \\ \text{in } \caln_m^c}} 
    Y_m(\bm W_{\caln_m} +e_{\bm n_m}) - \rho_m(\bm W_{\caln_m}) Y_m(\bm W_{\caln_m}) \Bigg],
\intertext{where here we used that the treatment assignment has a fixed number of treated units, so $N^c$ is not affected by the choice of $\bm w$,}
    &= \sum_{\bm w_{\caln_m}} Y_m(\bm W_{\caln_m}) \Bigg[
    \frac{1}{\binom{N^c}{K}} \sum_{k = 0}^K \binom{N^c - N^c_m}{K - k} \sum_{\bm w_{\caln_m}^{*k}} \pi_{\caln_m}(\bm w_{\caln_m}^{*k}) - \rho_m(\bm W_{\caln_m}) \pi_{\caln_m}(\bm w_{\caln_m})
    \Bigg],
\end{align*}
where the last equation is acquired by realizing that each potential outcome $Y_m(\bm w_{\caln_m})$ appears in the summation as itself, or as $Y_m(\bm w_{\caln_m}^{*k} +e_{\bm n_m})$ for values of $k = 0, 1, \dots, K$, any vector $\bm w_{\caln_m}^{*k}$ and appropriately chosen $\bm n_m$ of length $k$.

Solving for $\rho_m(\bm w_{\caln_m})$ when setting the equation above equal to zero, the weights in the theorem are the only weights that lead to unbiased estimation of $\overline Y^{+K}$.
\end{proof}

\section{Application results under alternative graph structure}
\label{supp_sec:application-second-degree}

Here, we discuss results from our application under a second-degree bipartite graph.

\subsection{The second-degree bipartite graph}

For the second-degree network, an HSR line $i$ is connected to a prefecture $j$ if there exists a prefecture $j'$ which is first-degree connected with $i$ and $j'$ shares some stations with line $j$. 
We report descriptive statistics for the graph in \cref{tab:graph_description-2nd}. (Quantities in parentheses are the corresponding descriptive statistics from the first-degree graph in \cref{sec:application}.) 
9\% (1.43\%) of all intervention and outcome unit pairs are connected. The average number of prefectures connected to an HSR line is 23.93 (3.81) and the average number of HSR lines connected to a prefecture is 13.5 (2.14). As expected, the second-degree graph is much denser than the first-degree graph.

\begin{table}[!t]
    \centering
    \begin{tabular}{lc}
        \hline
        \% of graph entries equal to 1  & 9.00 \\
        average size of intervention sets & 13.50 \\
        average size of outcome sets & 23.93 \\
        \% of HSR lines connected to only two prefectures & 0.67 \\
        \% of prefecture connected to only one HSR line & 0.75 \\
        maximum number of HSR lines connected to a prefecture & 40 \\
        maximum number of prefectures connected to a HSR line & 63 \\
        number of outcome units with all treated HSR lines & 10\\
        number of outcome units with all control HSR lines & 5\\
        \hline
    \end{tabular}
    \caption{Descriptive statistics of the second-degree bipartite graph between intervention units (HSR lines) and outcome units (prefectures).}
    \label{tab:graph_description-2nd}
\end{table}

Since the second-degree graph is more dense, it also leads to a higher number of outcome units with positivity violations, reported in \cref{tab:posviolations-2nd}.
%As expected, more violations are observed in the denser, second-degree graph, compared to the first-degree graph.
With the second-degree graph,  127 units are excluded from the estimation of the all-or-none effect because of positivity violation under the complete randomization design. From the remaining outcome units, only 6 units have all treated connected lines and 3 units have all connected lines under control. Therefore, only 9 units are used in total for the estimation of the all-or-none effect.

\begin{table}[!t]
    \centering
    \resizebox{\textwidth}{!}{
    \begin{tabular}{c|ccccccc}
        \makecell{Design} & All-or-None & \makecell{All-vs-\\Status Quo} & \makecell{Status Quo-vs-\\None} & +1 & Stochastic 1 & Stochastic 2 & Stochastic 3\\
        \hline
%        CRD (1) & 24 & 0 & 24 & 0 & 0 & 0 & 0\\
        CRD & 127 & 7 & 127 & 7 & 7 & 7& 0\\
        % \hline
        % Bernoulli (1) & 18 & 0 & 18 & 0 & 0 & 0& 0\\
        Bernoulli & 52 & 0 & 52 & 0 & 0 & 0& 0\\
    \end{tabular}}
    \caption{Number of outcomes units violating positivity for the various estimands under different experimental designs and the second-degree bipartite graph.}
    \label{tab:posviolations-2nd}
\end{table}

\subsection{Results}

\cref{tab:tests2degree} include the results from the randomization tests under the second-degree bipartite graph, for both experimental designs, and all three test statistics (two at the outcome unit level, and one at the interventional unit level). The corresponding results based on the first-degree graph are given in \cref{tab:tests1degree}. P-values from the randomization tests based on the second-degree graph are consistently higher than the corresponding ones based on the first-degree graph. This can be explained if the first-degree graph more accurately describes the interference patterns in our application.

\begin{table}[!t]
    \centering
    \begin{tabular}{ccccc}
      Design & Unit level & \makecell{Test\\statistic} &\makecell{outcomes are \\the percentage\\ (\%) variation}  & \makecell{outcomes are \\the log\\ difference} \\
         
        \hline
       & Outcome & $W^{tot}$ & 0.156&0.163\\
      CRD & Outcome & $ \overline W$ & 0.133&0.139 \\
        & Intervention & Difference & 0.088&0.081 \\
        \hline
       & Outcome & $W^{tot}$ & 0.178&0.189 \\
      Bernoulli & Outcome &  $\overline W$ & 0.131&0.136 \\
       & Intervention & Difference & 0.248&0.240 \\
        \hline
    \end{tabular}
    \caption{Randomization tests for Fisher's sharp null. Results correspond to the two definitions of the outcome along the columns, and the two designs (CRD, Bernoulli) and test statistics along the rows. Results correspond to the second-degree bipartite graph.}
    \label{tab:tests2degree}
\end{table}

%     \begin{table}[h]
%     \centering
%     \begin{tabular}{ccc}
%         &\makecell{outcomes are \\the percentage\\ (\%)variation}  & \makecell{outcomes are \\the log\\ difference} \\

%         \hline
%         $W^{tot}$+CRD+2nd degree&0.156&0.163\\
%         $W^{bar}$+CRD+2nd degree&0.133&0.139\\
%         Intervention+CRD+2nd degree&0.088&0.081\\
%         $W^{tot}$+Bernoulli+2nd degree&	0.178&0.189\\
%         $W^{bar}$+Bernoulli+2nd degree&0.131&0.136\\
%         Intervention+Bernoulli+2nd degree&0.248&0.240\\
%         \hline
%     \end{tabular}
%     \caption{Randomization tests for Fisher's sharp null - 2nd degree graph}
%     \label{tab:tests2degree}
% \end{table}

\begin{table}[!t]
    \centering
    \begin{tabular}{cccccccc}
         \makecell{Design} & All-or-None & \makecell{All-vs-\\Status Quo} & \makecell{Status Quo-vs-\\None} & +1 & $\tau^{(1)}_\text{stoch}$ & $\tau^{(2)}_\text{stoch}$ & $\tau^{(3)}_\text{stoch}$\\
         \hline
         \\
         \multicolumn{8}{c}{Outcome: Difference of log employment between 2007 and 2016} \\
        \cmidrule{2-7}  
         CRD &0.258 & -0.222 &  0.322 & 0.016&-0.158&0.046&0.198\\
         & (0.199) & (0.032) & (0.140) &&&\\
         \hline
         Bernoulli & 0.135&-0.233&0.329& 0.017&-0.162&0.001&0.111\\
         & (0.079) & (0.030) & (0.007) &&(0.103)&(0.246)&(0.104)\\
          \hline \\
         \multicolumn{8}{c}{Outcome: Percent variation of employment between 2007 and 2016} \\
        \cmidrule{2-7} 
         CRD &26.915 & -29.167 &  35.200 & 2.196 &-21.166&8.146&20.600\\
         & (20.423) &    (4.098) &   (12.163) \\
         \hline
         Bernoulli & 14.209 &  -30.473 &  39.576 &    2.233 &-21.416 & 1.409&11.781 \\
         & (7.768) &    (3.825)  &  (6.141) &&(13.695)&(36.579)&(12.072)\\
        \end{tabular}
    \caption{Estimates and standard errors (in parentheses) for the effects (columns) of HSR line completion on employment. The outcome is measured as difference in log employment, and as percent variation in employment between 2007 and 2016 (big rows). Results are reported under the different experimental designs (rows). Estimates are obtained removing units that violate positivity for each of the estimands separately. The results correspond to the second-degree bipartite graph.}
    
    \label{tab:application-estimates-2nd}
\end{table}

\begin{table}[!t]
    \centering
    \begin{tabular}{cccccccc}
         \makecell{Design} & All-or-None & \makecell{All-vs-\\Status Quo} & \makecell{Status Quo-vs-\\None} & +1 & $\tau^{(1)}_\text{stoch}$ & $\tau^{(2)}_\text{stoch}$ & $\tau^{(3)}_\text{stoch}$\\
         \hline
         \\
         \multicolumn{8}{c}{Outcome: Difference of log employment between 2007 and 2016} \\
        \cmidrule{2-7}  
        CRD  & 0.258 & -0.064 &  0.322 & 0.008&-0.075 &0.045&0.198\\
         & (0.199) & (0.065) & (0.140) &&&
        \\
         \hline
         Bernoulli & 0.135&-0.194&0.329& 0.015&-0.137&0.087&0.111\\
         & (0.079) & (0.039) & (0.070) &&(0.108)&(0.273)&(0.114)
         \\
          \hline \\
         \multicolumn{8}{c}{Outcome: Percent variation of employment between 2007 and 2016} \\
        \cmidrule{2-7} 
          CRD  &26.915 & -8.286 &  35.200 &  1.001 &-9.898   &5.069&20.600
          \\
          & (20.423) &    (8.401) &   (12.163) \\
         \hline
         Bernoulli & 14.209 &  -25.367 &  39.576 &    2.024 &-18.120 & 13.186 &11.781 \\
         & (7.768) &    (4.900)  &  (6.141)&&(14.599)& (40.286)&(13.037)
        \end{tabular}
    \caption{Estimates and standard errors (in parentheses) for the effects (columns) of HSR line completion on employment. The outcome is measured as difference in log employment, and as percent variation in employment between 2007 and 2016 (big rows). Results are reported under the different experimental designs (rows). Estimates are obtained removing units that violate positivity for any estimand, resulting in 110 units under CRD and and 185 under Bernoulli design. The results correspond to the second-degree bipartite graph.}
    
    \label{tab:application-estimates-union-2nd}
\end{table}

For the estimates of the different causal estimands, qualitative results are comparable with those under the first-degree bipartite graph. The main difference lies in the all-vs-status quo effects that are negative under the second-degree bipartite graph, suggesting that the completion of the remaining lines would have a detrimental effect on employment growth.

For the results under the second-degree bipartite graph, the estimators for the estimands under stochastic interventions are missing estimates of the variance upper bound, whereas the same quantity is available for the first-degree graph. That is because the terms $\lambda, \lambda^\tau$ in the variance upper bounds in \cref{subsec:stochastic-estimation} require enumeration of all possible treatment vectors for each outcome unit's intervention set. For the first-degree graph, the maximum size of intervention sets is 9 (see \cref{tab:graph_description}), whereas for the second-degree graph, the maximum size of intervention sets is 40 (see \cref{tab:graph_description-2nd}), rendering the enumeration of all possible treatment vectors for each outcome unit computationally infeasible.

\end{document}